\documentclass[11pt,a4paper]{article}
\usepackage{longtable}
\usepackage{amsmath}
\usepackage{amssymb}
\usepackage{amsfonts}
\usepackage{graphicx}
\usepackage{bm}
\usepackage{footmisc}
\usepackage{afterpage}
\usepackage{float}
\usepackage{lscape}
\usepackage{longtable}
\usepackage{float}

\newcommand*{\no}{\noindent}
\newcommand*{\bea}{\begin{eqnarray}}
\newcommand*{\eea}{\end{eqnarray}}
\newcommand*{\be}{\begin{equation}}
\newcommand*{\ee}{\end{equation}}
\newcommand*{\pd}{\partial}
\newcommand*{\pdm}{\pd_{\mu}}

\newcommand*{\pref}[1]{(\ref{#1})}

\newcommand*{\mn}{{\mu\nu}}

\newcommand*{\nn}{\nonumber}

\newcommand*{\tr}{\mathrm{tr}}

\newcommand{\bma}{\begin{pmatrix}}
\newcommand{\ema}{\end{pmatrix}}

\usepackage[square,comma,sort&compress,numbers]{natbib}

\title{Dependence of the propagators on the sampling of Gribov copies inside the first Gribov region of Landau gauge}

\author{Axel Maas\\
Institute of Physics, NAWI Graz, University of Graz,\\
Universit\"atsplatz 5, A-8010 Graz, Austria}

\begin{document}

\maketitle

\begin{abstract}

Beyond perturbation theory the number of gauge copies drastically increases due to the Gribov-Singer ambiguity. Any way of treating them defines, in principle, a new, non-perturbative gauge, and the gauge-dependent correlation functions can vary between them. Herein various such gauges will be constructed as completions of the Landau gauge inside the first Gribov region. The dependence of the propagators and the running coupling on these gauges will be studied for SU(2) Yang-Mills theory in two, three, and four dimensions using lattice gauge theory, and for a wide range of lattice parameters. While the gluon propagator is rather insensitive to the choice, the ghost propagator and the running coupling show a stronger dependence. It is also found that the influence of lattice artifacts is larger than in minimal Landau gauge.

\end{abstract}

\section{Introduction}

Gauge-dependent correlation functions have proven a valuable tool in constructing gauge-invariant physics \cite{Alkofer:2000wg,Fischer:2006ub,Binosi:2009qm,Maas:2011se,Boucaud:2011ug,Vandersickel:2012tg,Pawlowski:2005xe,Gies:2006wv}. However, as their name implies, they depend on the chosen gauge. Already in perturbation theory, this dependence is qualitative, and can be even more so beyond perturbation theory. Hence, as long as gauge-dependent quantities are combined or compared it is necessary to do so in the same gauge.

This appears trivial at first, and indeed is so in perturbation theory \cite{Henneaux:1992ig}. Beyond perturbation theory, this is no longer the case. Due to the Gribov-Singer ambiguity \cite{Gribov:1977wm,Singer:1978dk,vanBaal:1997gu,vanBaal:1991zw,Dell'Antonio:1991xt}, gauges can no longer be specified only by the same conditions as in perturbation theory, and supplemental conditions are required. The reason is that beyond perturbation theory, which is essentially a small field expansion, gauge copies differing by non-infinitesimal gauge transformations exist. Thus, after fixing a perturbative gauge condition, usually many more gauge copies remain, the so-called Gribov copies \cite{Gribov:1977wm}, forming the residual gauge orbit \cite{Maas:2011se}.

This necessitates to somehow specify how to treat these additional copies. This problem differs for every perturbative gauge condition, and so far no fully general and practical prescription is known. Hence, in the following only the case emerging from the perturbative Landau gauge will be considered.

Thus, after fixing the perturbative Landau gauge, there remains a set of Gribov copies. It needs to be specified how to treat them. Just like in the case of perturbation theory \cite{Bohm:2001yx}, any such treatment is essentially a prescription on how to average correlation functions over the residual gauge orbit. The weight can be anything from an average over the full or residual gauge orbit \cite{Hirschfeld:1978yq,Kalloniatis:2005if,vonSmekal:2007ns,vonSmekal:2008en,Fischer:2008uz,Maas:2012ct,Parrinello:1990pm,Fachin:1991pu,Fachin:1993qg,Serreau:2012cg,Serreau:2013ila,Mehta:2009zv,Neuberger:1986xz,vonSmekal:2008es}, a subset of the residual gauge orbit \cite{Maas:2011se,Maas:2013vd} or a $\delta$-function-like weight to select a single representative for each gauge orbit \cite{Maas:2011se,Cucchieri:1997dx,Vandersickel:2012tg,Dudal:2014rxa,Dudal:2008sp,Boucaud:2011ug,Schaden:2013ffa,Bogolubsky:2005wf,Bogolubsky:2007bw,Bornyakov:2008yx,Bornyakov:2010nc,Dudal:2009xh,Maas:2009se,Maas:2008ri,Maas:2009ph,Silva:2004bv,Schaden:2014bea,Sternbeck:2012mf,Tissier:2011ey,Zwanziger:1993dh,LlanesEstrada:2012my,Henty:1996kv}. The latter is the choice most similar to the perturbative Landau gauge \cite{Bohm:2001yx}.

Of course, the question remains, whether this is indeed necessary, and whether there are differences for the gauge-dependent correlation functions between the different choices. Furthermore, constructing explicit weights is non-trivial in the continuum \cite{Maas:2011se,Fischer:2008uz,Serreau:2012cg,Serreau:2013ila}, while on a lattice an explicit average can be performed \cite{Maas:2011se,Parrinello:1990pm,Fachin:1991pu,Fachin:1993qg}. Thus, to compare continuum and lattice results requires to understand the relation between the lattice version of a gauge and the continuum version of a gauge.

Hence, in the following a particular subclass of gauges which average over part of the residual gauge orbit, the so-called first Gribov region, will be considered. The precise choices studied here will be presented in section \ref{s:gauges}. The elementary propagators, the gluon and ghost propagator, as well as the running coupling, being gauge-dependent \cite{Deur:2016tte}, in the miniMOM scheme \cite{vonSmekal:2009ae} will be determined for these gauges. Their dependence on the gauge choices will then be analyzed in sections \ref{s:gluon}, \ref{s:ghost}, and \ref{s:alpha}. This will be done in two, three, and four dimensions and for a range of lattice parameters, as these objects tend to be rather sensitive to lattice artifacts \cite{Maas:2011se}. The technical details of the lattice simulations can be found in section \ref{s:tech} and appendix \ref{a:tech}.

Note that the calculation of the propagators in these diverse gauges is very expensive. Thus, even though this is the first systematic investigation of a large class of gauges simultaneously, it cannot be considered to be a final answer, especially in four dimensions. This is highlighted by the study of lattice artifacts in section \ref{s:artifacts}. Still, the results are quite intriguing, and should motivate further investigations, once substantially more computational resources for this purpose become available. This is summarized in section \ref{s:sum}.

Some preliminary results can be found in \cite{Maas:2009se,Maas:2011ba,Maas:2011se,Maas:2013vd,Maas:2016frv}, and the following relies heavily on results published in \cite{Maas:2015nva}.

\section{Technical setup}\label{s:tech}

In the following the standard SU(2) Wilson action is employed. The simulations are performed for $d=2,3,4$ on symmetric lattices of size $N^d$ for bare gauge couplings $\beta$, using a mixture of overrelaxation and heat-bath sweeps \cite{Cucchieri:2006tf}. The lattice spacing is set by assigning the string tension a value of $(440$ MeV$)^2$, as described in \cite{Cucchieri:2006tf,Maas:2008ri,Maas:2014xma}. The sets of lattice parameters can be found in appendix \ref{a:tech} in table \ref{tcgf}.

To obtain Gribov copies, every decorrelated configuration is gauge-fixed to a gauge copy which satisfies the perturbative Landau gauge condition and in addition is inside the first Gribov region, i.\ e.\ having a positive-semidefinite spectrum of the Faddeev-Popov operator
\be
M^{ab}=-\pdm D_\mu^{ab}=-\pd^2\delta^{ab}+gf^{abc} A_\mu^c=-D_\mu^{ab}\pdm\nn.
\ee
\no This is done using an adaptive stochastic overrelaxation algorithm \cite{Cucchieri:2006tf}. This gauge-fixing is repeated for every configuration a number of times listed in table \ref{tcgf} in appendix \ref{a:tech}, initialized each time with a random seed \cite{Cucchieri:1997dx}. It is assumed that this algorithm finds every Gribov copy inside the first Gribov region with the same probability. Though there are no indications to the contrary, there is neither for this algorithm, nor for any unbiased\footnote{It should be noted that there exists a number of algorithms optimized to fix to the so-called absolute Landau gauge \cite{Silva:2004bv,Oliveira:2003wa,Yamaguchi:1999hp,Yamaguchi:1999hq,Yamaguchi:1998yz,Bogolubsky:2005wf,Bogolubsky:2007bw,Silva:2004bv,Oliveira:2003wa,Maas:2008ri,Sternbeck:2007ug,Bornyakov:2009ug,Bornyakov:2013pha,Fachin:1991pu,Fachin:1993qg}, the definition of which will be given in section \ref{s:absolute}.} other algorithms \cite{Cucchieri:1995pn}, yet a proof available demonstrating that this assumption is valid.

Note that every gauge orbit has at least one gauge copy satisfying these conditions in principle \cite{Dell'Antonio:1991xt} and actually usually much more than one in practice \cite{Maas:2015nva}. Thus, the algorithm is guaranteed to succeed.

For every Gribov copy the color-averaged and, for the gluon, space-time-index averaged gluon propagator $D$ and ghost propagator $D_G$ are calculated, using the methods described in \cite{Cucchieri:2006tf}. Since here only the low-momentum behavior will turn out to be interesting, all propagators are only calculated along edge momenta, i.\ e.\ parallel to one of the lattice axises, with only one non-vanishing component. These show the least sensitivity at low momenta to lattice artifacts \cite{Maas:2014xma}.

There is, of course, a probability that the same Gribov copy is found more than once. Based on the results in \cite{Maas:2015nva}, two Gribov copies will be considered distinct if they have numerically different values for the two quantities
\bea
F&=&-\frac{1}{VdN_c}\int d^d x A_\mu^a A_\mu^a=1-\frac{1}{VdN_c}\sum_{x,\mu}\tr U_\mu(x)\label{f}-{\cal O}(a^2)\\
b&=&\tilde{Z}_3\frac{1}{dN_c}\sum_i G\left(p(i)_\text{min}^2\right)\label{b},
\eea
\no where in the first line $U$ are the link variables, $N_c=2$ is the number of colors, and $A_\mu^a$ are the gauge fields. Note that in the actual lattice calculation the sum over the traces of $U_\mu$ is used to define $F$. The negative ${\cal O}(a^2)$ correction is needed to remove the lattice correction to get the continuum expression. In the second line \pref{b} $G=p^2D_G$ is the ghost dressing function. For the lattice calculations, it was determined as described in \cite{Cucchieri:2006tf}. The $p(i)_\text{min}^2$ are the lowest, non-zero momenta along a coordinate axis $i$ on a given lattice. The actual value of this momentum is listed in table \ref{tcgf} for every lattice setting. $\tilde{Z}_3$ is the ghost renormalization constant, which is usually determined at $\mu=2$ GeV, but is fixed for every set of lattice parameters. At this momenta it will be found that the effect of selecting different gauges becomes negligible compared to the statistical error. A detailed motivation why this is a necessary, but probably not sufficient, criterion for the distinction of Gribov copies can be found in \cite{Maas:2015nva}.

Note that the number of actual distinct Gribov copies found in this way can potentially differ between different gauge orbits. This distribution is rather peaked, with a width decreasing quickly with physical volume, especially in dimensions larger than two \cite{Maas:2015nva}. This slight mismatch of the number of Gribov copies found can thus be considered an additional finite-volume artifact. The dependence of the results on the number of sampled distinct Gribov copies in sections \ref{s:gluon}-\ref{s:alpha} show no strong dependence towards the edge of the search space on this problem, and it is therefore a minor effect. Of course, averages over Gribov copies will still incorporate the difference in number.

\section{Definition of gauges}\label{s:gauges}

In the sections \ref{s:gluon}-\ref{s:alpha} the correlation functions will be determined for a set of different gauges, which are defined in the following.

\subsection{Minimal Landau gauge}

The first is the minimal Landau gauge, which is the one used in most (lattice) calculations. It is based on the random choice of a Gribov copy \cite{Maas:2011se}. This should be equivalent to an average with flat weight over the first Gribov region, if no bias exists \cite{Maas:2011se,Maas:2013vd,Langfeld:2004vu}. Both cases will be compared here, and indeed they agree. This supports the assumption that there is no bias build into the algorithm used to search for Gribov copies, though this is at best circumstantial evidence.

This gauge therefore probes the average structure of the first Gribov region. In terms of the coordinates $F$ and $b$, this will be dominated by the most populated area of the first Gribov region. Since the Gribov region shows a broad, but at larger volumes singly-peaked, structure in the interior in terms of the two coordinates \cite{Maas:2015nva}, this will generate also average values for the gauge-fixed correlation functions. Since in the following often ratios of the results in minimal Landau gauge and in other gauges will be computed, results in minimal Landau gauge will be denoted by a subscript ``m''.

\subsection{Absolute and inverse Landau gauge}\label{s:absolute}

The second is the so-called absolute Landau gauge \cite{Maas:2011se}, based\footnote{In the Abelian case, where there are no continuum Gribov copies, this is the best choice to avoid the impact of lattice Gribov copies  \cite{Bornyakov:1993yy,deForcrand:1994mz}.} on the absolute minimum of \pref{f}, see e.\ g.\ \cite{Cucchieri:1997dx,Vandersickel:2012tg,Dudal:2014rxa,Dudal:2008sp,Boucaud:2011ug,Schaden:2013ffa,Bogolubsky:2005wf,Bogolubsky:2007bw,Bornyakov:2008yx,Bornyakov:2010nc,Dudal:2009xh,Fachin:1991pu,Fachin:1993qg,Fischer:2008uz,Maas:2009se,Maas:2008ri,Maas:2009ph,Schaden:2014bea,Sternbeck:2012mf,Tissier:2011ey,Zwanziger:1993dh,LlanesEstrada:2012my,Henty:1996kv}. An interesting alternative to it will be as the third option the inverse Landau gauge, which attempts to maximize \pref{f} among all possible minima \cite{Maas:2009se,Silva:2004bv}. It therefore searches for the shallowest possible minimum, which is not necessarily unique. Note that it needs to be a minimum to stay within the first Gribov region, as the Hessian of \pref{f} is just the Faddeev-Popov operator. There is no known particular structure associated with this condition, but it is a possible definition of a gauge. In both cases, if multiple minima fulfill the condition an average over them will be performed.

Of course, any such extremalization can be approximative at best, as any available gauge-fixing algorithm cannot guarantee to find all, and thus also the global, extrema. In the present case, this will be performed by choosing the most extreme copy among the ones generated, the so-called multi-restart approach \cite{Cucchieri:1997ns,Bakeev:2003rr,Cucchieri:1997dx}. For alternative ways to fix to this gauge see \cite{Maas:2008ri,Bogolubsky:2005wf,Bogolubsky:2007bw,Silva:2004bv,Oliveira:2003wa,Yamaguchi:1999hp,Silva:2004bv,Oliveira:2003wa,Sternbeck:2007ug,Fachin:1991pu,Fachin:1993qg}.

\subsection{Extreme Landau-$b$ gauges}

In analogy, the so-called maximal and minimal Landau-$b$ gauges will be constructed by extremalizing \pref{b} \cite{Maas:2009se,Maas:2011se,Dudal:2014rxa}. Again, they will be created by choosing the extreme Gribov copies within the generated set, but this time with respect to the coordinate $b$. Note that these gauges should, in the continuum and infinite-volume limit, coincide with gauges which extremalize the lowest eigenvalue of the Faddeev-Popov operator \cite{Sternbeck:2012mf}, as then $b$ appears to be essentially determined by this lowest eigenvalue \cite{Cucchieri:2008fc}. On small volumes this is not the case \cite{Sternbeck:2005vs}.

\subsection{Weighted gauges}

Due to the trapezoidal shape of the first Gribov region in the coordinates $F$ and $b$ \cite{Maas:2015nva} and the general convexity \cite{Zwanziger:1993dh}, the four extremal gauges defined above will probe particular boundaries along the coordinate axes.

Though these extreme constructions are interesting concerning the boundaries of the first Gribov region, the so-called Gribov horizon, they require in the present definition to select Gribov copies essentially by hand\footnote{In case more than one Gribov copy satisfies the criteria, the first one generated will be picked, thus generating, as in the minimal Landau gauge case, a flat average over all such Gribov copies. As noted in \cite{Maas:2015nva}, this happens much more likely for gauges selected with respect to $F$ than for $b$. Requiring both $F$ and $b$ to be different leads for the present lattice setups only at the sub-per-mill level to this problem.}. Though in a numerical lattice calculation this is straightforward \cite{Maas:2015nva}, this is far from trivial in continuum approaches \cite{Maas:2011se}. There are some investigations which suggest that it may be possible to approximate some of the boundaries in the continuum by the inclusion of auxiliary fields \cite{Vandersickel:2012tg,Zwanziger:2011yy,Zwanziger:1993dh} or boundary conditions \cite{Fischer:2008uz,Maas:2011se}. However, a proven construction without approximations would be desirable.

Since Gribov copies are nothing but ordinary gauge copies, one possibility which suggests itself is to extend the concept of covariant gauges in perturbation theory \cite{Bohm:2001yx} and average over Gribov copies \cite{Hirschfeld:1978yq,Fujikawa:1978fu}, with some suitable weight function \cite{Maas:2011se,Maas:2013vd,Fachin:1991pu,Fachin:1993qg,Mehta:2009zv,Neuberger:1986xz,vonSmekal:2008en,vonSmekal:2008es,vonSmekal:2007ns,Serreau:2012cg,Parrinello:1990pm,Serreau:2013ila}.

To test this idea, this will be done here with the weight function
\bea
w(\xi,\zeta)&=&\Theta(-\pdm D_\mu^{ab})\times\label{zetaxi}\\
&&\times\exp\left({\cal N}+\frac{\xi}{V}\int d^dxd^dy\pdm^x\bar{c}^a(x)\pdm^yc^a(y)-\frac{\zeta}{V}\int d^dx A_\mu^a A_\mu^a\right)\nn,
\eea
\no which is inserted in addition to the usual Landau-gauge term into the path integral. The functional $\Theta$-function ensures to be within the first Gribov region. It is interpreted as a product of ordinary $\theta$-functions for all eigenvalues of the Faddeev-Popov operator as its argument. It is also required to satisfy $\Theta(0)=1$, to count Gribov copies on the boundary with the same weight as those in the interior\footnote{In \cite{Vandersickel:2012tg,Zwanziger:2011yy,Zwanziger:1993dh}, the approximation was to replace this $\theta$-functions by $\delta$-functions, to restrict to Gribov copies on the Gribov horizon. This ignores the possibility that gauge orbits could intersect the Gribov horizons multiple times, including the possibility of different numbers of intersections per gauge orbit.}. The parameters $\xi$ and $\zeta$ act as additional gauge-parameters. ${\cal N}$ is a, potentially orbit-dependent, normalization to count Gribov copies. This constant is necessary to ensure that gauge-invariant observables remain so, since then
\be
\langle{\cal O}\rangle=\frac{\int D A D g{\cal O} e^{i(S+S_{\text{gf}})}}{\int D A D g e^{i(S+S_\text{gf})}}\overset{{\cal O}\text{ gauge-invariant}}{=}\frac{\int D A {\cal O} e^{iS}\int Dg e^{iS_\text{gf}}}{\int D A e^{iS}\int Dg e^{iS_{\text{gf}}}}\nn
\ee
\no holds. If the number of Gribov copies inside the first Gribov region is orbit-in\-de\-pen\-dent, or at least differs only be a measure-zero amount, the normalization constant in this expression drops out. Whether this is the case is not clear, as counting the Gribov copies is highly non-trivial, due to their large number \cite{Maas:2015nva,Hughes:2012hg,Maas:2011ba}. At any rate, in a  numerical simulation including only a finite, and potentially varying, number of Gribov copies, as the present one, the number of copies will in general differ \cite{Maas:2015nva}. Therefore the explicit inclusion of the normalization is necessary \cite{Maas:2011ba}. On the lattice, \pref{zetaxi} takes the form, up to a rescaling of the gauge parameter\footnote{Note that, strictly speaking, the ghost propagator on a given configuration is not rotationally and translationally symmetric. However, it has even on a single configuration formally the same limit as the momentum goes to zero, and any deviations compared to the actual boundary integral in \pref{zetaxi} vanishes in the infinite-volume integral.},
\bea
w(\xi,\zeta)&=&\Theta(-\pdm D_\mu^{ab})\exp\left({\cal N}+\xi b-\zeta F\right)\label{latweight}.
\eea
\no The $\theta$-function is implemented by the fact that the gauge-fixing algorithm only yields gauge copies inside the first Gribov region, by construction \cite{Cucchieri:1995pn}. On every found Gribov copy the values for $b$ and $F$ are then calculated according to \pref{f} and \pref{b} for each Gribov copy. The normalization constant, where needed, could then be calculated explicitly by summing $w$ over all Gribov copies and setting ${\cal N}$ to become one. This yields, for any value of $\xi$ and $\zeta$, the weight factor. For the propagators, as described next, this is not explicitly necessary.

To perform the averages for the propagators as a function of $\zeta$ and $\xi$, the propagators will be calculated for every distinct Gribov copy. The results for any of the propagators $D$ will then be averaged with the weight \pref{latweight} for every gauge orbit, normalized by \pref{latweight},
\be
\langle\langle D\rangle\rangle=\left\langle\frac{\sum_i D_i w(\xi,\zeta)_i}{\sum_j w(\xi,\zeta)_j}\right\rangle=\frac{1}{\sum_\beta}\sum_\alpha\frac{\sum_i D_{i\alpha} w(\xi,\zeta)_{i\alpha}}{\sum_j w(\xi,\zeta)_{j\alpha}}\nn,
\ee
\no where the Latin indices count the Gribov copies and Greek indices count orbits, and indicate that the quantity is evaluated on the corresponding orbit and Gribov copy.

This construction has a number of relations to the extreme gauges discussed before. Sending $\xi$ and $\zeta$ to $\pm\infty$ recreates the extreme gauges \cite{Maas:2011se,Fachin:1991pu,Fachin:1993qg,Parrinello:1990pm,Maas:2013vd,Serreau:2013ila,Maas:2016frv}. Furthermore, the minimal Landau gauge corresponds to $\xi=\zeta=0$ \cite{Maas:2011se,Maas:2013vd,Langfeld:2004vu}, and thus represents also a fixed point of this class of gauges, similar to the Landau gauge in perturbation theory \cite{Bohm:2001yx}. This statement will be checked in sections \ref{s:gluon}-\ref{s:alpha}, and seems to be fulfilled within the numerical accuracy.

Finally, it has been conjectured \cite{Maas:2011se} that because the ghost term is a surface term \cite{Garcia:1996np}, this gauge condition should translate into a boundary condition for functional equations \cite{Fischer:2008uz,Maas:2011se,Maas:2013vd,Maas:2009se}. This would allow a simple implementation of such gauges in functional equations. A similar simple translation is not known for the second part of the gauge-fixing term \cite{Fischer:2008uz,Maas:2008ri}.

This class of gauges samples the first Gribov region in the $F$-$b$ plane with different emphasis. They should therefore provide a good idea of the variability of the propagators and running coupling inside the first Gribov region. For simplicity, in the following always either $\xi$ or $\zeta$ will be set to zero. As the dependence on $\zeta$ will be found to be quite weak, this appears a reasonable restriction.

\section{The gluon propagator}\label{s:gluon}

\begin{figure}
\includegraphics[width=\linewidth]{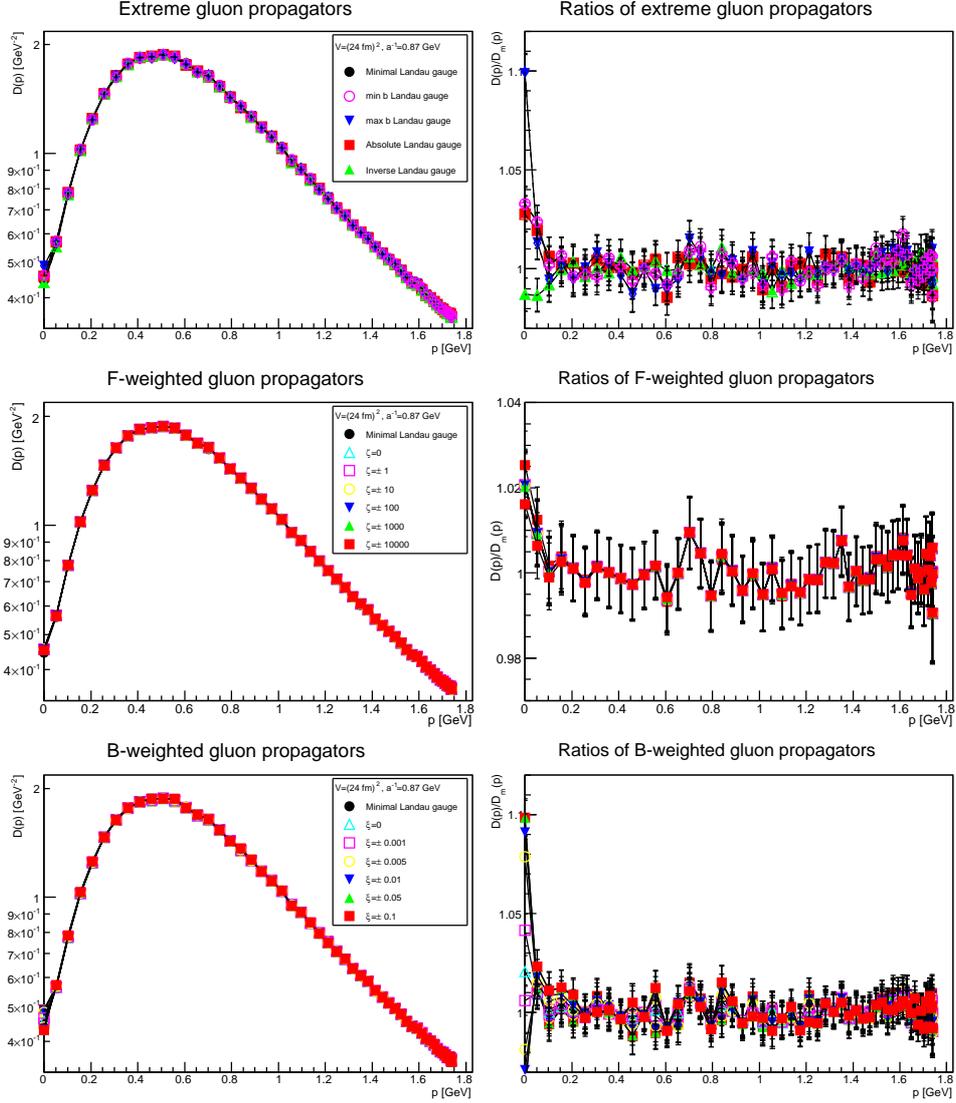}
\caption{\label{fig:2dgp}The gluon propagator for different gauges (left panels) and its ratio with the minimal-Landau-gauge gluon propagator (right panels) in two dimensions. The top panels show the extreme gauges, while the bottom panels show the averaged gauges. The dashed curves are the extrapolations to an infinite number of Gribov copies, see text. Note the different scales on the right-hand side.}
\end{figure}

\begin{figure}
\includegraphics[width=\linewidth]{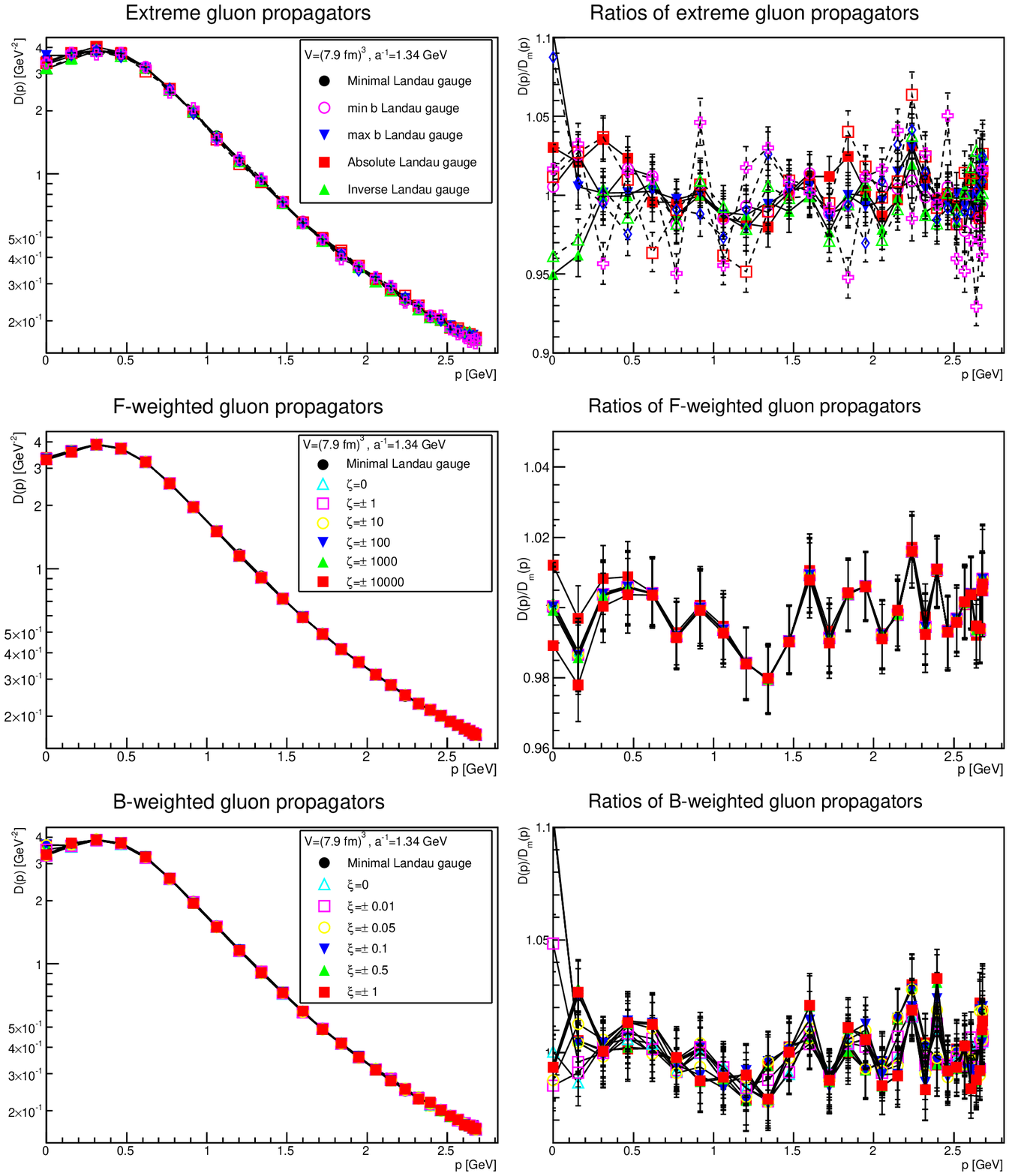}
\caption{\label{fig:3dgp}The gluon propagator for different gauges (left panels) and its ratio with the minimal-Landau-gauge gluon propagator (right panels) in three dimensions. The top panels show the extreme gauges, while the bottom panels show the averaged gauges. The dashed curves are the extrapolations to an infinite number of Gribov copies, see text. Note the different scales on the right-hand side.}
\end{figure}

\begin{figure}
\includegraphics[width=\linewidth]{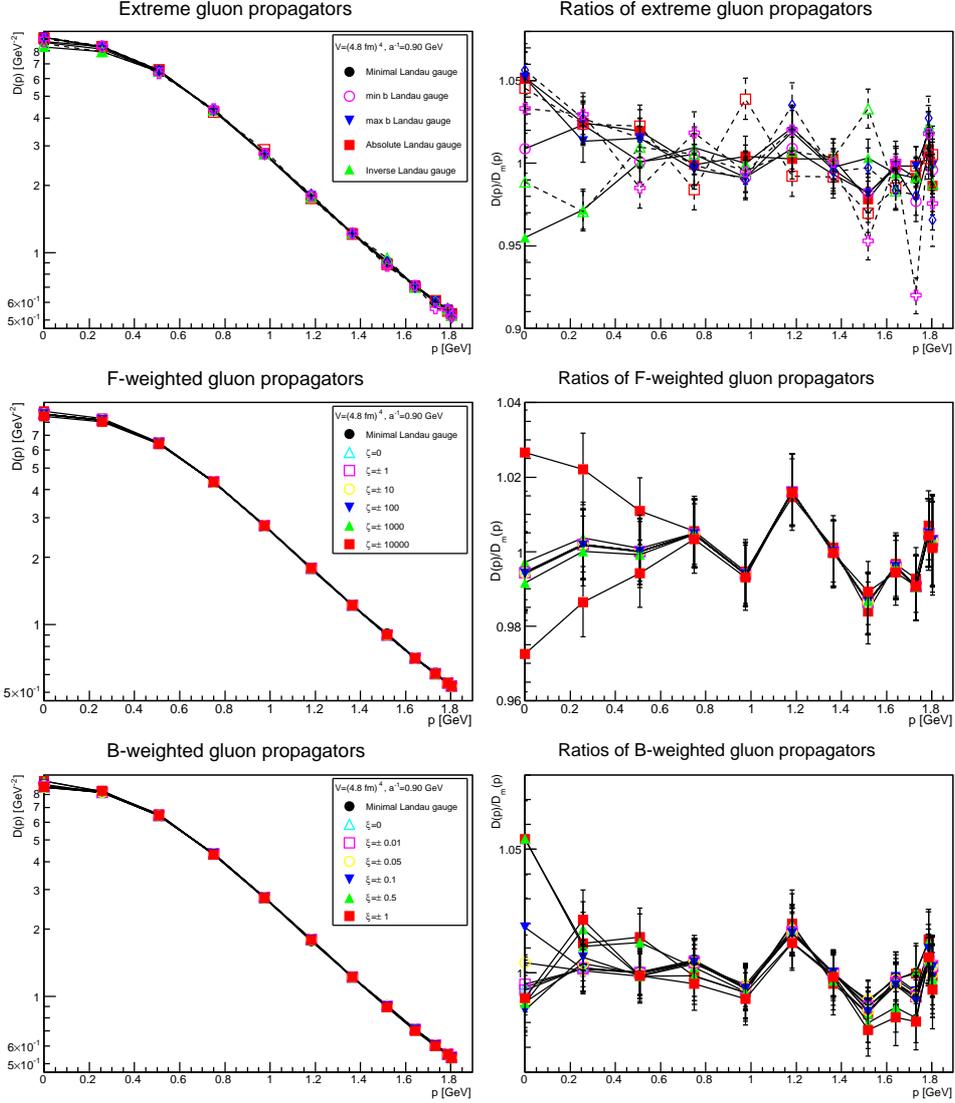}
\caption{\label{fig:4dgp}The gluon propagator for different gauges (left panels) and its ratio with the minimal-Landau-gauge gluon propagator (right panels) in four dimensions. The top panels show the extreme gauges, while the bottom panels show the averaged gauges. The dashed curves are the extrapolations to an infinite number of Gribov copies, see text. Note the different scales on the right-hand side.}
\end{figure}

The gluon propagator is shown in the extreme gauges and the averaged gauges in figures \ref{fig:2dgp}-\ref{fig:4dgp} for two, three, and four dimensions, respectively. In all cases the result for the largest volume are shown, given that the volume-dependence is the dominating lattice artifact for the gluon propagator \cite{Maas:2011se,Cucchieri:2007rg,Bogolubsky:2007ud,Bogolubsky:2009dc}.

Note that due to exceptional configurations \cite{Sternbeck:2005vs,Cucchieri:2006tf,Maas:2015nva} the values of $\xi$ are somewhat limited, as even long double precision is otherwise not enough to cope with the large numbers arising from the exponentiation. This problem does not arise for $\zeta$, as here no exceptional behavior is observed, and the values of $F$ are anyway much more limited than $b$. While the deviations at the largest accessible values of $\xi$ to the extreme gauges for the gluon propagator, and for the lowest momentum for the ghost propagator are small, this leads to deviations for the ghost at momenta slightly larger than the minimal one, and thus also for the running coupling. This will be seen in sections \ref{s:ghost} and \ref{s:alpha}. This is the only deviation from the expected behavior for $\zeta\to\pm\infty$ and $\xi\to\pm\infty$ as well as $\xi=\zeta=0$. This will therefore not be further discussed here.

The results in figures \ref{fig:2dgp}-\ref{fig:4dgp} show almost no statistically significant deviations between the different gauges. The very few points were they occur are at small momenta, but the onset momentum increases the higher the dimensionality. Still, even when ignoring statistical effects, the deviations are at most at the level of a few percent. Thus, also in the additional gauges the effects are not stronger as to what has been seen in the comparison between minimal Landau gauge and absolute Landau gauge \cite{Bogolubsky:2005wf,Bogolubsky:2007bw,Bornyakov:2008yx,Bornyakov:2010nc,Cucchieri:1997dx,Maas:2008ri,Maas:2009ph} or other extreme gauges \cite{Silva:2004bv,Sternbeck:2012mf} before.

\begin{figure}
\includegraphics[width=0.5\linewidth]{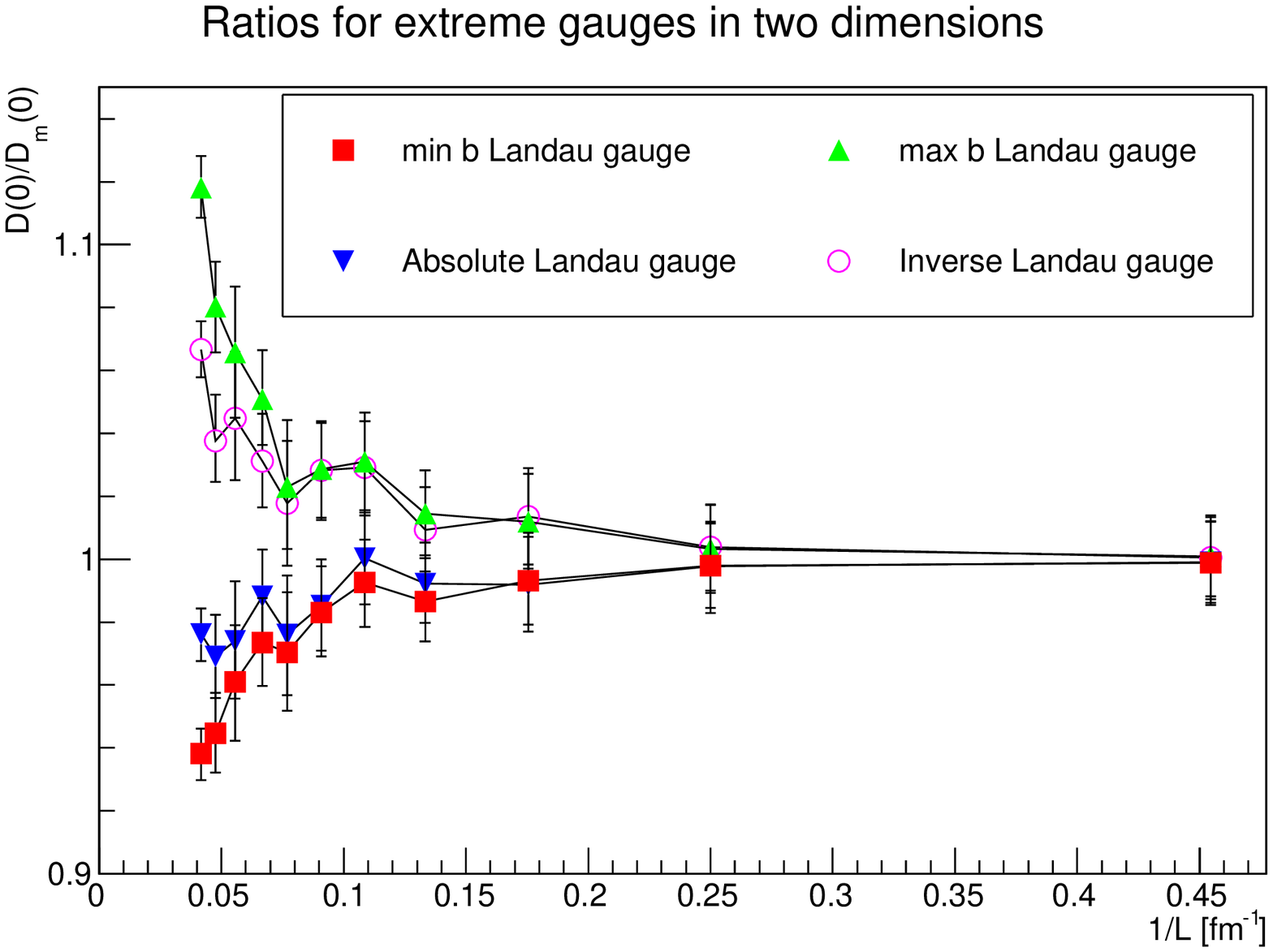}\includegraphics[width=0.5\linewidth]{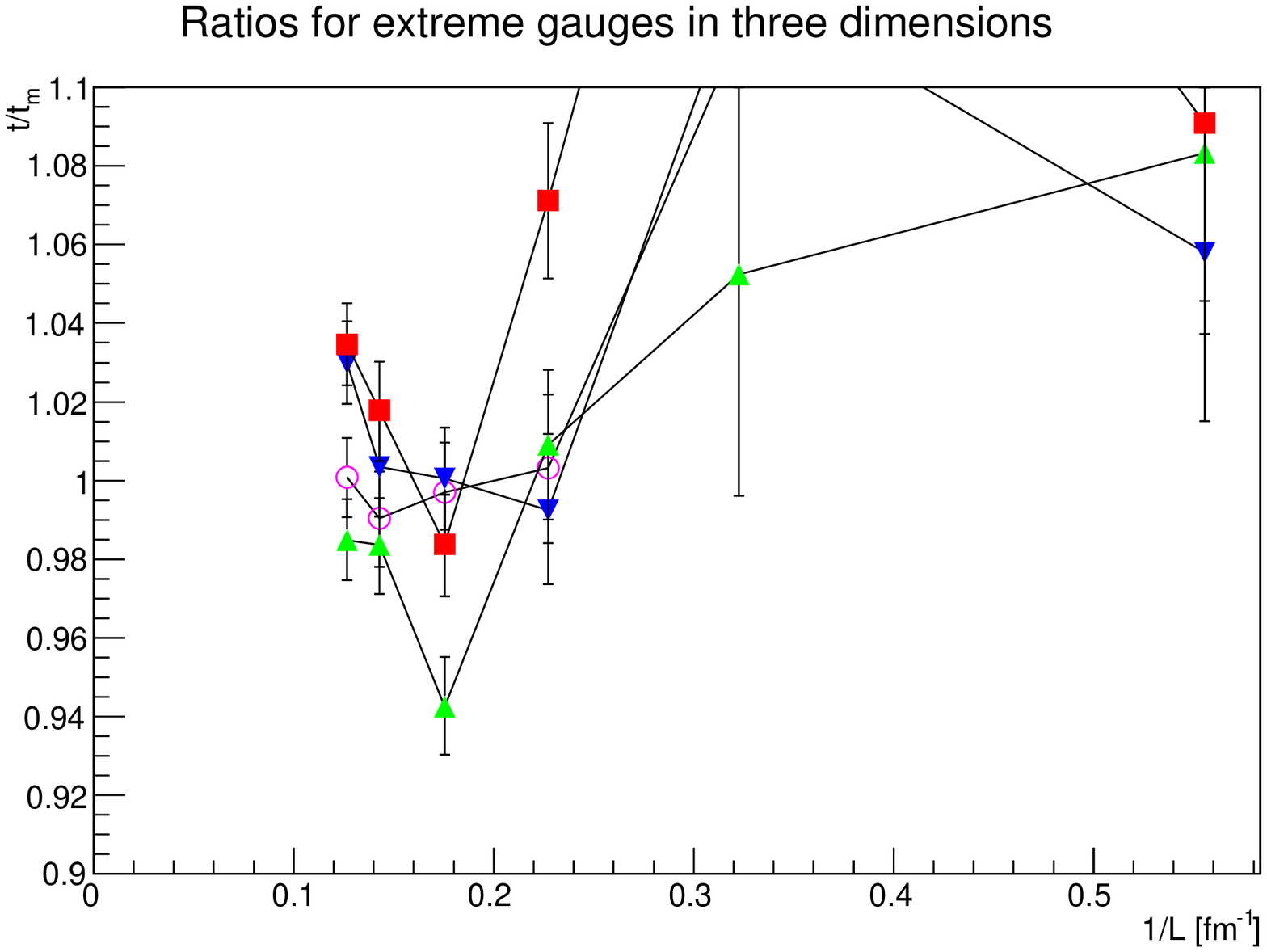}\\
\includegraphics[width=0.5\linewidth]{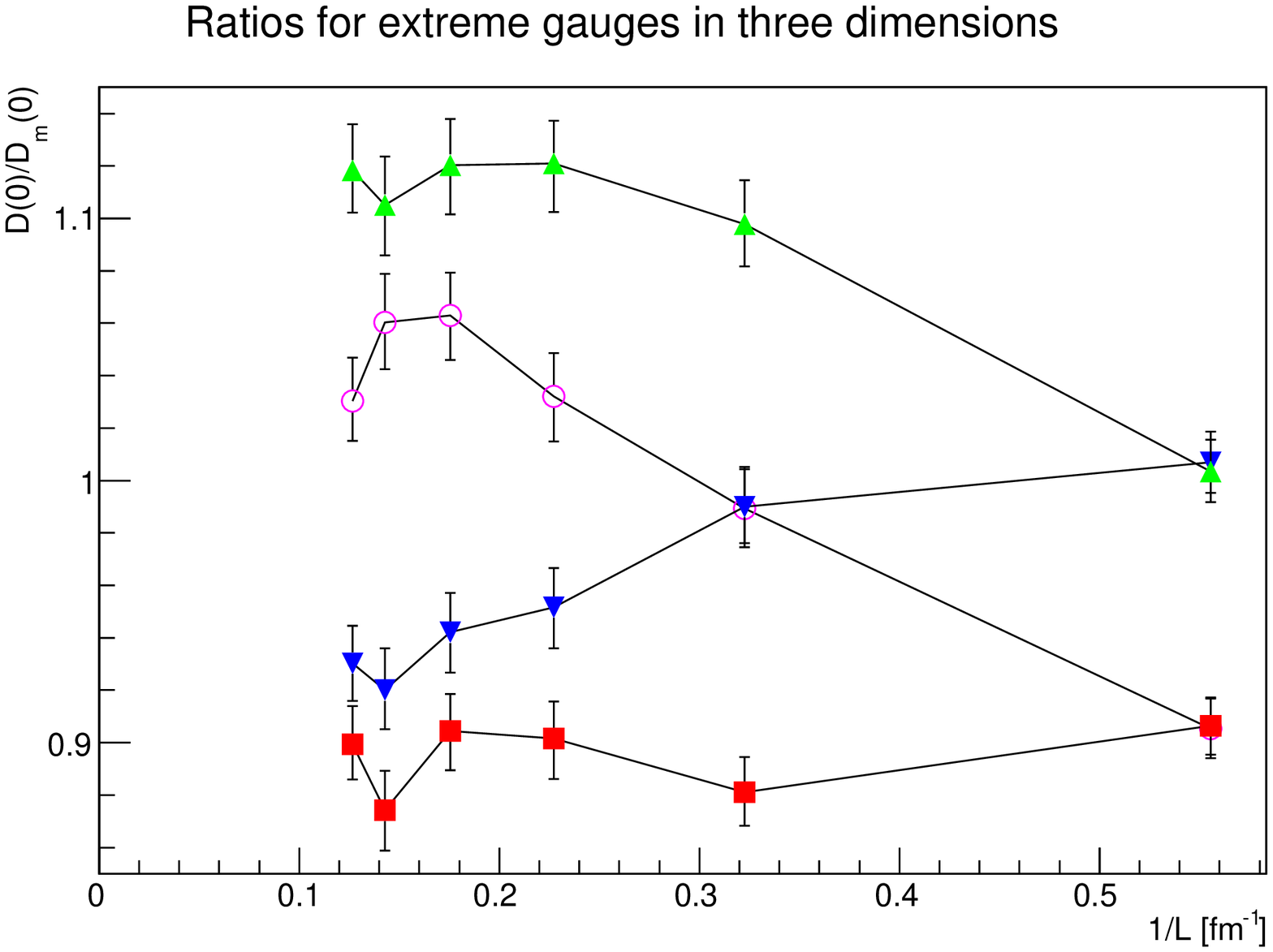}\includegraphics[width=0.5\linewidth]{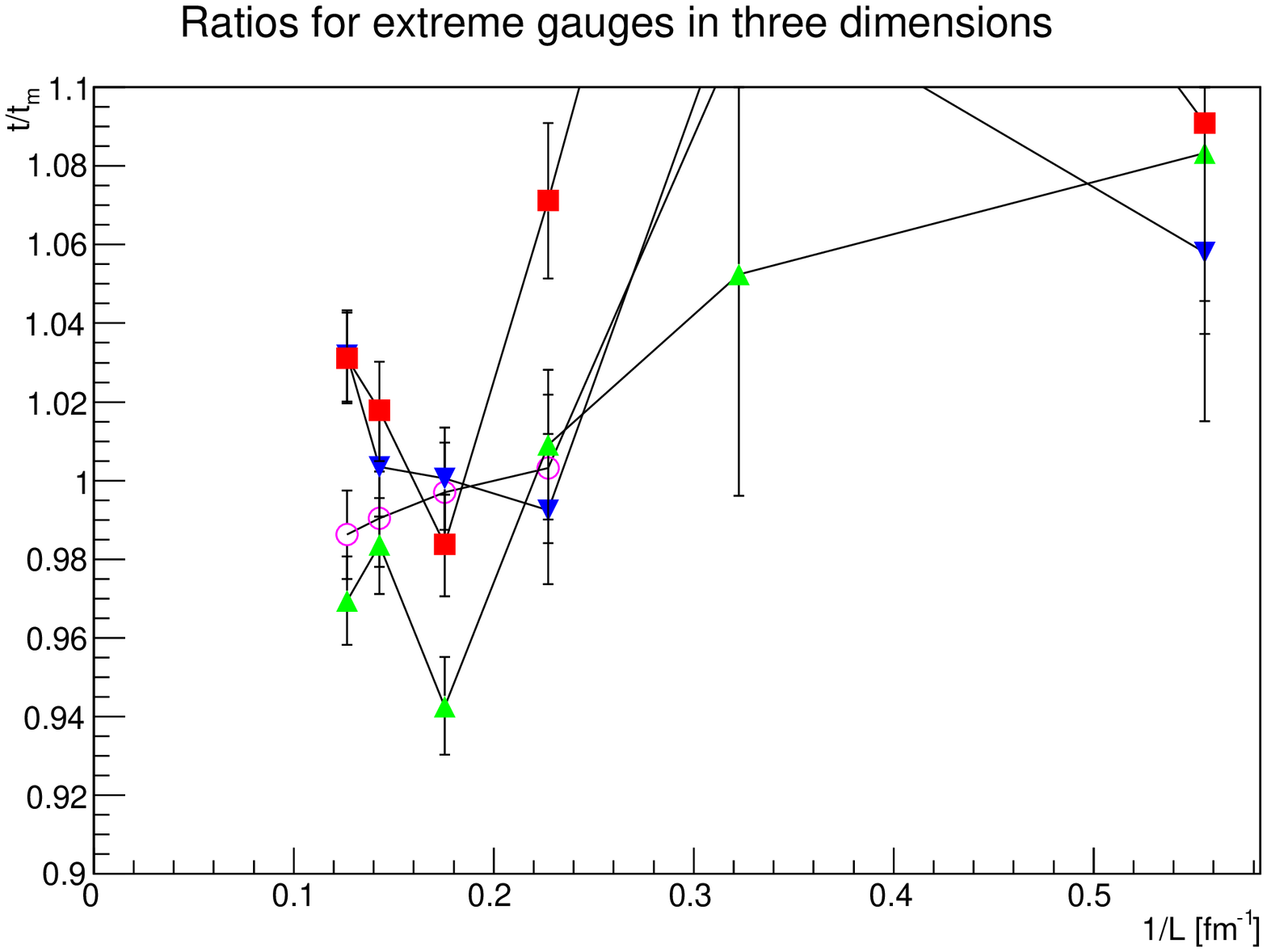}\\
\includegraphics[width=0.5\linewidth]{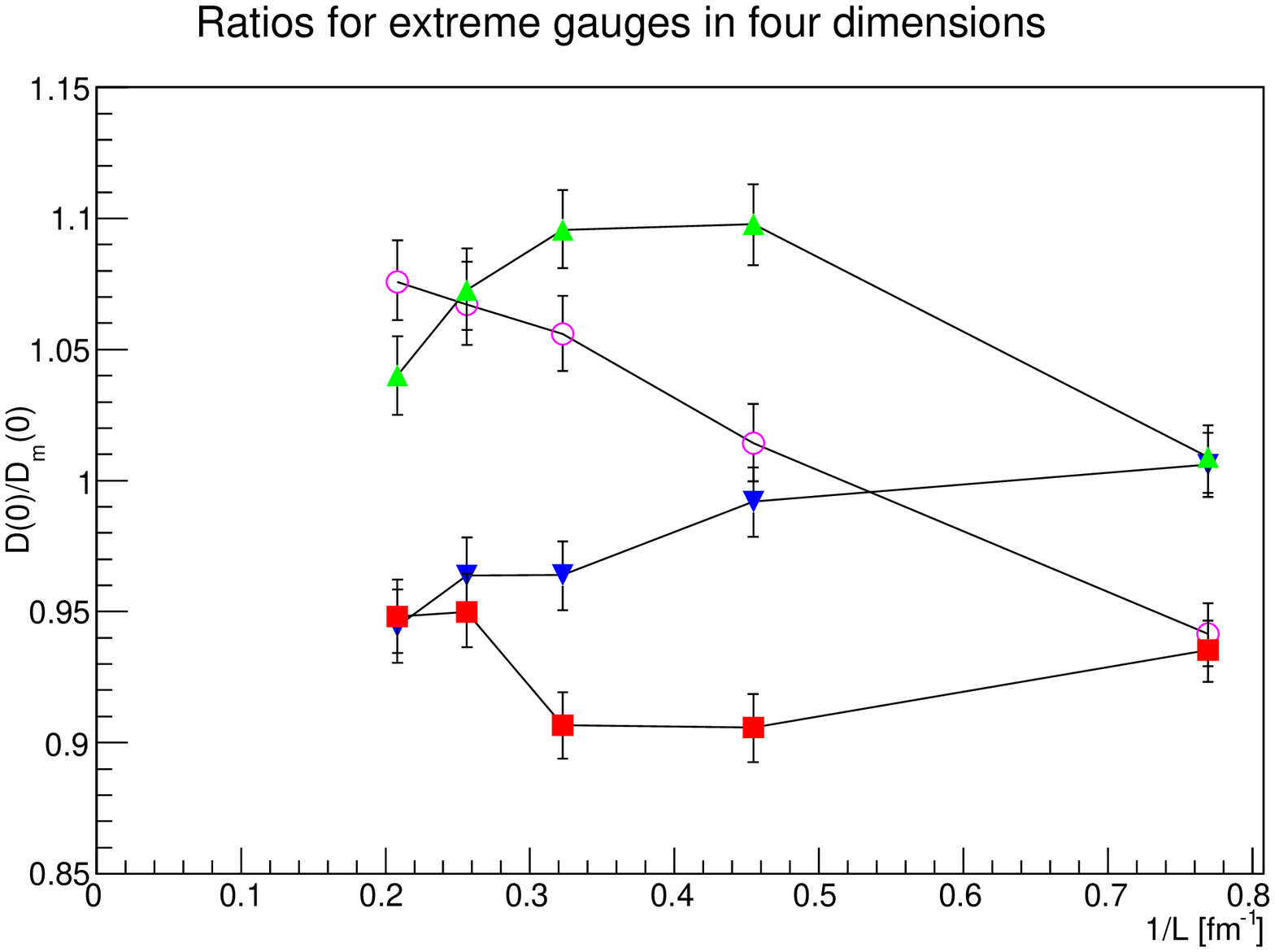}\includegraphics[width=0.5\linewidth]{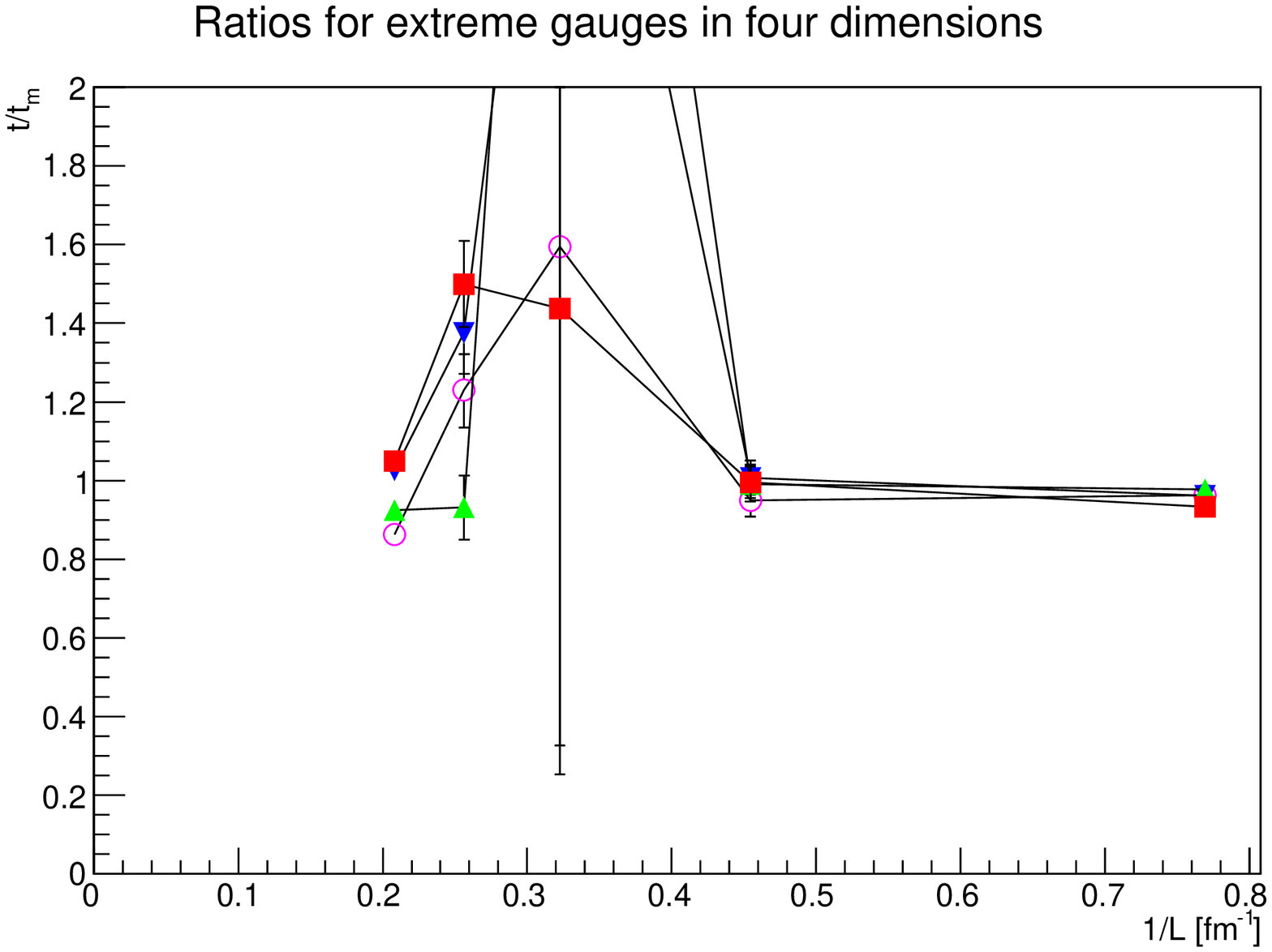}
\caption{\label{fig:gpir} Infrared properties of the gluon propagator for the extreme gauges, as ratios to the minimal Landau gauge case. The left panels show the gluon propagator at zero momentum and the right panels the effective infrared exponent. The top panels show the results in two dimensions, the middle panels in three dimensions, and the bottom panel in four dimensions. The results are always for the finest available discretizations.}
\end{figure}

This is also visible when considering the value of the gluon propagator at zero momentum. This value is shown as a function of volume for the extreme gauges normalized to the minimal Landau gauge value in figure \ref{fig:gpir}. While the deviations are statistically significant, they are small. However, there is a difference in the volume dependence between two and higher dimensions. In two dimensions at small volumes there is no difference, which is mainly due to the fact that essentially no Gribov copies are found \cite{Maas:2015nva}. Going to larger volumes, an ever-increasing deviation starts to appear. In higher dimensions, where there are already Gribov copies found in small volumes, an effect is already seen at small volumes. But at larger volumes the effect seems to saturate. It may even decrease, but this trend is not (yet?) statistically significant.

Interestingly, at small volumes the $\max b$ and the absolute Landau gauge give the same effect, as do the $\min b$ and the inverse Landau gauge. But at intermediate volumes the behavior of the two gauges based on $F$ cross, and in the end the behavior expected from the distribution of Gribov copies in the first Gribov region \cite{Maas:2015nva} emerges: The $\max b$ and the inverse Landau gauge are similar, and so is the absolute Landau gauge and the $\min b$ gauges. Still, the gauges triggering in $b$ yield a wider variation than those based on $F$. The reason for this is that a condition based on \pref{f} is sensitive to the integrated weight of the gluon propagator \cite{Maas:2008ri}. Because of the integral weight, small changes at mid momentum are as efficient as large changes at small momenta. Therefore, when studying small volumes, and therefore primarily large momenta, different modifications of the gluon propagator help to fulfill a condition on \pref{b} than on larger volumes.

Also shown in figure \ref{fig:gpir} are the ratios for the effective infrared exponent $t$ \cite{Fischer:2007pf,Maas:2007uv}. This exponent is determined in a fixed volume by fitting the low-momentum behavior of the dressing function to a power-law $p^{2t}$ \cite{Fischer:2007pf}. In practice, this was done by discarding the two lowest non-vanishing momentum points, to avoid a too large contamination by finite-volume effects. Then the next five highest points in momentum were used to fit a power-law and finally yielding the exponent \cite{Maas:2007uv}. This effective exponent is then a volume-dependent function.

In two dimensions, there is essentially no statistical relevant distinction of the infrared exponents. In three and four dimensions, this is different, and the differences seem to be large. However, the actual size should not be taken too important. At the available volumes, the exponents are still small \cite{Maas:2014xma}. Therefore small variations do already create sizable factors. Still, the deviations from the minimal Landau gauge are statistically significant, especially in four dimensions. Therefore the behavior is influenced. The net effect is, as seen from figures \ref{fig:2dgp}-\ref{fig:4dgp}, nonetheless small.

As noted above, only a small subset of Gribov copies are found. The investigations of the properties of the first Gribov region in \cite{Maas:2015nva} have shown that not all quantities reach their limiting value as a function of the number of Gribov copies with the subset included here. Since the gluon propagator is most affected in the far infrared and for extreme gauges, the quantity to estimate the effect is the ratio of the gluon propagator at zero momentum to the minimal Landau gauge version.

\begin{figure}
\includegraphics[width=\linewidth]{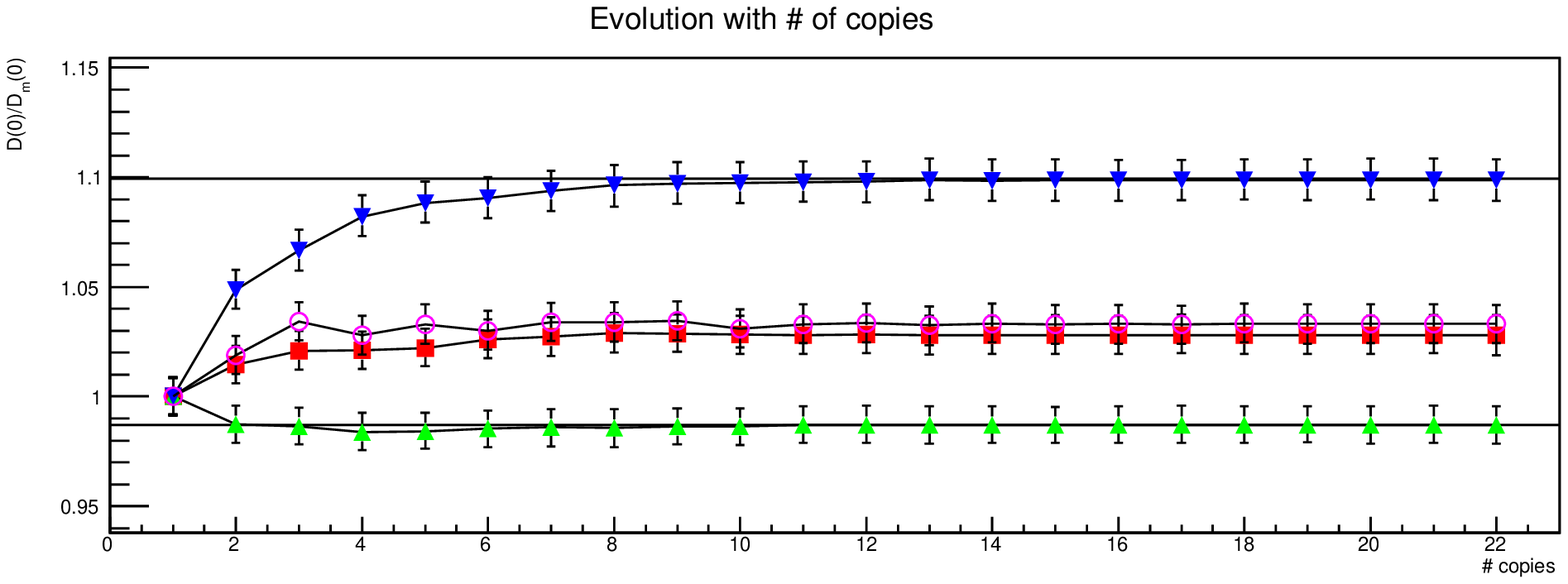}\\
\includegraphics[width=\linewidth]{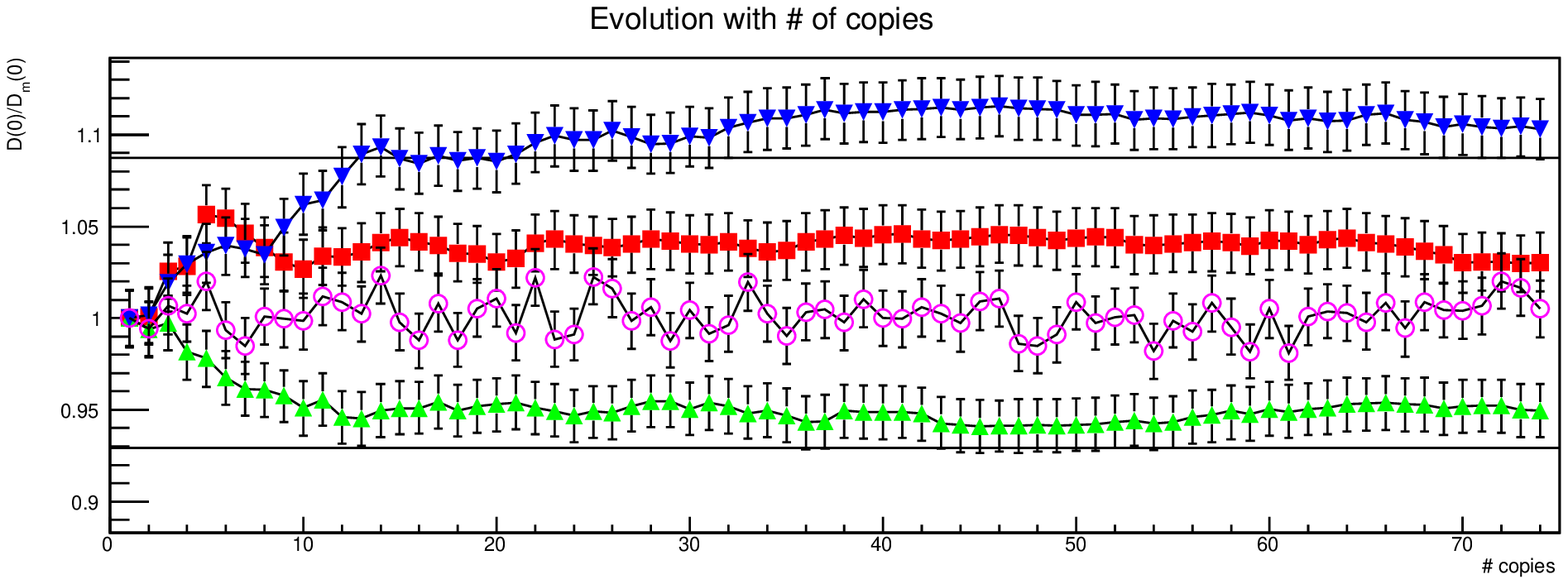}\\
\includegraphics[width=\linewidth]{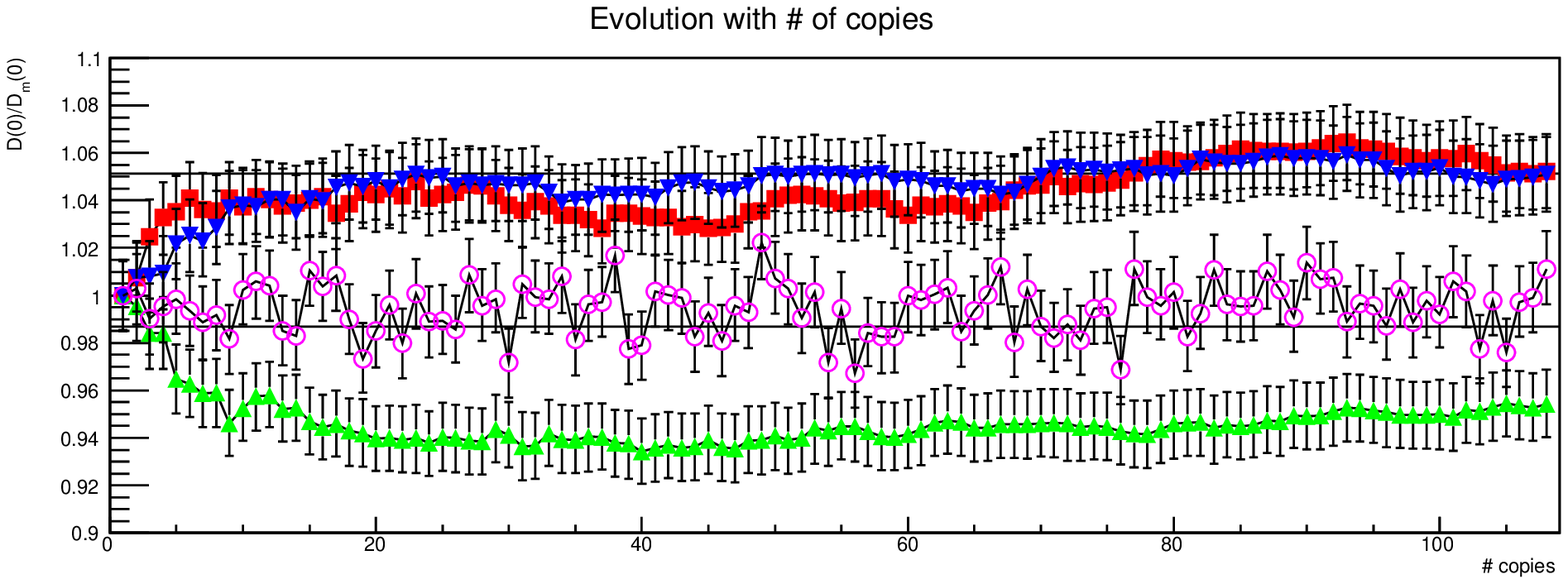}
\caption{\label{fig:gpge} Dependence of the ratio of the gluon propagator at zero momentum in the extreme gauges to the ones in minimal Landau gauge as a function of the number of included Gribov copies for the largest physical volumes at the finest available lattice spacings, with top, middle, and bottom panels being for two, three, and four dimensions, respectively. The lines are the extrapolations discussed in the text. The symbols have the same meaning as in figures \ref{fig:2dgp}-\ref{fig:4dgp}.}
\end{figure}

The result is shown in figure \ref{fig:gpge}. It is seen that the dependence is rather mild after some initial 10-20 Gribov copies. An extrapolation has been performed assuming a dependence
\be
a=a_0+\frac{a_1}{N}\label{extra},
\ee
\no where $a$ is in this case the gluon propagator at zero momentum, and $N$ is the number of Gribov copies included. To obtain a reasonable result, the constant $a_0$ is determined as an average over the fits from two adjacent values for the Gribov copies over the 30 (in two dimensions: 10) largest numbers of Gribov copies. In figure \ref{fig:gpge} the largest and smallest value among the four extreme gauges is also plotted. As is seen, the difference remains at a few percent level, and is consistent within statistical errors already after about 10 included Gribov copies.

In figures \ref{fig:2dgp}-\ref{fig:4dgp} the same extrapolation is done at every momentum value for the extreme gauges, and also displayed. Again, no substantial effect is seen. Thus, in total, the dependence of the gluon propagator on the selection of Gribov copies is small, at the few percent level, and even that only at momenta smaller than 200-300 MeV at most.

It is also useful to consider the Schwinger function, i.\ e.\ the position space gluon propagator
\be
\Delta(t)=\frac{1}{\pi}\int_0^\infty dp_0\cos(tp_0)D(p_0^2)=\frac{1}{a\pi}\frac{1}{N_t}\sum_{P_0=0}^{N_t-1}\cos\left(\frac{2\pi tP_0}{N_t}\right)D(P_0^2)\nn,\\
\ee
\no as well. This quantity is very sensitive even to small changes in the propagator \cite{Maas:2011se,Alkofer:2003jj}. Therefore, potentially even small changes may affect it. As it plays an important role to identify the analytic structure of the gluon propagator \cite{Maas:2011se,Alkofer:2003jj,Maas:2014xma}, any such changes would be important.

\begin{figure}
\includegraphics[width=\linewidth]{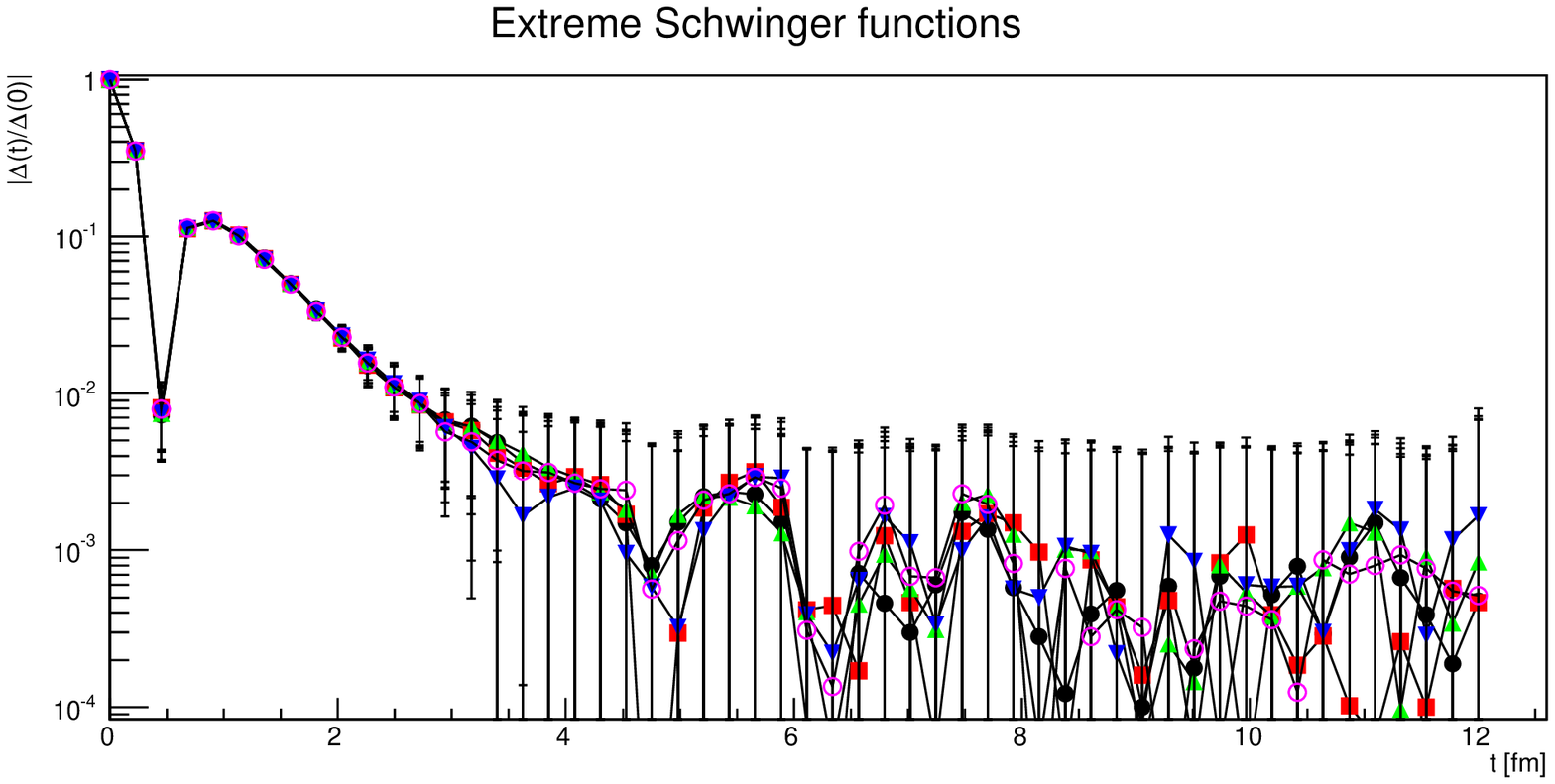}\\
\includegraphics[width=\linewidth]{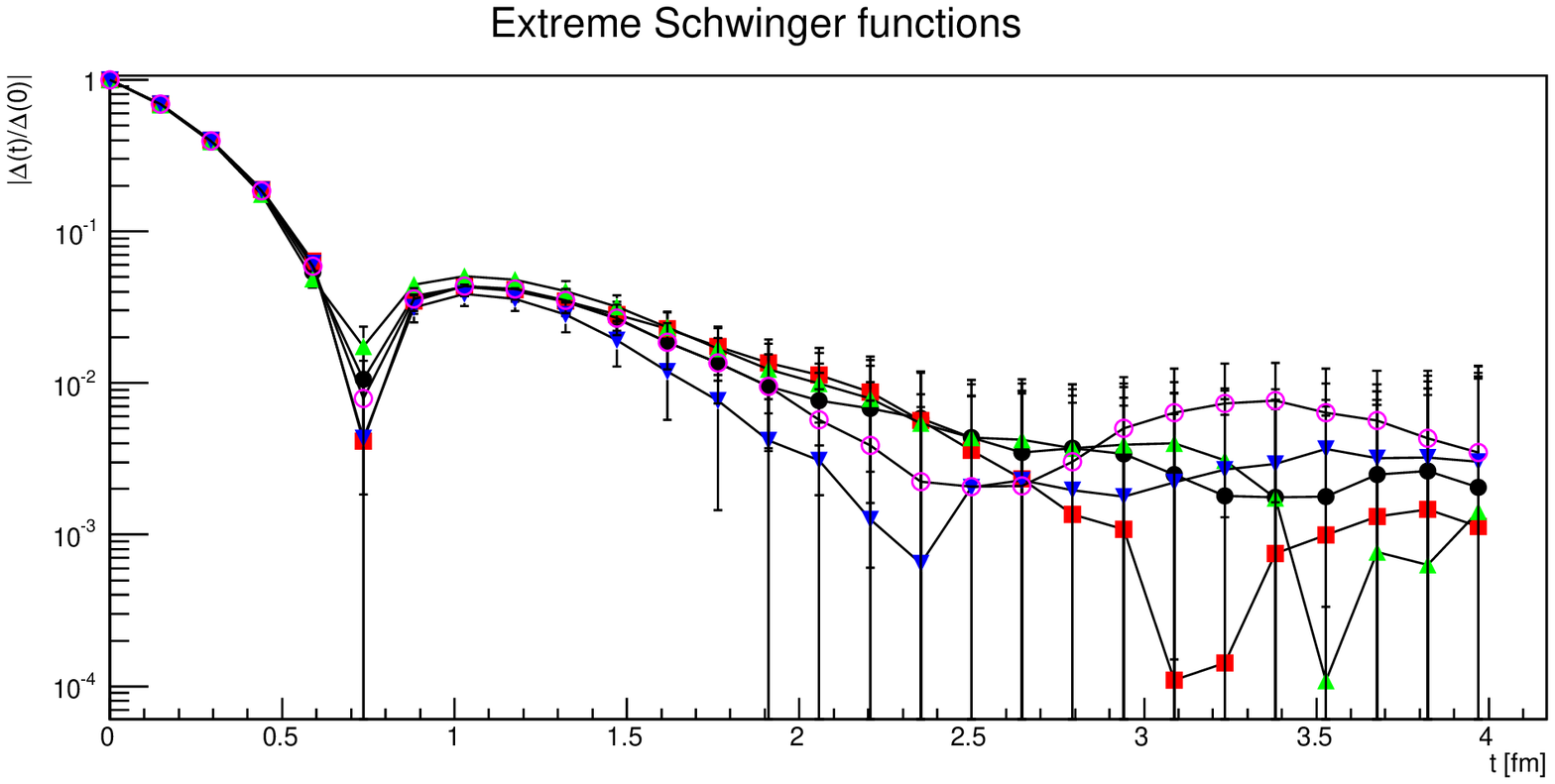}\\
\includegraphics[width=\linewidth]{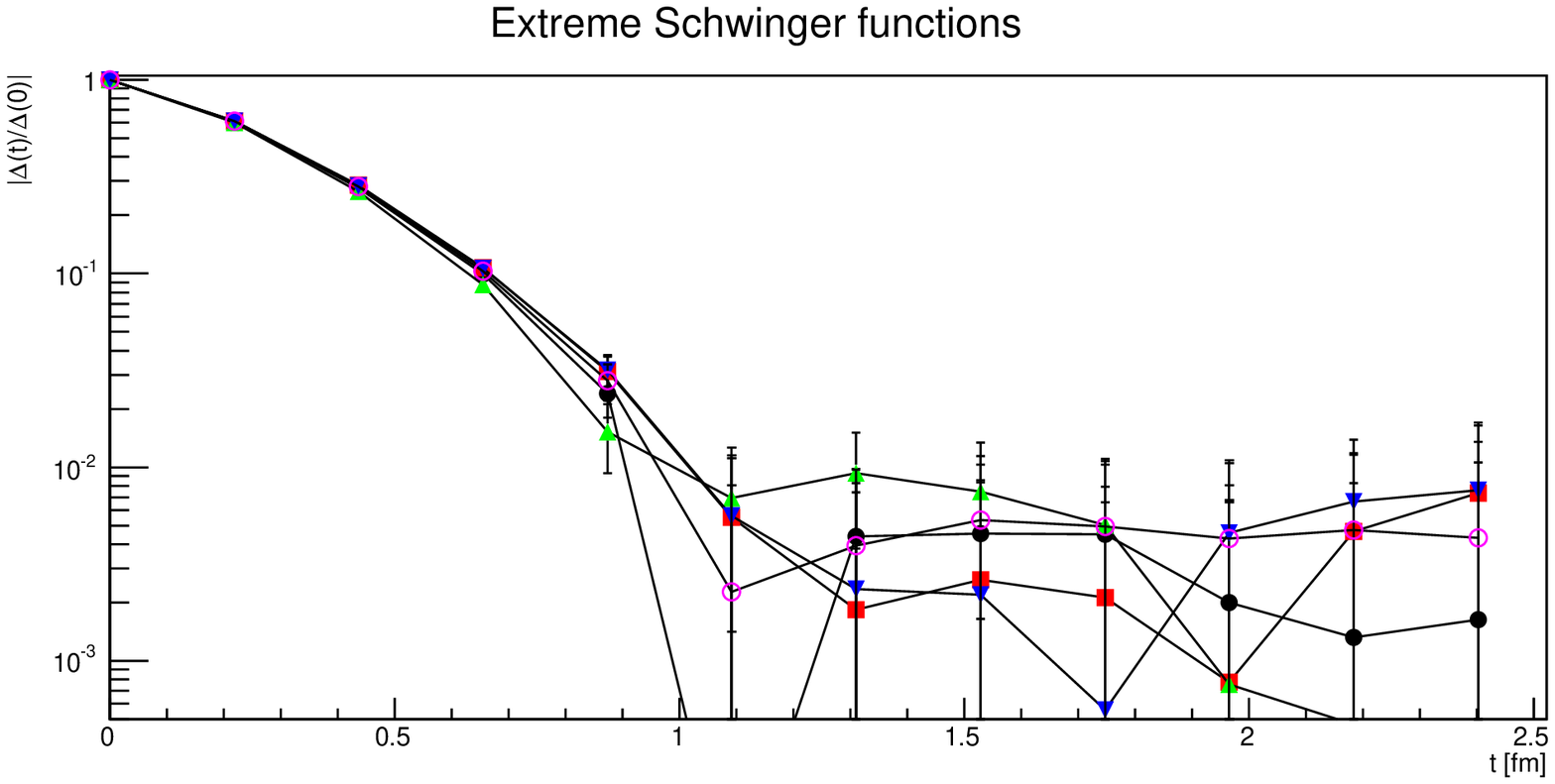}
\caption{\label{fig:schwinger} The Schwinger function for the extreme gauges for the largest physical volumes at the finest available lattice spacings, with top, middle, and bottom panels being for two, three, and four dimensions, respectively. The symbols have the same meaning as in figures \ref{fig:2dgp}-\ref{fig:4dgp}.}
\end{figure}

The results are shown in figure \ref{fig:schwinger}. While there appears some systematic deviations between the different extreme gauges, the effect is never beyond statistical doubt. In particular, the characteristic zero crossing \cite{Maas:2011se,Maas:2014xma,Alkofer:2003jj,Cucchieri:2003di} remains at the same time. Thus, also the Schwinger function is not significantly affected by the choice of Gribov copies.

\section{The ghost propagator}\label{s:ghost}

Already earlier results demonstrate that the ghost propagator will be substantially stronger affected by the choice of gauge than the gluon propagator \cite{Cucchieri:1997dx,Bogolubsky:2005wf,Bornyakov:2008yx,Maas:2008ri,Maas:2009se,Maas:2011ba,Maas:2011se,Sternbeck:2012mf,Maas:2013vd,Maas:2016frv}. And indeed, this is what will be confirmed in the following. Furthermore, the already substantial lattice artifacts due to finite volumes for the ghost propagator \cite{Maas:2011se,Maas:2014xma,Bogolubsky:2007ud,Bogolubsky:2009dc,Cucchieri:2008fc} are enhanced in gauges other than the minimal Landau gauge, as will be discussed in detail in section \ref{s:artifacts}. However, discretization artifacts seem to have the opposite effect as finite-volume artifacts, thereby leading partly to a cancellation. As a compromise, in the following the lowest non-zero momentum point will be dropped from the plots. Note that this had not yet been recognized in the preliminary results presented in \cite{Maas:2009se,Maas:2011ba,Maas:2011se,Maas:2013vd,Maas:2016frv}, though the effects there were still smaller as only smaller volumes were then available.

\begin{figure}
\includegraphics[width=\linewidth]{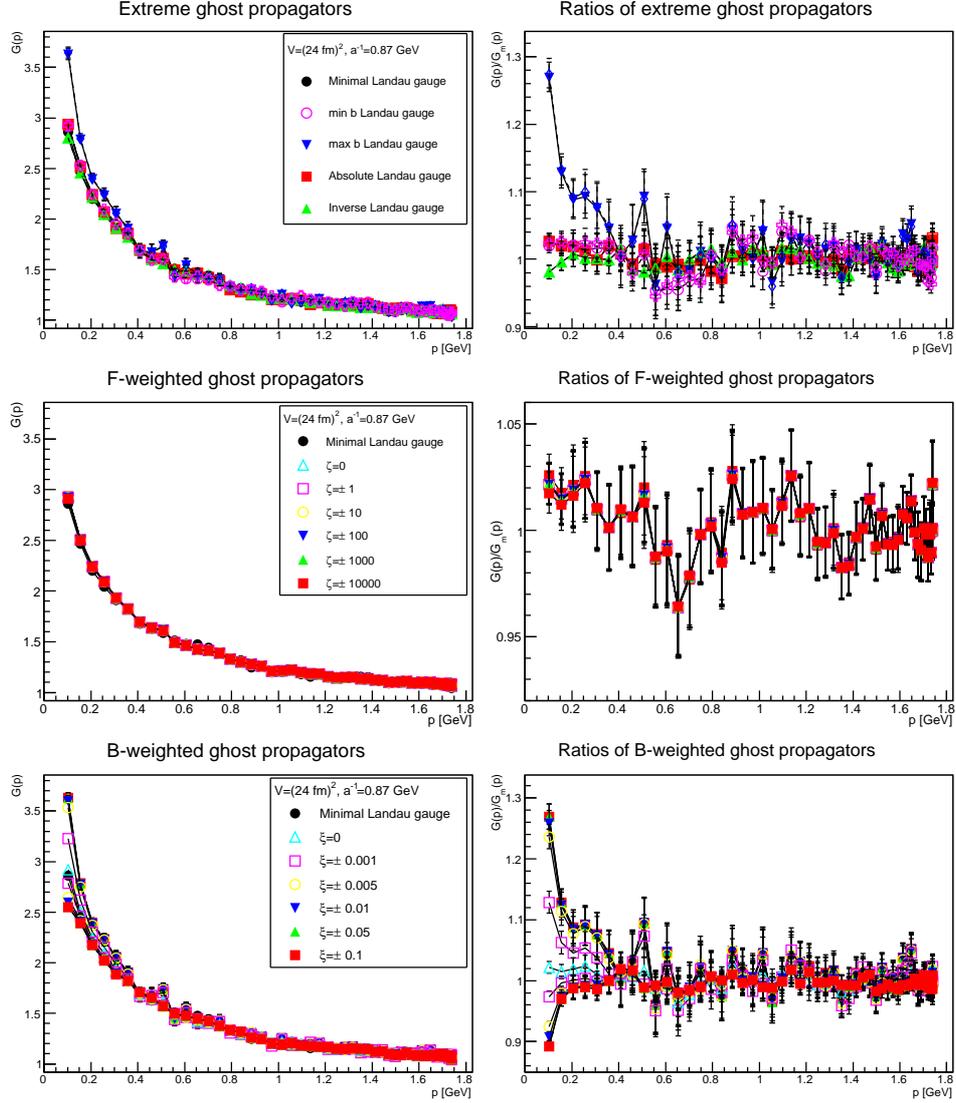}
\caption{\label{fig:2dghp} The ghost dressing function itself (left panels) and its ratio with the minimal-Landau-gauge ghost dressing function (right panels) in two dimensions. The top panels show the extreme gauges, while the bottom panels show the averaged gauges. The dashed curves are the extrapolations to an infinite number of Gribov copies, see text.}
\end{figure}

\begin{figure}
\includegraphics[width=\linewidth]{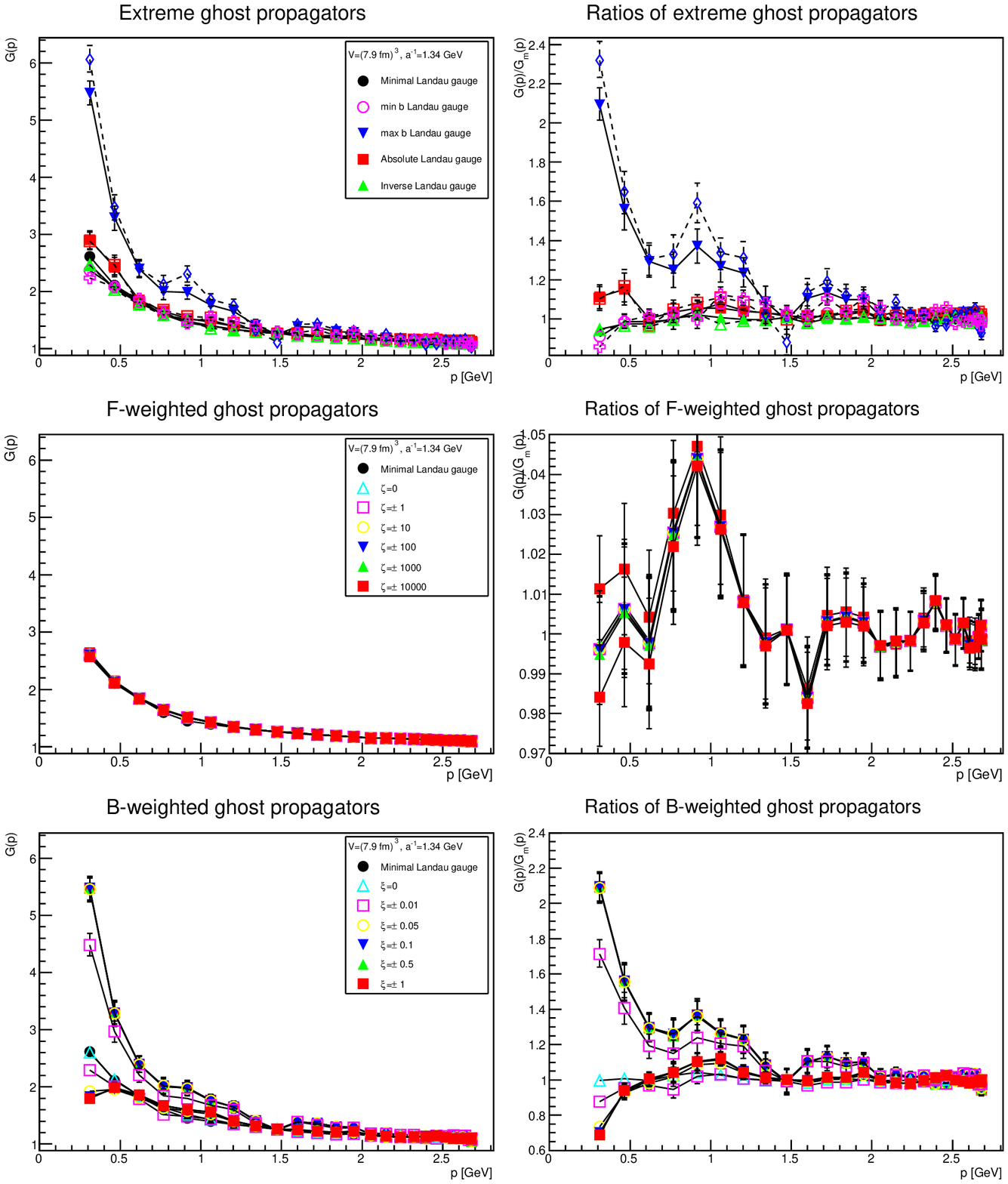}
\caption{\label{fig:3dghp} The ghost dressing function itself (left panels) and its ratio with the minimal-Landau-gauge ghost dressing function (right panels) in three dimensions. The top panels show the extreme gauges, while the bottom panels show the averaged gauges. The dashed curves are the extrapolations to an infinite number of Gribov copies, see text.}
\end{figure}

\begin{figure}
\includegraphics[width=\linewidth]{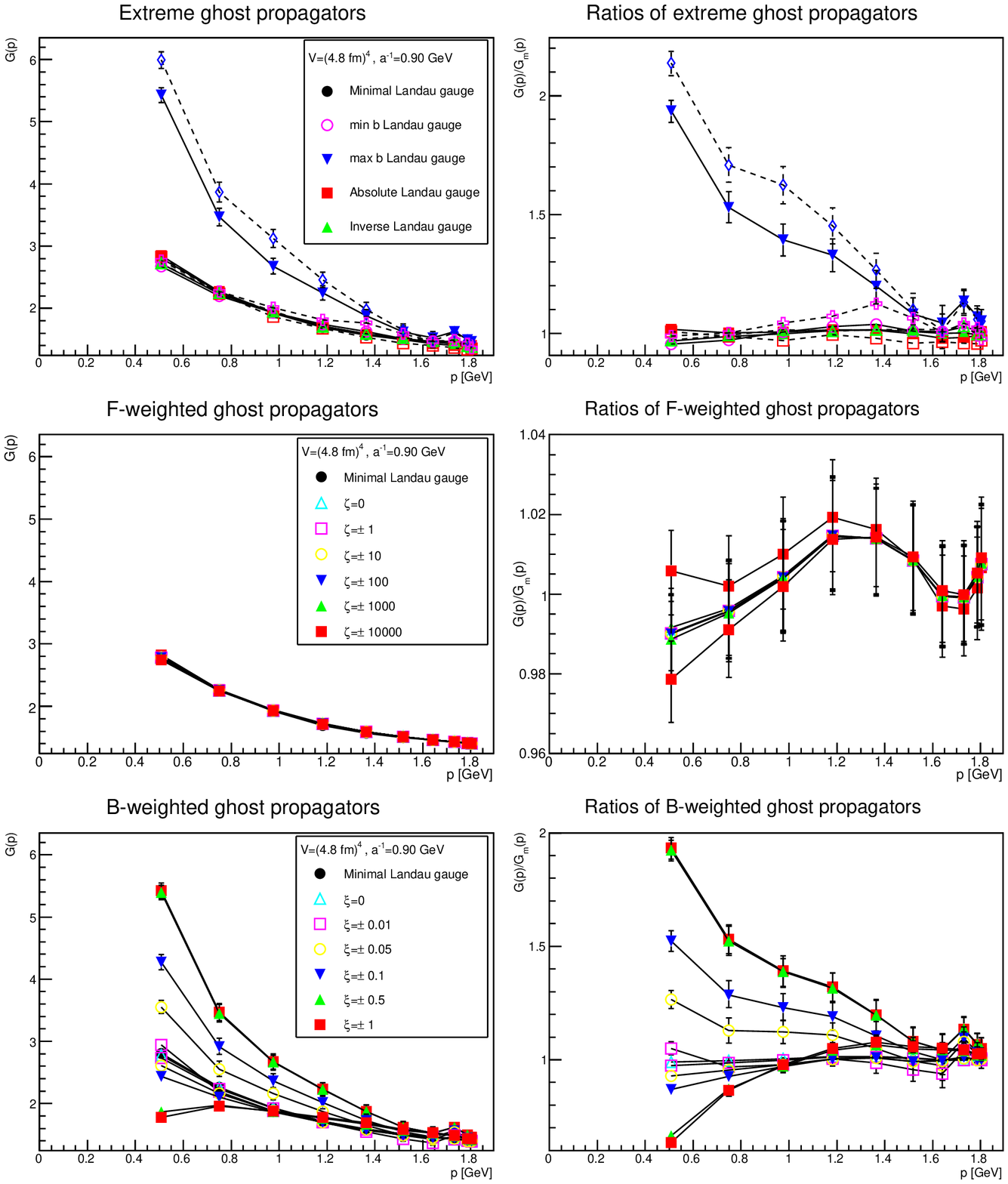}
\caption{\label{fig:4dghp} The ghost dressing function itself (left panels) and its ratio with the minimal-Landau-gauge ghost dressing function (right panels) in four dimensions. The top panels show the extreme gauges, while the bottom panels show the averaged gauges. The dashed curves are the extrapolations to an infinite number of Gribov copies, see text.}
\end{figure}

Even starting at such a larger momentum, much stronger effects are observed than for the gluon propagator. This is directly seen from the dressing function shown in figures \ref{fig:2dghp}-\ref{fig:4dghp}. Beyond any statistical doubt in three and four dimensions, and almost so in two dimensions, the ghost propagator starts to differ in the different gauges at momenta below 300 MeV, 600 MeV, and even 1.2 GeV in two, three, and four dimensions, respectively. The differences are large, even factors of more than two are reached. However, this large effects only occurs for the Landau-$b$ gauges, while there is still (almost) no change for the gauges weighted by $F$.

\begin{figure}
\includegraphics[width=0.5\linewidth]{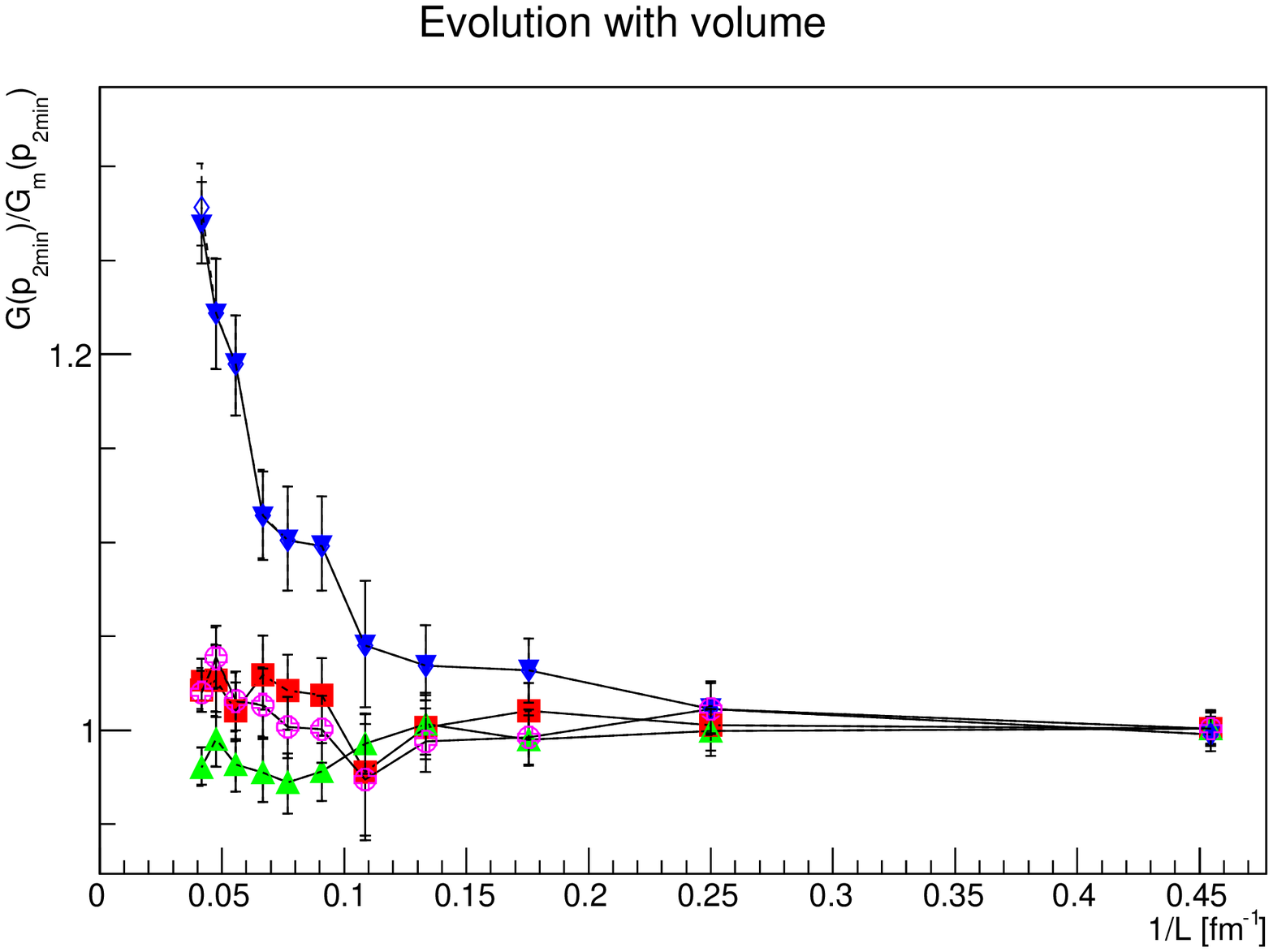}\includegraphics[width=0.5\linewidth]{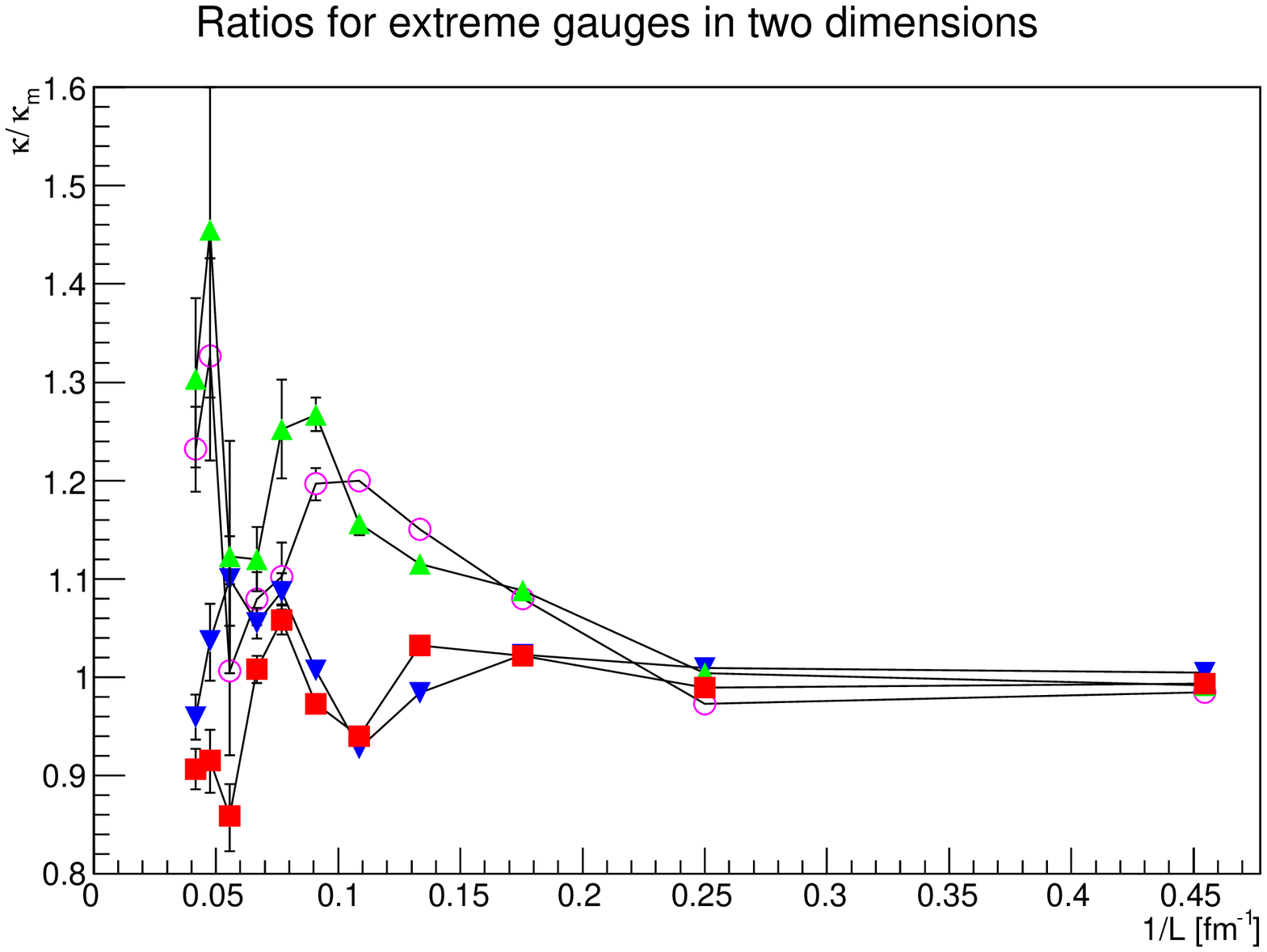}\\
\includegraphics[width=0.5\linewidth]{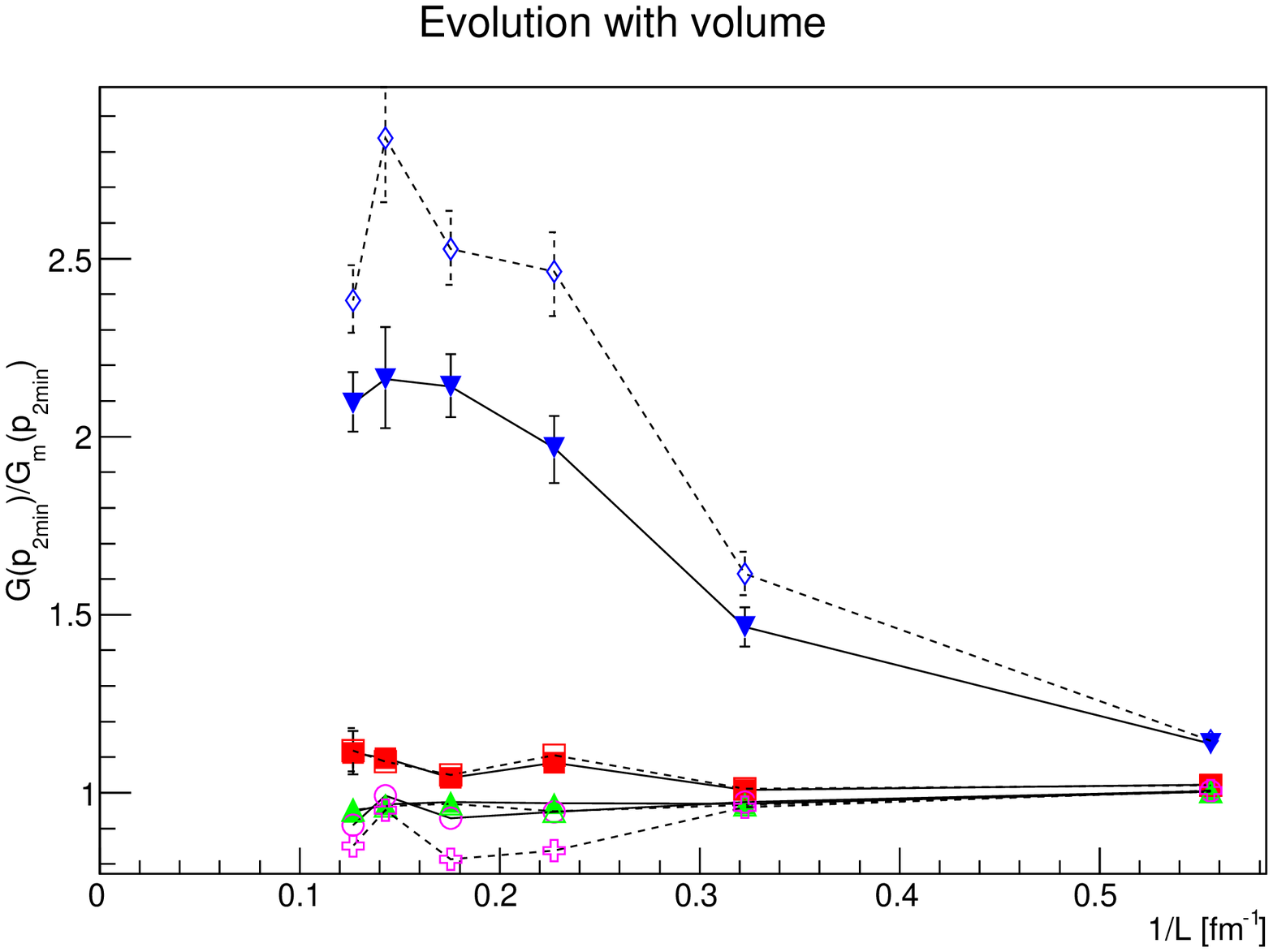}\includegraphics[width=0.5\linewidth]{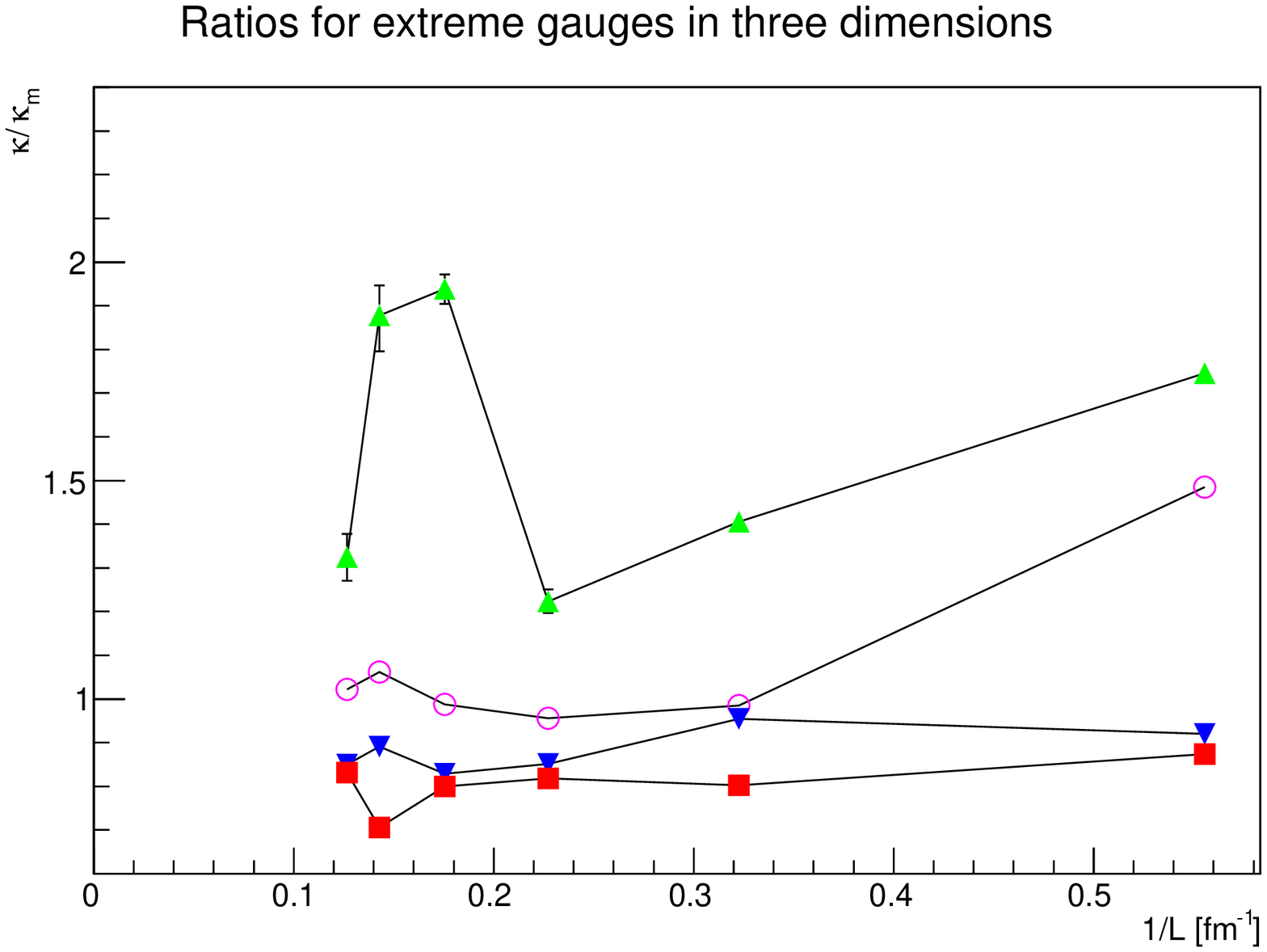}\\
\includegraphics[width=0.5\linewidth]{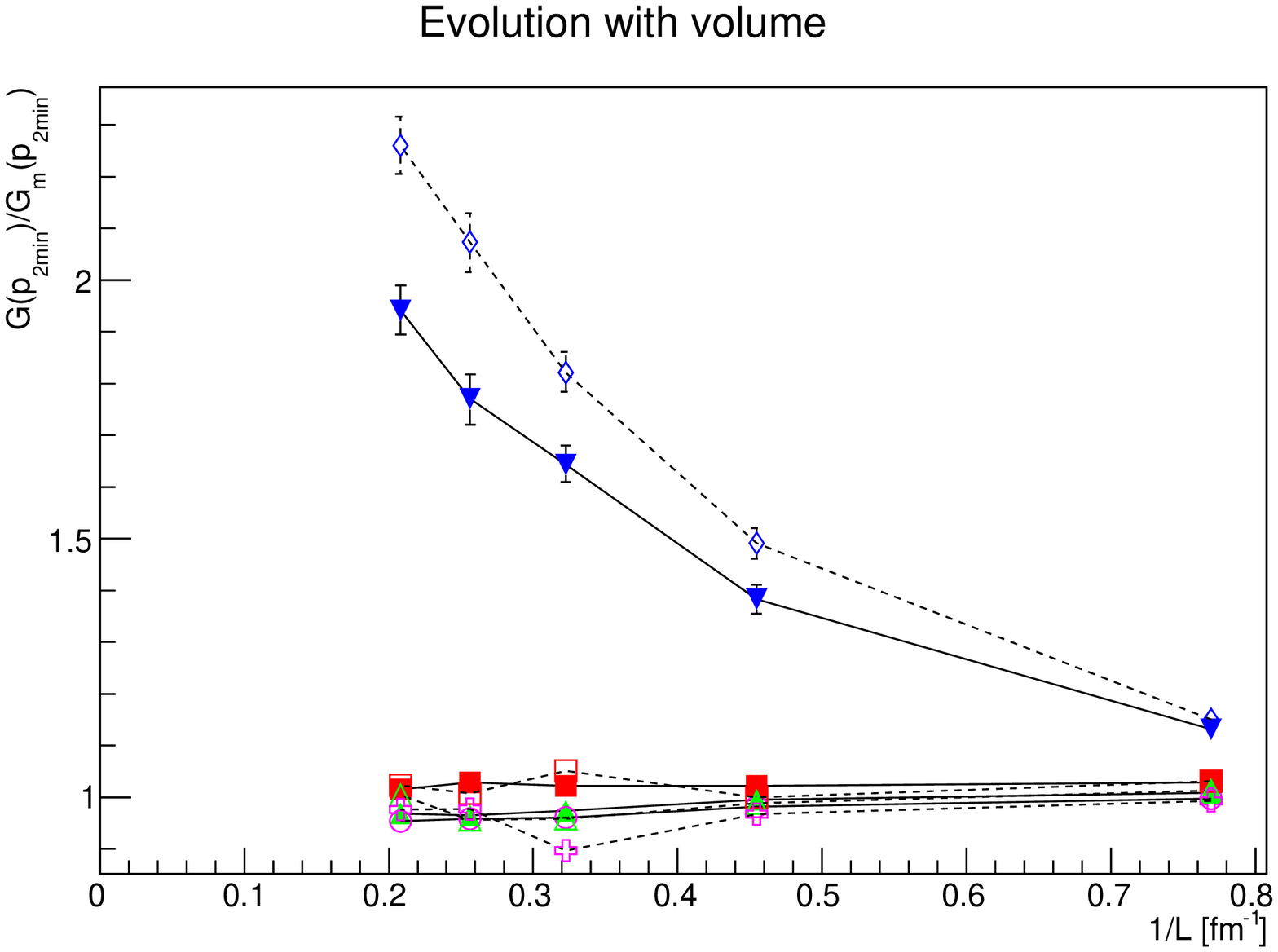}\includegraphics[width=0.5\linewidth]{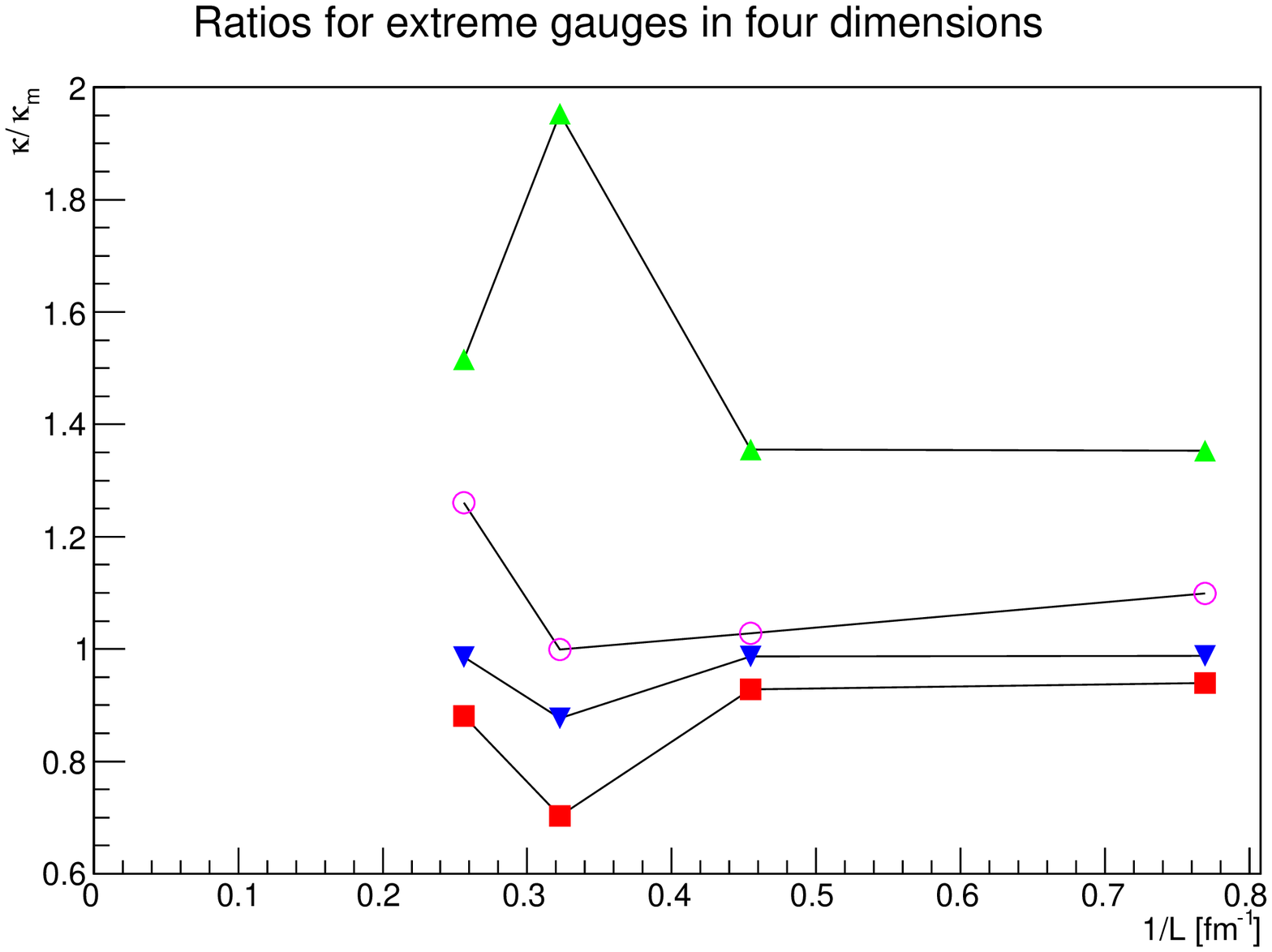}
\caption{\label{fig:ghpir} Infrared properties of the ghost dressing function for the extreme gauges, as ratios to the minimal Landau gauge case. The left panels show the ghost dressing function at the second-lowest accessible momentum, the right panels of the effective infrared exponent. The top panels show the results in two dimensions, the middle panels in three dimensions, and the bottom panel in four dimensions. The results are always for the finest available discretizations. Symbols have the same meaning as in figures \ref{fig:2dghp}-\ref{fig:4dghp}.}
\end{figure}

To study the volume-dependence, see also section \ref{s:artifacts}, the ratio of the ghost propagator to the one in minimal Landau gauge at the second-lowest momentum is shown in figure \ref{fig:ghpir}. The effective infrared exponent $\kappa$ of the ghost, which is again defined and obtained as in \cite{Maas:2007uv,Fischer:2007pf}, is also shown. Both quantities show quite interesting behaviors.

\begin{figure}
\includegraphics[width=\linewidth]{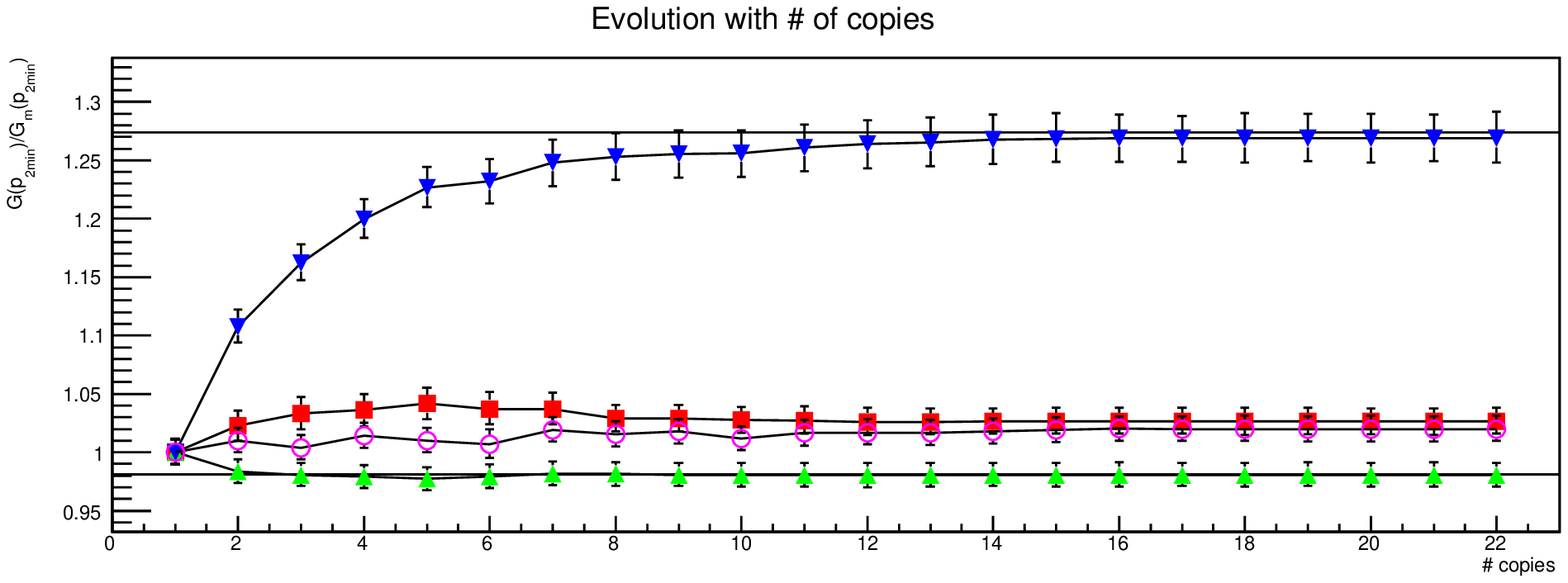}\\
\includegraphics[width=\linewidth]{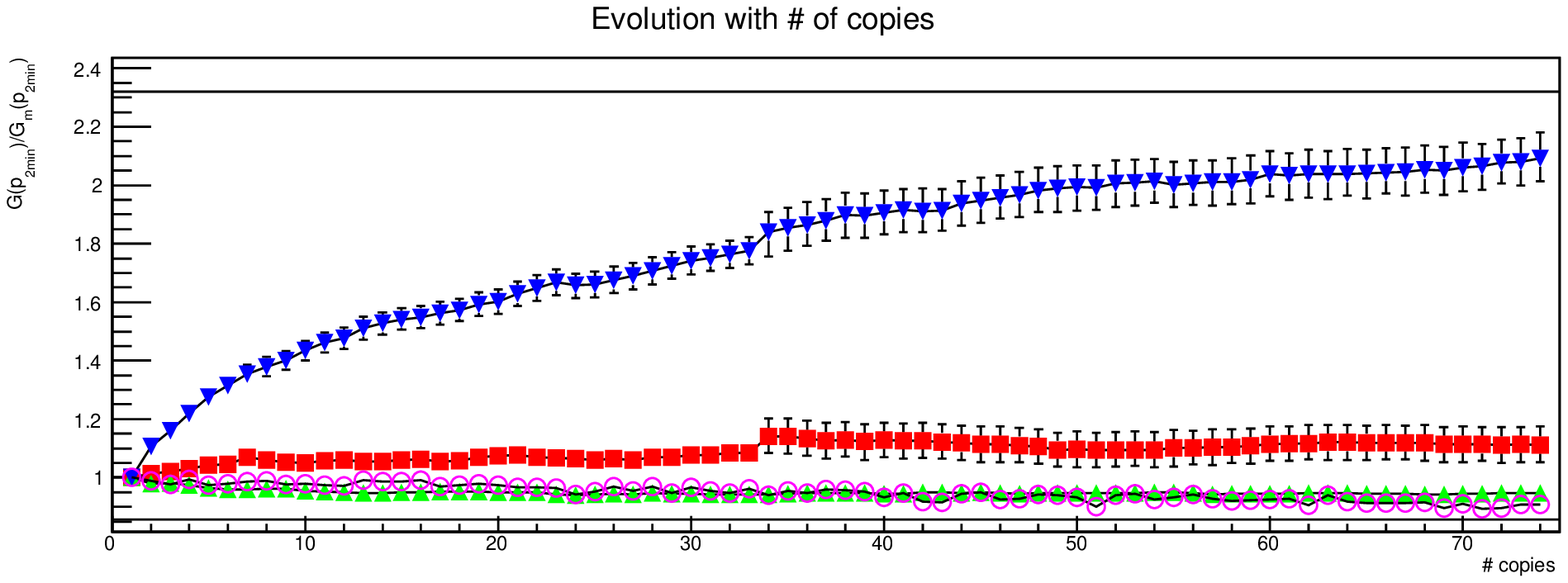}\\
\includegraphics[width=\linewidth]{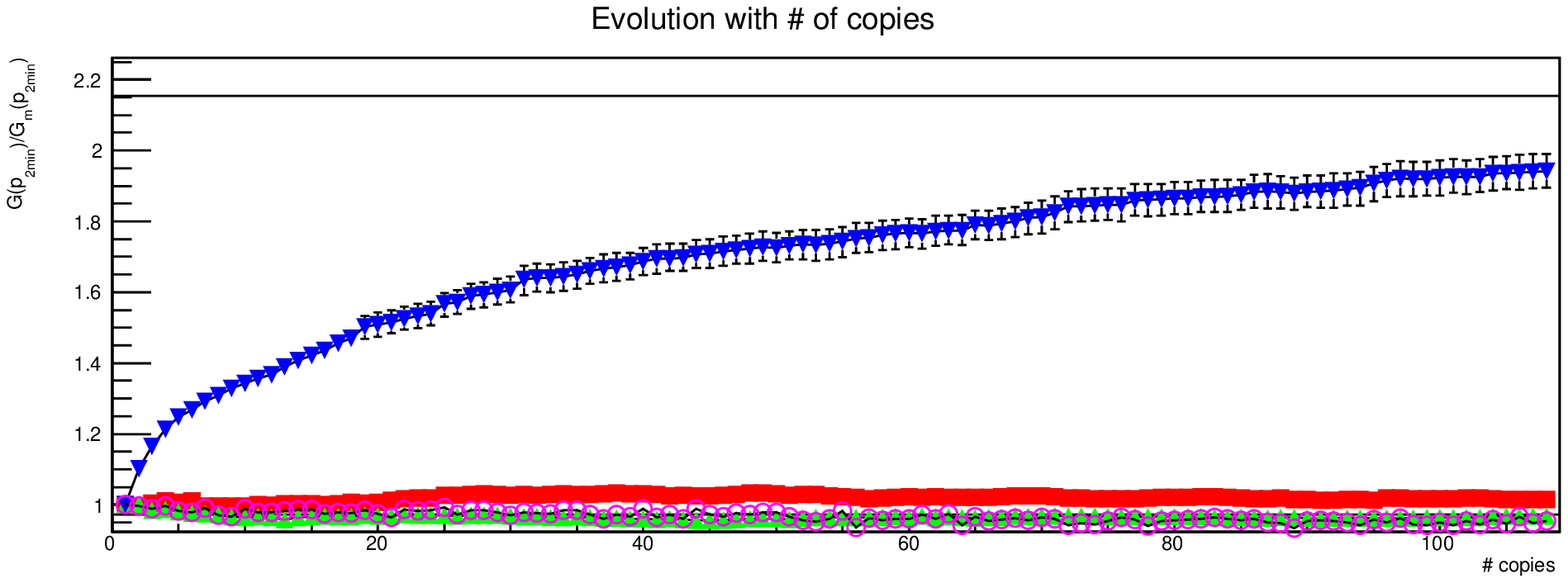}
\caption{\label{fig:ghpge} Dependence of the ratio of the ghost dressing function at the second-smallest momentum in the extreme gauges to the ones in minimal Landau gauge as a function of the number of included Gribov copies for the largest physical volumes at the finest available lattice spacings, with top, middle, and bottom panels being for two, three, and four dimensions, respectively. The lines are the extrapolations discussed in the text. The symbols have the same meaning as in figures \ref{fig:2dghp}-\ref{fig:4dghp}.}
\end{figure}

Consider first the ratio. In all dimensions an increase is seen, when the probing is done further and further in the infrared. However, again the effect is strongest for gauges biased by $b$. The effect is also strongest in four dimensions, and just above the level of statistical fluctuations in two dimensions. That is even more remarkable as the volumes in four dimensions are much smaller than in two dimension. Unfortunately, at these momenta an extrapolation in Gribov copies using \pref{extra} is affected by relatively large errors, see figure \ref{fig:ghpge}, as the deviation are somewhat small, and thus statistical errors are more important. It is thus unclear whether the situation changes if many more Gribov copies could have been taken into account. At any rate, it is visible that the asymptotic behavior is only (almost) reached in two dimensions with the sample size of Gribov copies available.

\begin{figure}
\includegraphics[width=\linewidth]{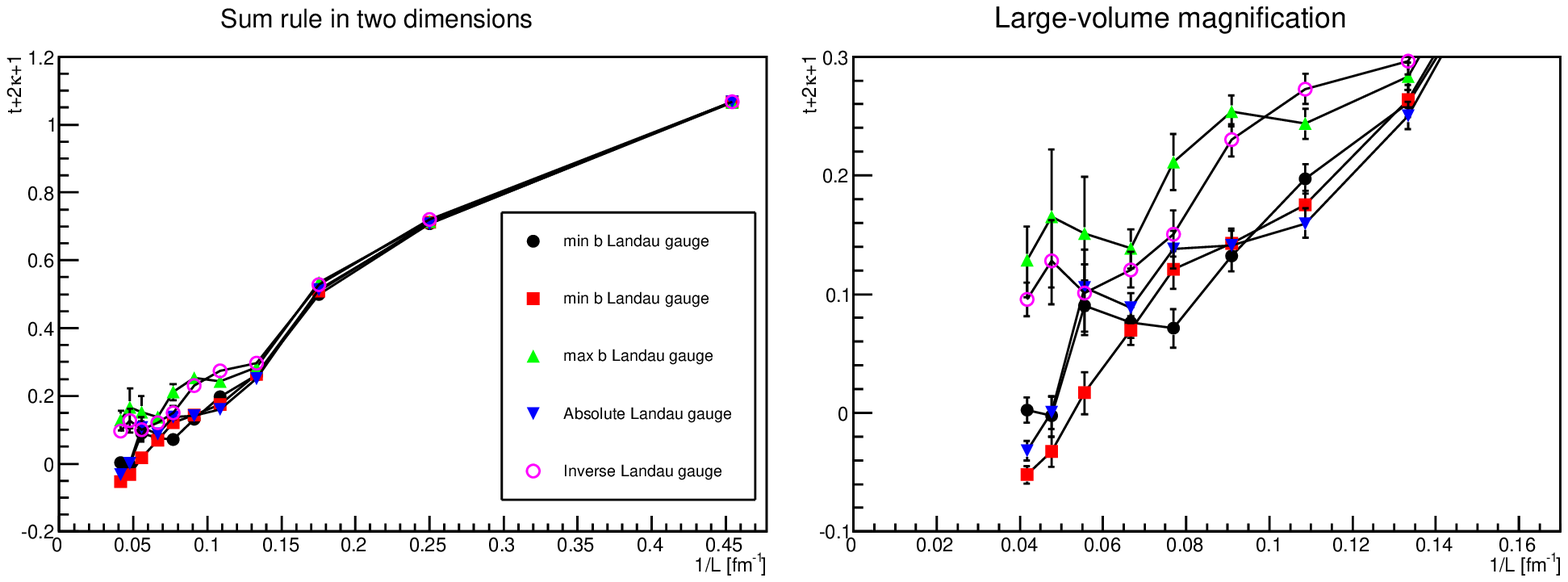}
\caption{\label{fig:sumrule}The sumrule $t+2\kappa+1=0$ for the infrared exponents in two dimensions for the extreme gauges. The right-hand-side shows a magnification at large volumes.}
\end{figure}

Next, consider the infrared effective exponent. As is seen in figure \ref{fig:ghpir}, the impact on the effective infrared exponent is more substantial, but less monotonous. Surprisingly, it is not the $\max b$-gauge, which is affected most strongly, but rather the  inverse Landau gauge. In this case two dimensions is especially interesting. The ghost propagator seems to diverge more and more the larger $\xi$. Also the effective infrared exponent, which is obtained in the same manner as for the gluon propagator, seems to differ from the minimal Landau gauge value, though the statistical errors are not small. This result is especially remarkable since the minimal Landau gauge results satisfy \cite{Maas:2011se,Maas:2007uv,Cucchieri:2007ta,Cucchieri:2007uj,Cucchieri:2016jwg}, or at least almost satisfy \cite{Maas:2014xma}, the predicted relation $t+2\kappa+1=0$ \cite{Lerche:2002ep,Zwanziger:2001kw}. This relation is tested in two dimensions in figure \ref{fig:sumrule}. While the results in minimal Landau gauge are close to fulfilling the sumrule \cite{Maas:2014xma}, the other gauges do not show any tendency to do so. However, the statistical uncertainty is too large to make a definite statement, whether they may indeed violate the sumrule. This is particularly interesting, as it is rather hard to avoid fulfilling the qualitative behavior in general and this sumrule in particular in studies using functional methods \cite{Huber:2012td,Dudal:2012td}.

It needs to be emphasized that the effective infrared exponents are determined dropping the two lowest non-zero momenta, and should thus be comparatively stable against lattice artifacts.

In dimensions higher than two, there is also substantial difference in the effective infrared exponent, as is seen in figure \ref{fig:ghpir}. However, the actual value in these dimensions in minimal Landau gauge is very likely zero \cite{Maas:2011se,Cucchieri:2007ta,Cucchieri:2007uj,Cucchieri:2016jwg,Maas:2014xma}. Therefore, even small deviations may not imply more than small quantitative deviations. Still, any non-zero value in another gauge would here lead to a qualitative different behavior, as was speculated in \cite{Fischer:2008uz,Maas:2009se,Maas:2008ri} to be perhaps possible. This seems not be the case, but this is better visible when studying the running coupling.

\section{The running coupling}\label{s:alpha}

The renormalization-scale invariant running coupling in the miniMOM scheme is defined as \cite{vonSmekal:1997vx,vonSmekal:2009ae}
\be
\alpha(p^2)=\frac{\alpha(\mu^2)}{(d-1)(N_c^2-1)^3}p^6P_\mn D_\mn^{aa}(p^2,\mu^2)(D_G^{aa}(p^2,\mu^2))^2\nn,
\ee
\no and can be related to the standard $\overline{\text{MS}}$ up to four-loop order in perturbation theory \cite{vonSmekal:2009ae}. It shares all the standard features of running couplings in gauge theories, in particular, it is gauge-dependent \cite{Deur:2016tte} and scheme-dependent \cite{Deur:2016tte,Maas:2011se}. It can thus be expected that it will also be sensitive to the selection of Gribov copies, which indeed it is. From the results presented in sections \ref{s:gluon} and \ref{s:ghost}, it is also possible to infer that it will be mainly dominated, in an amplified way, by the dependence of the ghost propagator on the selection of Gribov copies, as the gluon propagator is essentially unchanged. As before, the lowest non-zero momentum point is dropped in the following, because of the finite-volume effects.

The importance of investigating this running coupling is that the dimensionless combination $p^{4-d}\alpha$ was found to be (almost) compatible with a constant in the infrared in two dimensions \cite{Maas:2007uv,Maas:2014xma}, while it vanishes in the infrared in three and four dimensions \cite{Maas:2011se,Maas:2014xma,Cucchieri:2007ta,Cucchieri:2007uj,Cucchieri:2016jwg,vonSmekal:2009ae,Sternbeck:2012qs,Boucaud:2011ug}. The decisive question, posed in \cite{Fischer:2008uz,Maas:2009se,Maas:2008ri}, is whether this behavior can be altered by modifying the way Gribov copies are accounted for. According to \cite{Maas:2011se,Fischer:2008uz,Maas:2012ct}, this should not be the case, as long as only Gribov copies inside the first Gribov region are sampled, and this would only change if all Gribov regions would be sampled. In fact, it has been argued that the result should be the same for any sampling of Gribov copies inside the first Gribov region \cite{Vandersickel:2012tg,Zwanziger:1993dh}. This is therefore a very central, qualitative question.

\begin{figure}
\includegraphics[width=\linewidth]{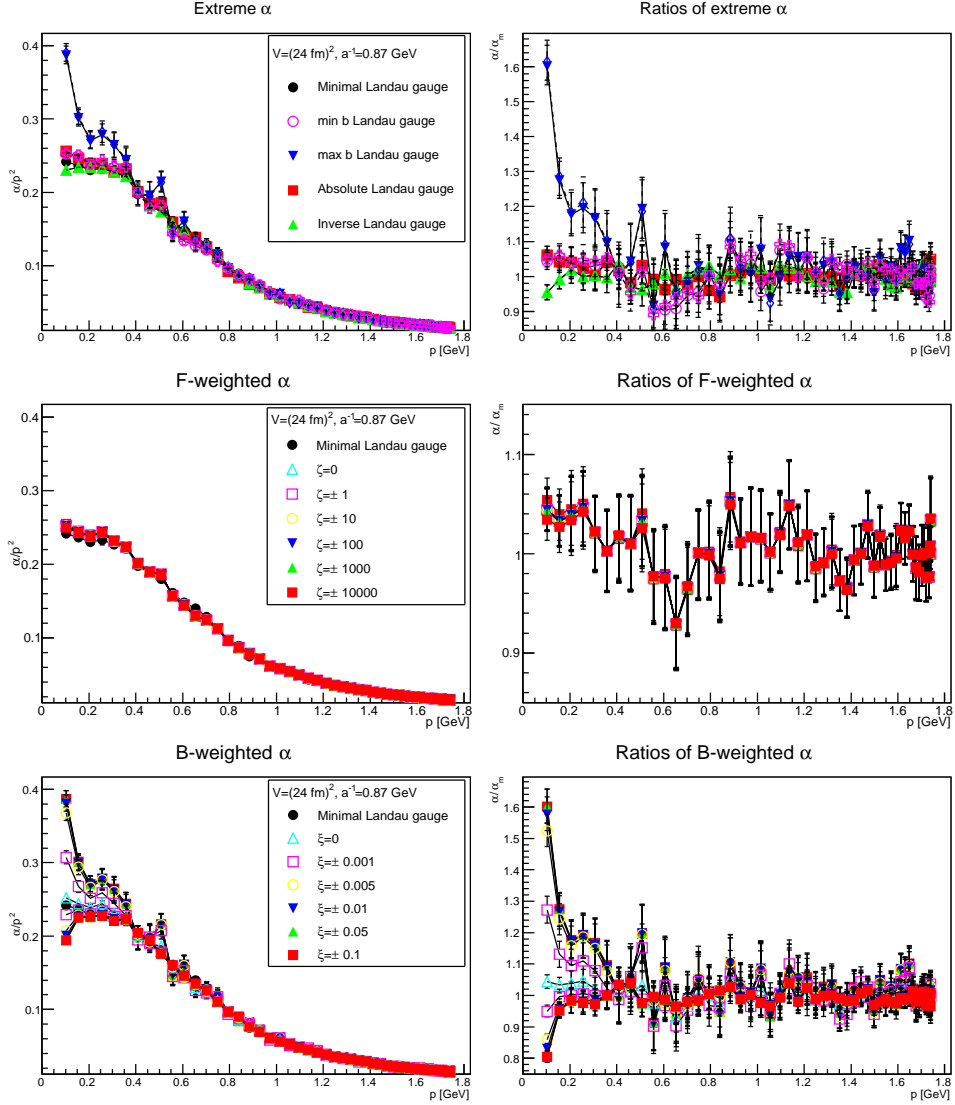}
\caption{\label{fig:2da} The running coupling itself (left panels) and its ratio with the minimal-Landau-gauge running coupling (right panels) in two dimensions. The top panels show the extreme gauges, while the bottom panels show the averaged gauges. The dashed curves are the extrapolations to an infinite number of Gribov copies, see text.}
\end{figure}

\begin{figure}
\includegraphics[width=\linewidth]{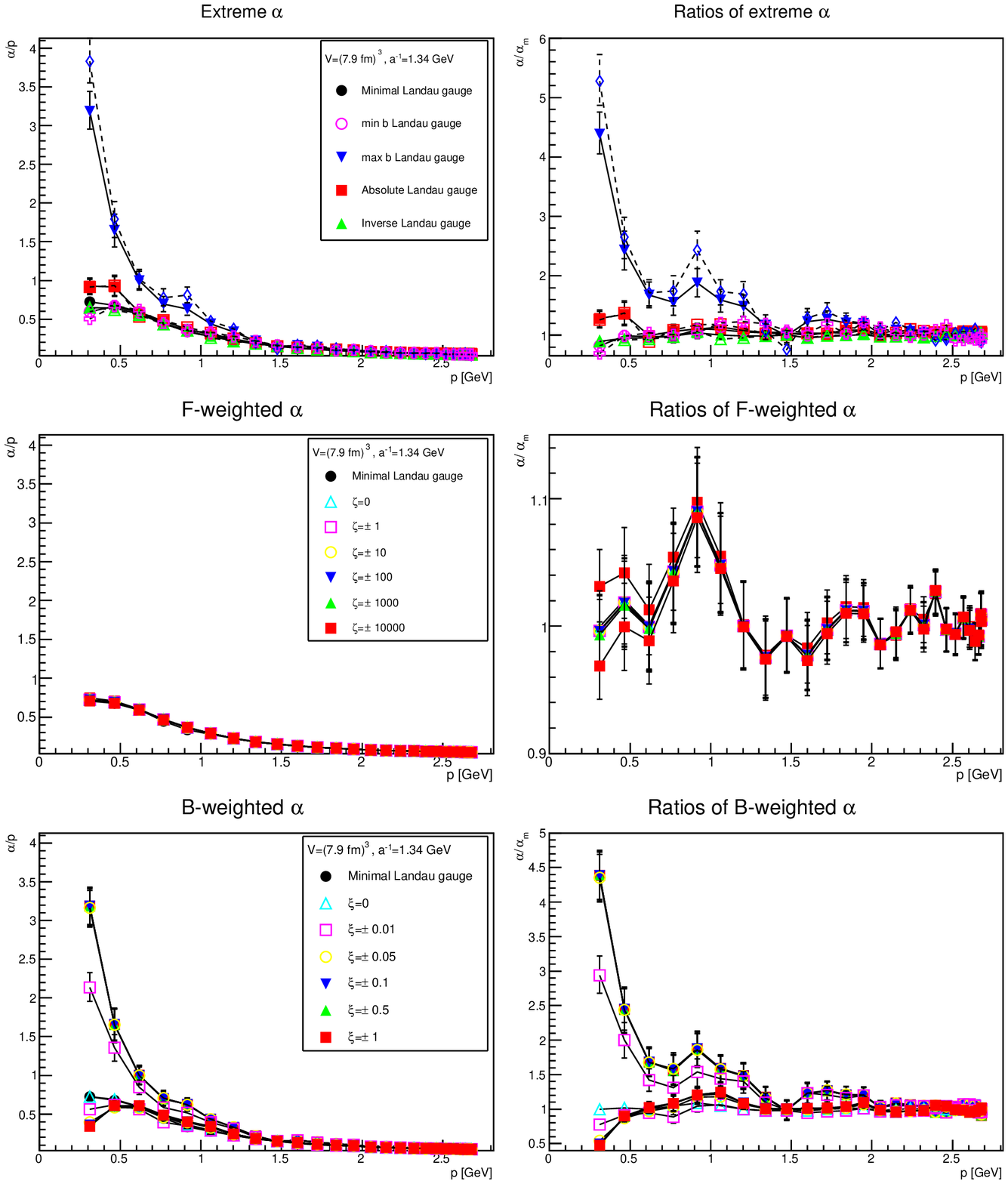}
\caption{\label{fig:3da} The running coupling itself (left panels) and its ratio with the minimal-Landau-gauge running coupling (right panels) in three dimensions. The top panels show the extreme gauges, while the bottom panels show the averaged gauges. The dashed curves are the extrapolations to an infinite number of Gribov copies, see text.}
\end{figure}

\begin{figure}
\includegraphics[width=\linewidth]{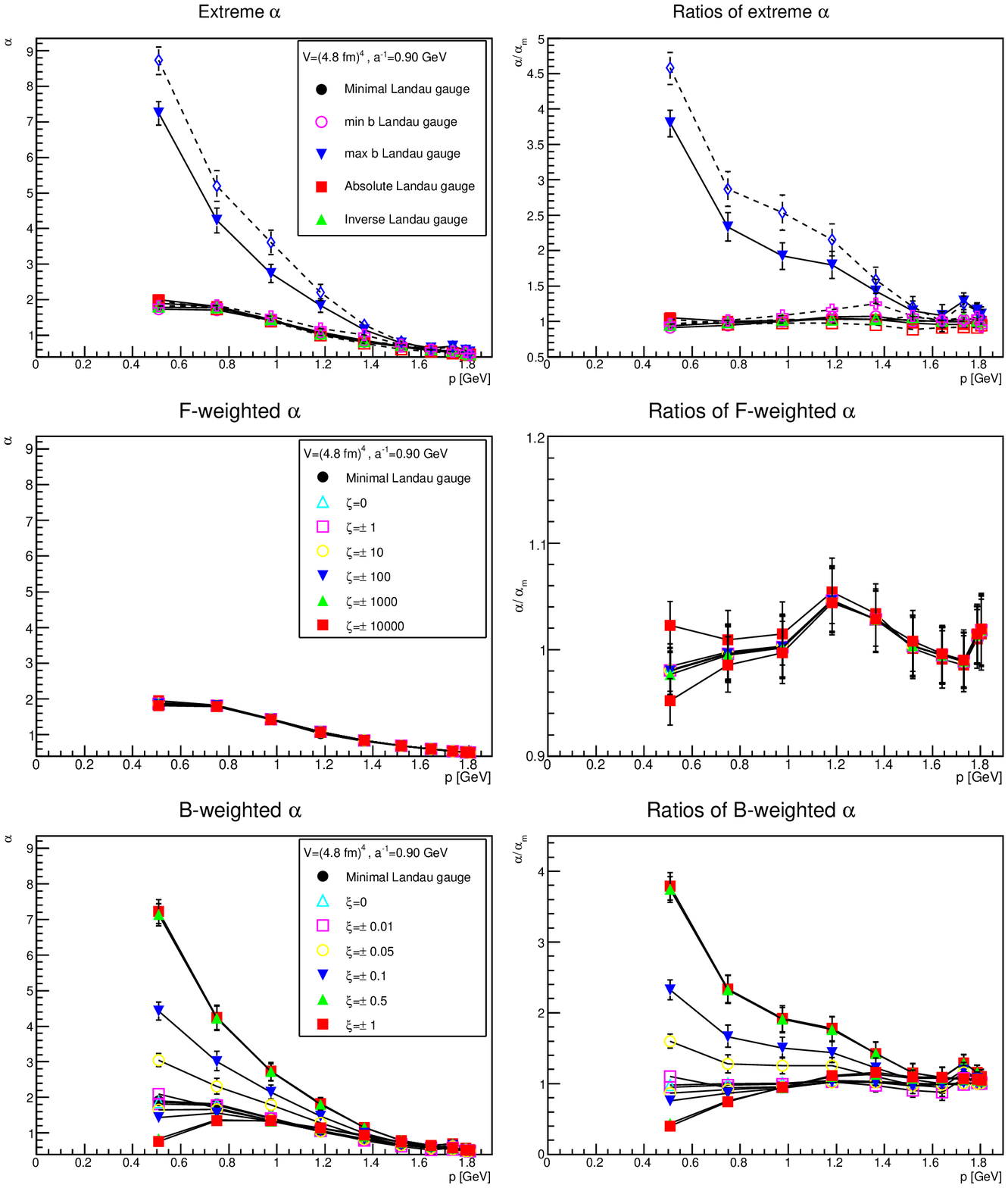}
\caption{\label{fig:4da} The running coupling itself (left panels) and its ratio with the minimal-Landau-gauge running coupling (right panels) in four dimensions. The top panels show the extreme gauges, while the bottom panels show the averaged gauges. The dashed curves are the extrapolations to an infinite number of Gribov copies, see text.}
\end{figure}

\begin{figure}
\includegraphics[width=\linewidth]{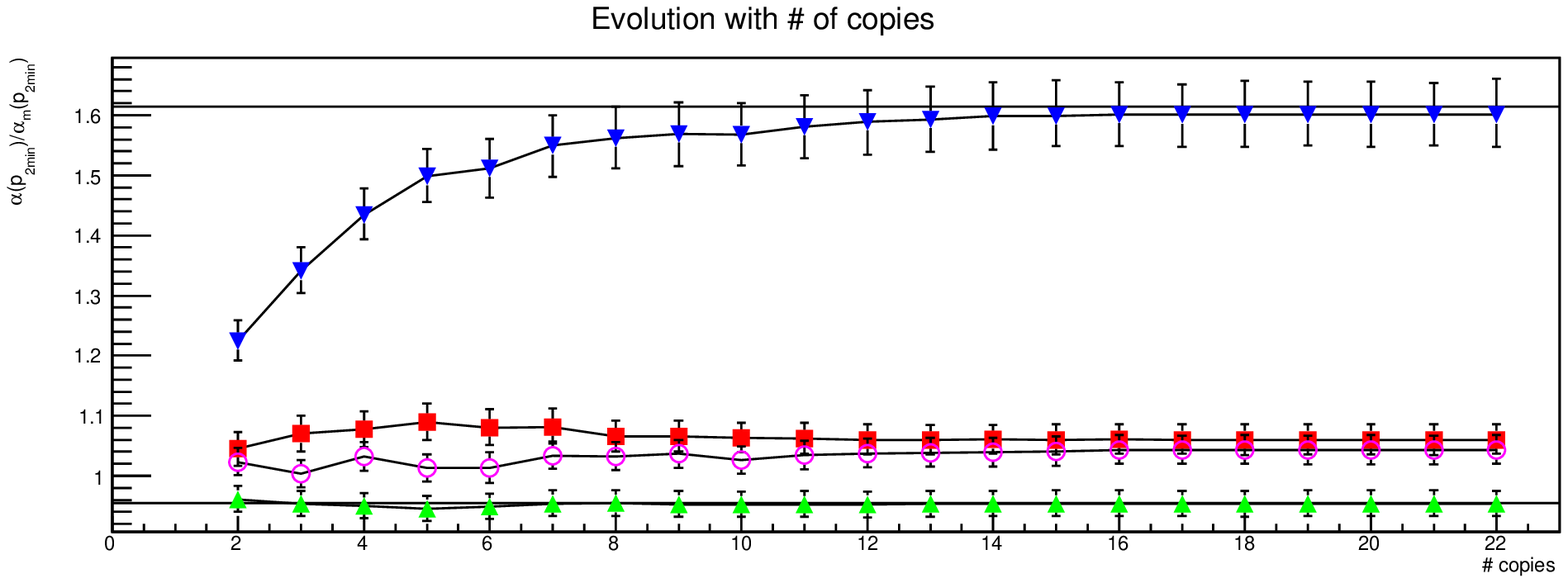}\\
\includegraphics[width=\linewidth]{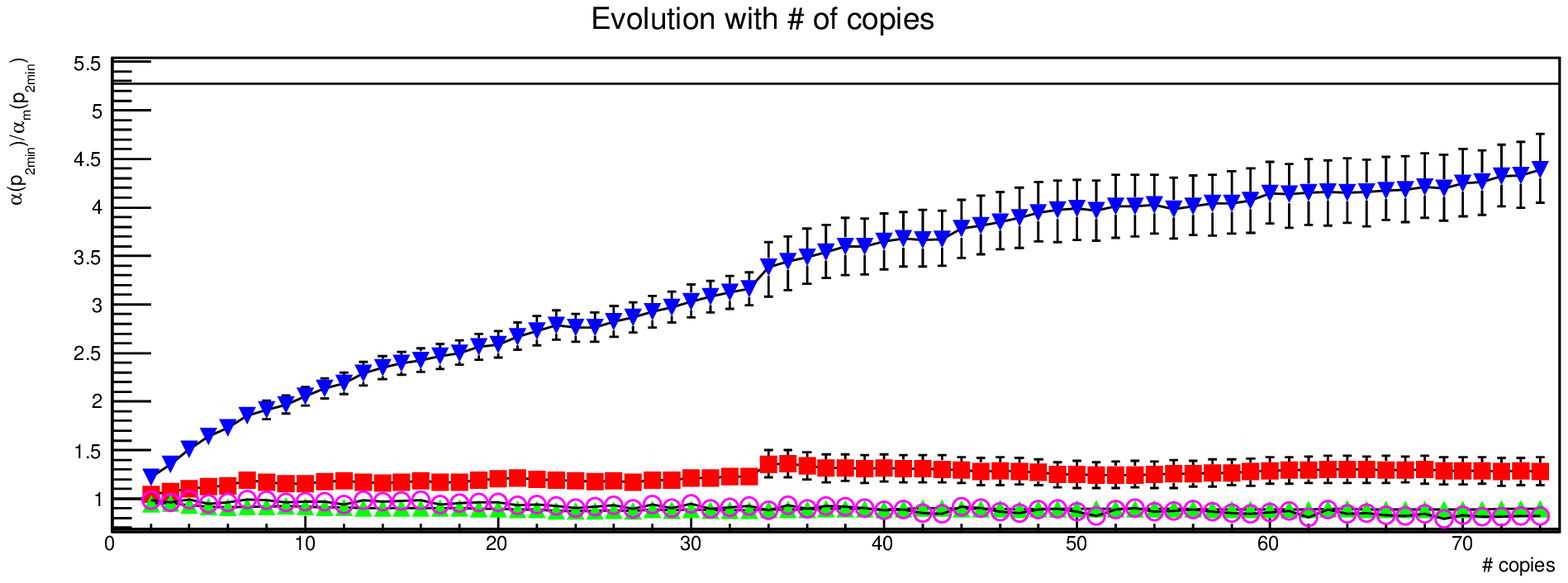}\\
\includegraphics[width=\linewidth]{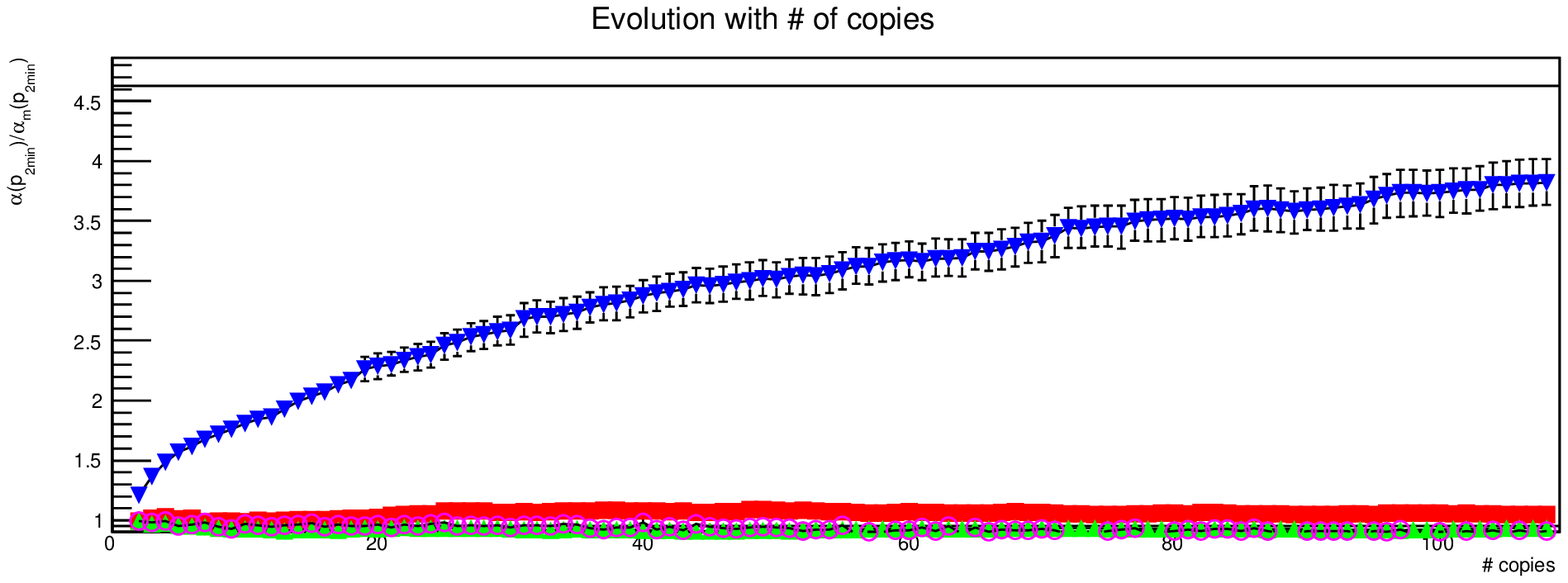}
\caption{\label{fig:alphage} Dependence of the ratio of the running coupling at the smallest included momentum in the extreme gauges to the ones in minimal Landau gauge as a function of the number of included Gribov copies for the largest physical volumes at the finest available lattice spacings, with top, middle, and bottom panels being for two, three, and four dimensions, respectively. The lines are the extrapolations discussed in the text. The symbols have the same meaning as in figures \ref{fig:2da}-\ref{fig:4da}.}
\end{figure}

The results are shown in figures \ref{fig:2da}-\ref{fig:4da} for two, three, and four dimensions, respectively. The extrapolations, based on \pref{extra} are also shown in these figures. The development with the number of included Gribov copies at the lowest included momentum is shown in figure \ref{fig:alphage}. 

In four dimensions the results show always the same qualitative behavior as in minimal Landau gauge, even after including the Gribov-copy dependency. However, the results cannot reach far enough into the infrared to see the characteristic bending-over of the coupling towards the infrared for every choice of gauge. However, as will be seen in section \ref{s:artifacts}, even though finite-volume effects tend to enhance the running coupling in the infrared, the qualitative behavior does not change. Thus, the running coupling remains infrared vanishing, and the qualitative feature of the propagators remain the same in the first Gribov region. However, the rate substantially depends on the choice of gauge, and there is no sign that all gauges show a common quantitative behavior.

The situation in two dimensions is quite different. While a plateau develops for some momenta below roughly 400 MeV, though the value of the plateau is gauge-dependent, there seems to be some non-universal behavior in the deep infrared. However, this affects in a statistically significant way almost always only the lowest few momentum points for the more extreme Landau-$b$ gauges to the max$b$ gauge. Given that this region is strongly affected by finite-volume artifacts \cite{Maas:2007uv}, see also section \ref{s:artifacts}, this may therefore not be real. Note that in two dimensions again the limiting behavior as a function of Gribov copies is still (almost) reached, as a consequence of that it is reached for the ghost dressing function. 

In three dimensions the situation is somewhat more involved. It seems that the behavior is qualitatively different for the different gauges. Especially, the running coupling seems still to rise for the max$b$ gauge, while it vanishes for the min$b$ gauge. The other gauges show a behavior in between. Given that the volume is just so border-line to where the bending over in three dimensions becomes visible \cite{Maas:2014xma}, this may actually be a finite-volume artifact.

Neither in three nor in four dimensions is seen any tendency to a common result for all possible gauges.

All of this together strongly suggests to study the volume-dependence of the gauge dependence, most notably of the max$b$ gauge. This is is subject of section \ref{ss:volume} below. The results are in line, as far as they can be tracked, with the interpretation above.

Thus, the cautious interpretation of the results seems to be that no qualitative change due to the way the Gribov copies are treated can be obtained. But quantitative differences seem to persist. This is in-line with the conjectures in  \cite{Maas:2011se,Fischer:2008uz,Maas:2012ct}. Note that the investigations here are the only large-scale investigations of the Gribov-copy dependence of the running coupling. However, since the gluon propagator is only weakly affected, the behavior of the running coupling is anyway driven by the ghost propagator. And for this the results found here are in line with the older results of \cite{Cucchieri:1997dx,Bogolubsky:2005wf,Bornyakov:2008yx,Maas:2008ri,Maas:2009se,Maas:2011ba,Maas:2011se,Sternbeck:2012mf,Maas:2013vd,Maas:2016frv}.

\section{Lattice artifacts}\label{s:artifacts}

As the results in \cite{Maas:2015nva} already suggest, the most relevant lattice artifact is the size of the physical volume. Since the strongest impact of Gribov copies has been found for the extreme Landau-$b$ gauges and for the running coupling, the best choice to study systematic effects will be these. Thus, in the following the impact of the discretization and the physical volume for the ratio of the running coupling between the minimal Landau gauge and the extreme Landau-$b$ gauges will be studied. As the behaviors found indicate the presence of substantial lattice artifacts, also some lattice setups were included for which it was not possible to obtain sufficient statistics to meet the demands of previous sections, but help already somewhat in the following. These lattice settings are marked in table \ref{tcgf} in the configurations column with an asterisk. The requirement to obtain sufficient statistics for them were however such that a full calculation was too expensive at the current time.

\subsection{Discretization}

Since the strongest effects are expected at low momenta, it is quite demanding to study discretization effects. As a compromise, here only a relative ratio of about 2 between the largest and smallest value of $a$ will be covered, to still be able to include reasonably large physical volumes. This gives a range between $a^{-1}\approx0.9$ GeV to $a^{-1}\approx2.5$ GeV. As here the lattice artifacts should be studied in detail, all momenta are included. Thus, the plots reach further intro the infrared than in the previous sections.

\begin{figure}
\includegraphics[width=\linewidth]{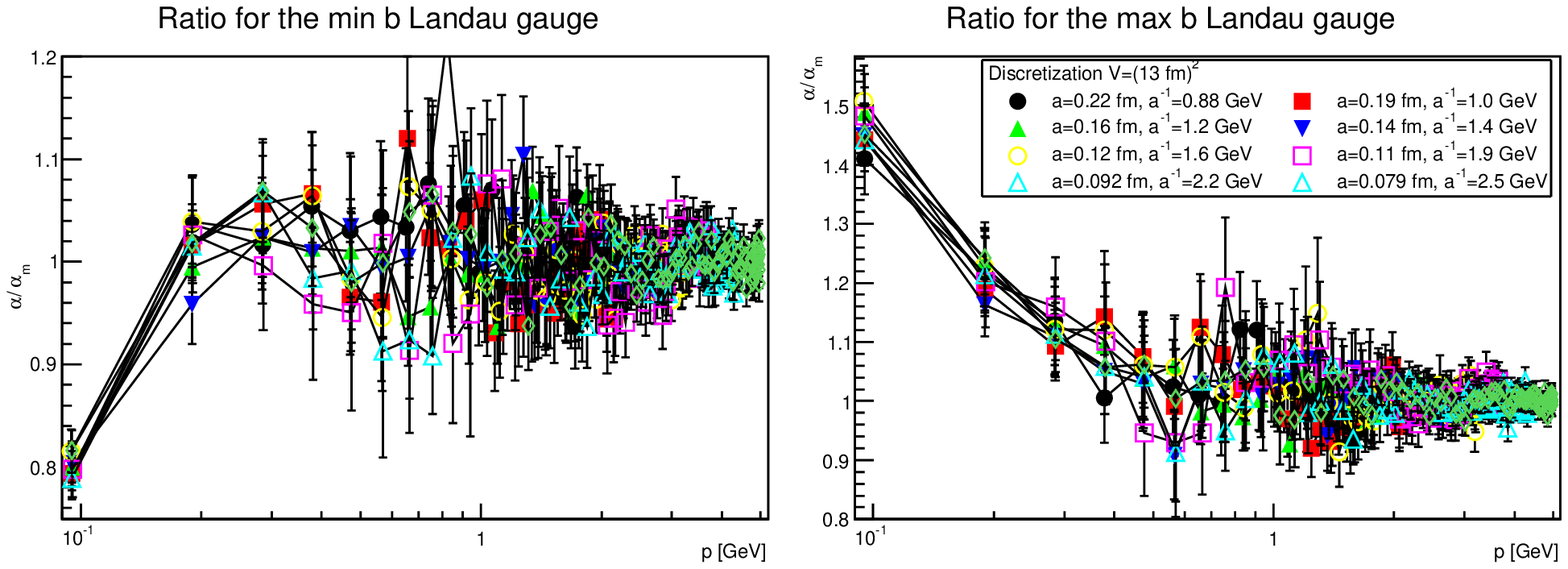}\\
\includegraphics[width=\linewidth]{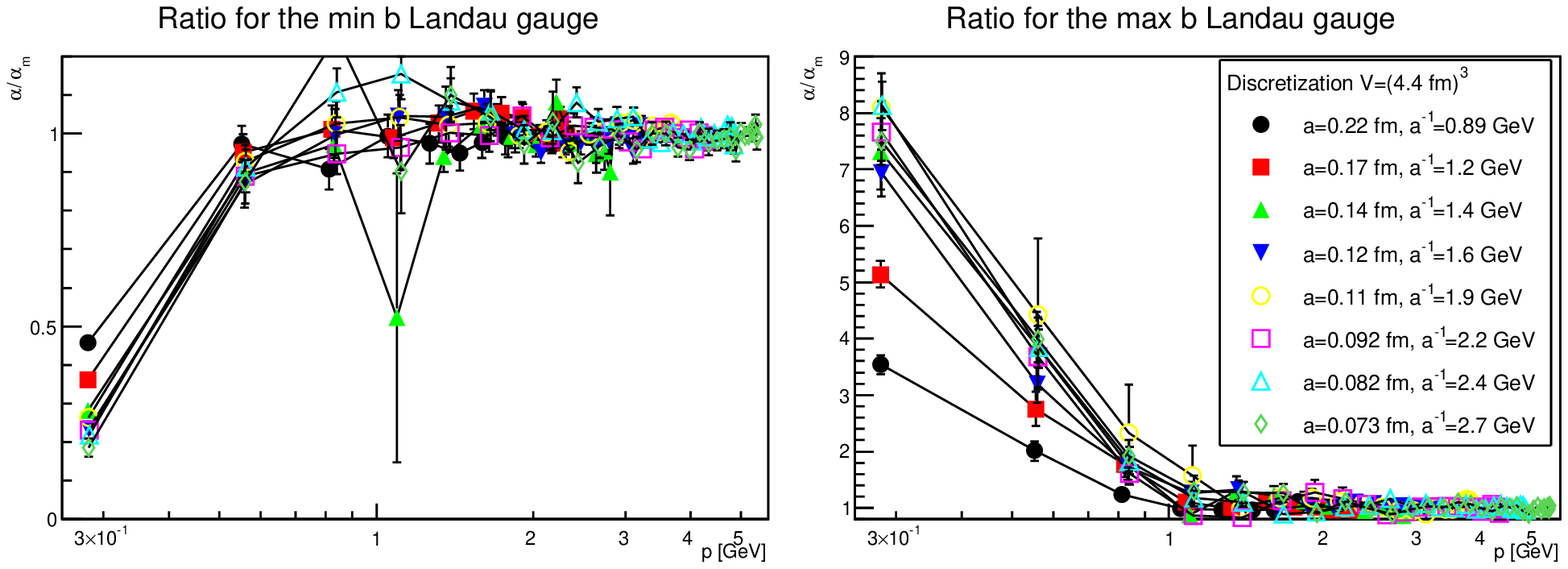}\\
\includegraphics[width=\linewidth]{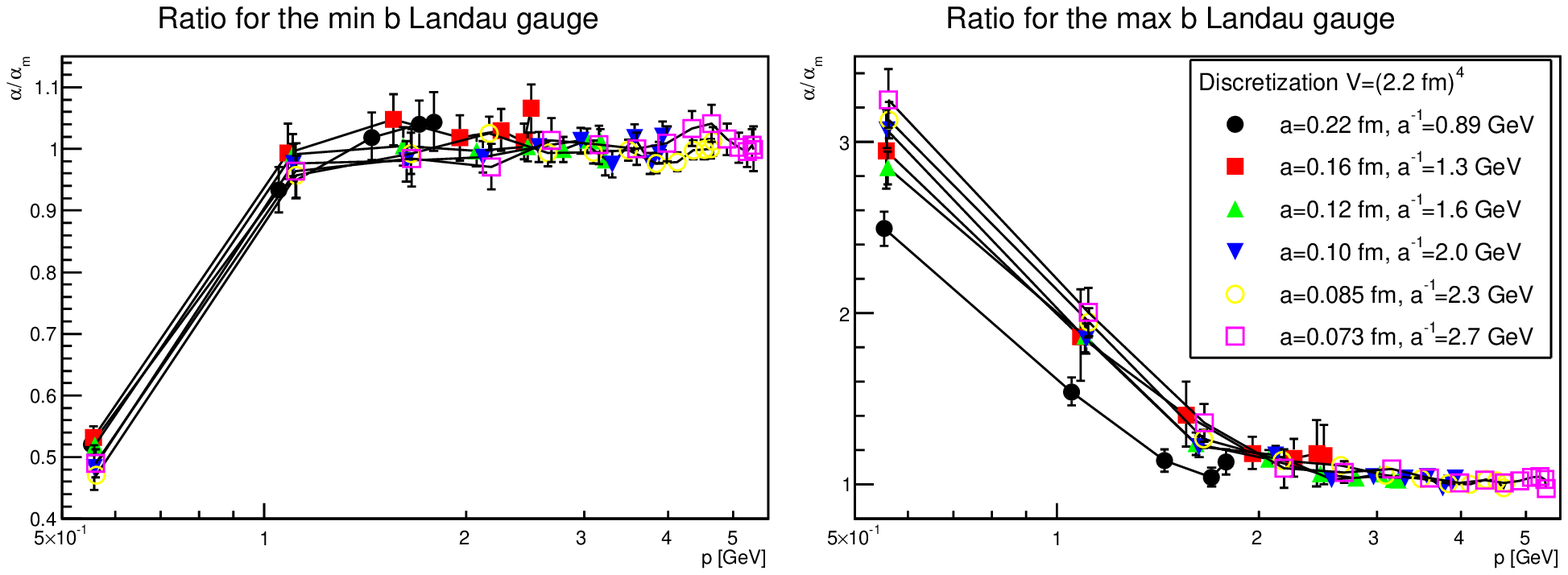}
\caption{\label{fig:alphaa} Dependence of the ratio of the running coupling the min$b$ gauge (left panels) max$b$ gauge (right panels) to the minimal Landau gauge for different discretization values at fixed physical volumes. The top, middle, and bottom panels shows results for two, three, and four dimensions, respectively.}
\end{figure}

The results are shown in figure \ref{fig:alphaa}. The results follow the pattern for the Gribov copies themselves found in \cite{Maas:2015nva}. In two dimensions, there are no statistically significant effects. In three and four dimensions, there is some effect when moving from the coarsest discretization to finer ones, but this slows down substantially above roughly $a^{-1}\approx1.5$ GeV. These effects are smaller for the min$b$ Landau gauge than for the max$b$ Landau gauge. Especially, only for the max$b$ Landau gauge an effect is seen for momenta larger than the smallest possible one on the given lattice. There may therefore be also a non-trivial interplay with finite-volume effects to be studied next. Note, however, that with finer lattices the deviations from minimal Landau gauge increase.

\begin{figure}
\includegraphics[width=\linewidth]{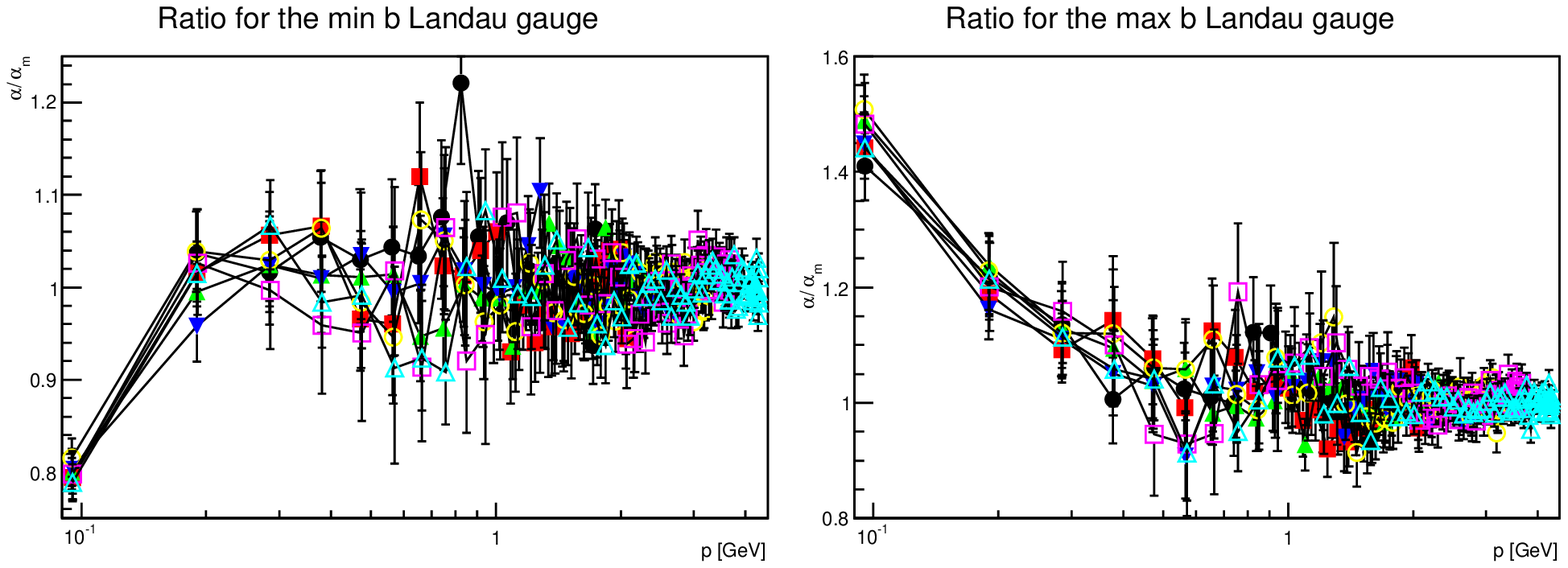}\\
\includegraphics[width=\linewidth]{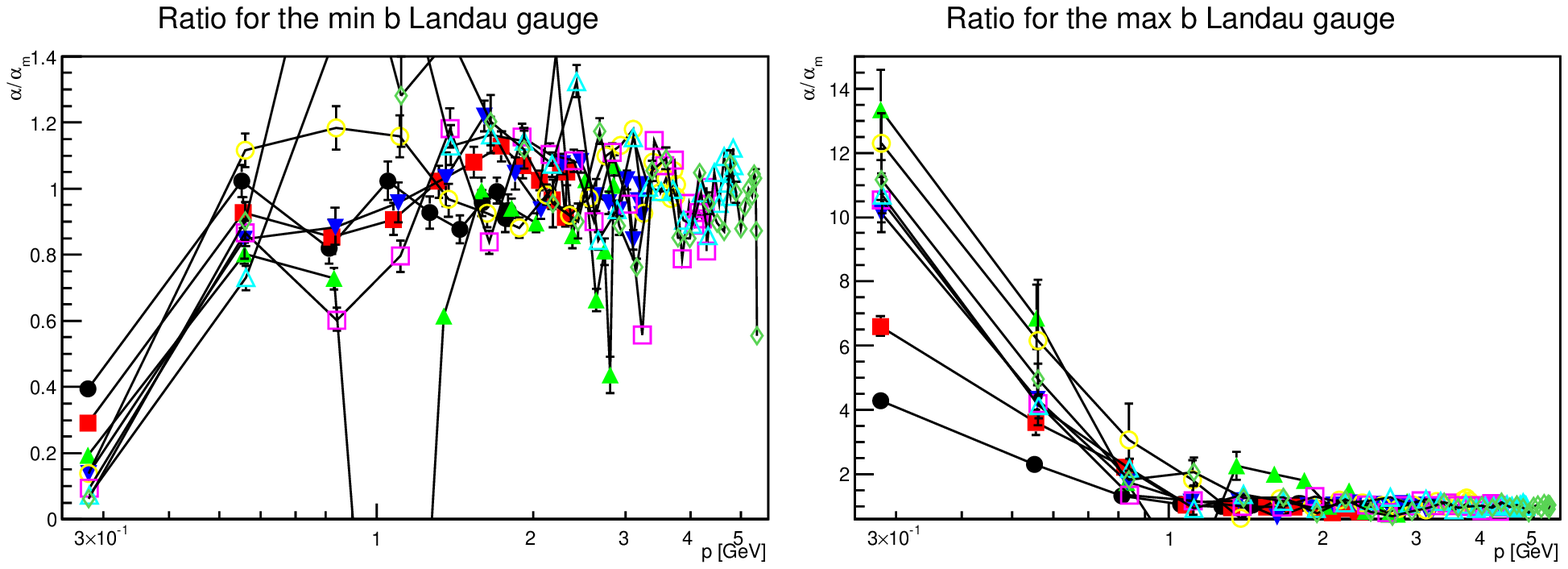}\\
\includegraphics[width=\linewidth]{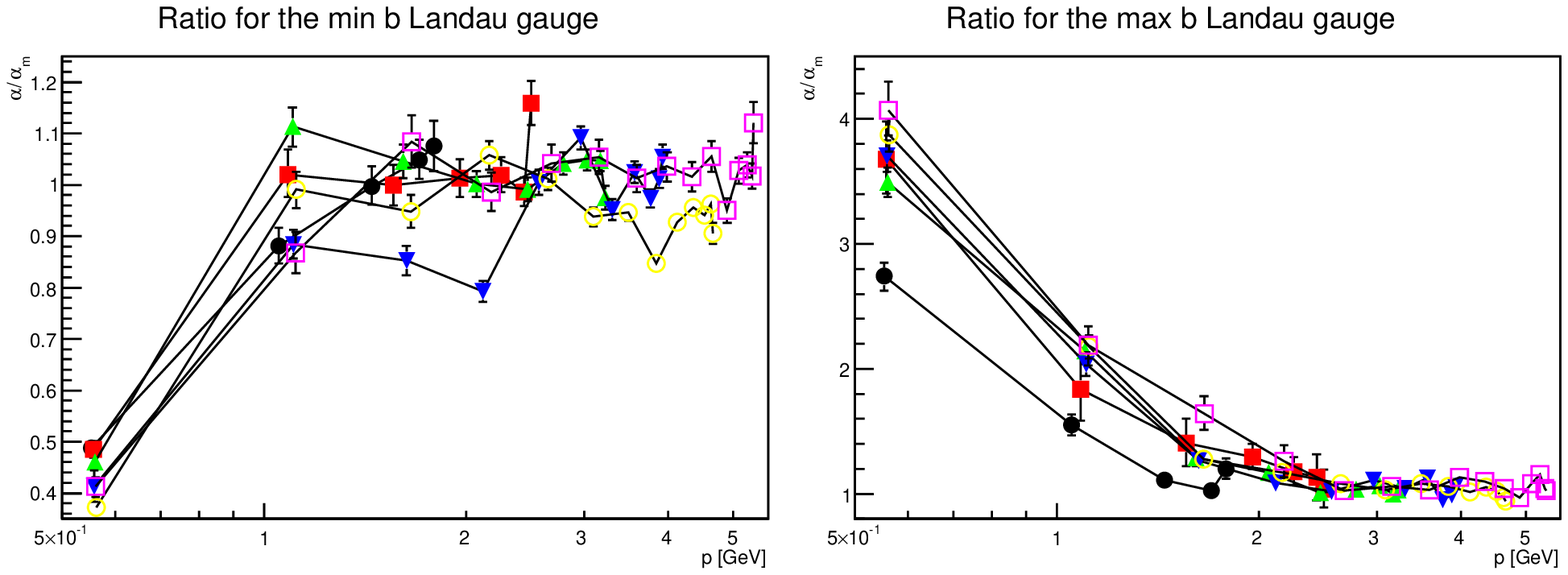}
\caption{\label{fig:alphaage} Same as figure \ref{fig:alphaa}, but extrapolated in Gribov copies as described in section \ref{s:gluon}.}
\end{figure}

Since, however, the number of Gribov copies increases with higher dimensions and finer discretizations, some of the effects may be due to capturing less and less Gribov copies. To study this, the situation is shown in figure \ref{fig:alphage} for the extrapolation in Gribov copies. Note that here, and in the next section, sometimes lattice settings have been used where only a more limited number of Gribov copies have been obtained. Therefore only the Gribov copies $N_g-10$ to $N_g$, rather than the higher numbers of the previous sections, have been used for the extrapolation. As is visible from \pref{fig:alphage}, the overall effect increases somewhat, but it also increases the statistical uncertainty, making the effect harder to judge.

\subsection{Volume}\label{ss:volume}

\begin{figure}
\includegraphics[width=\linewidth]{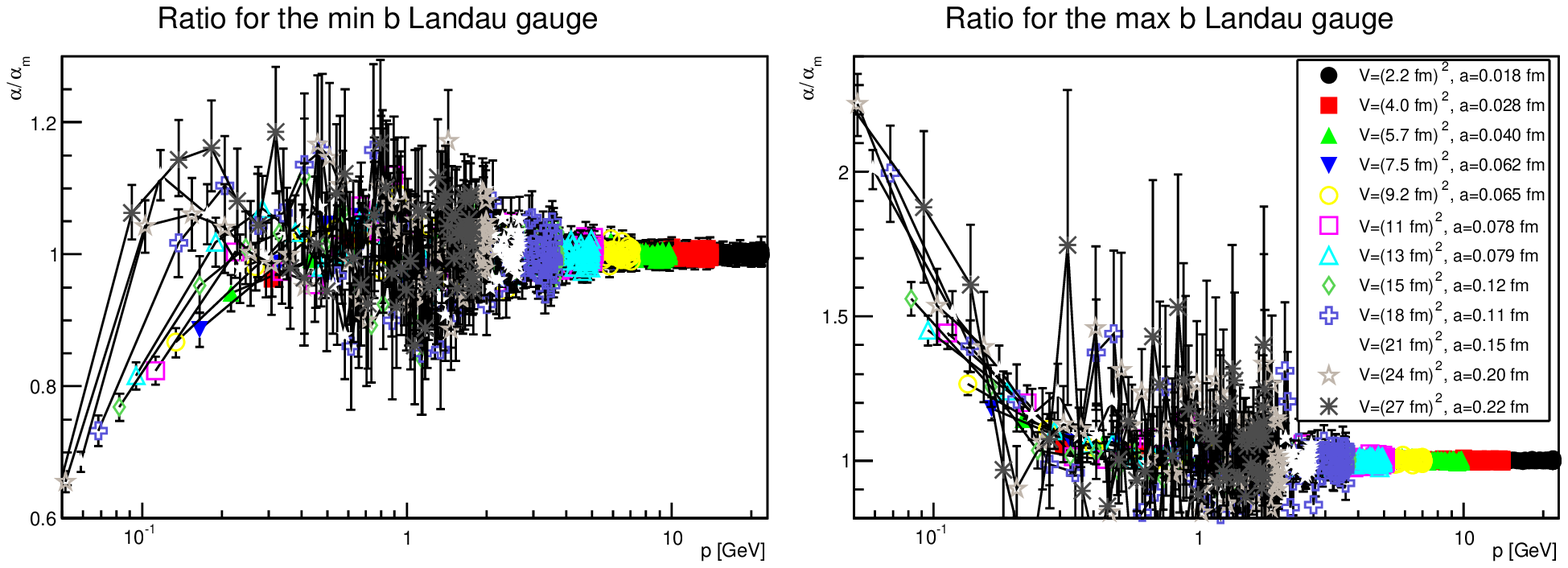}\\
\includegraphics[width=\linewidth]{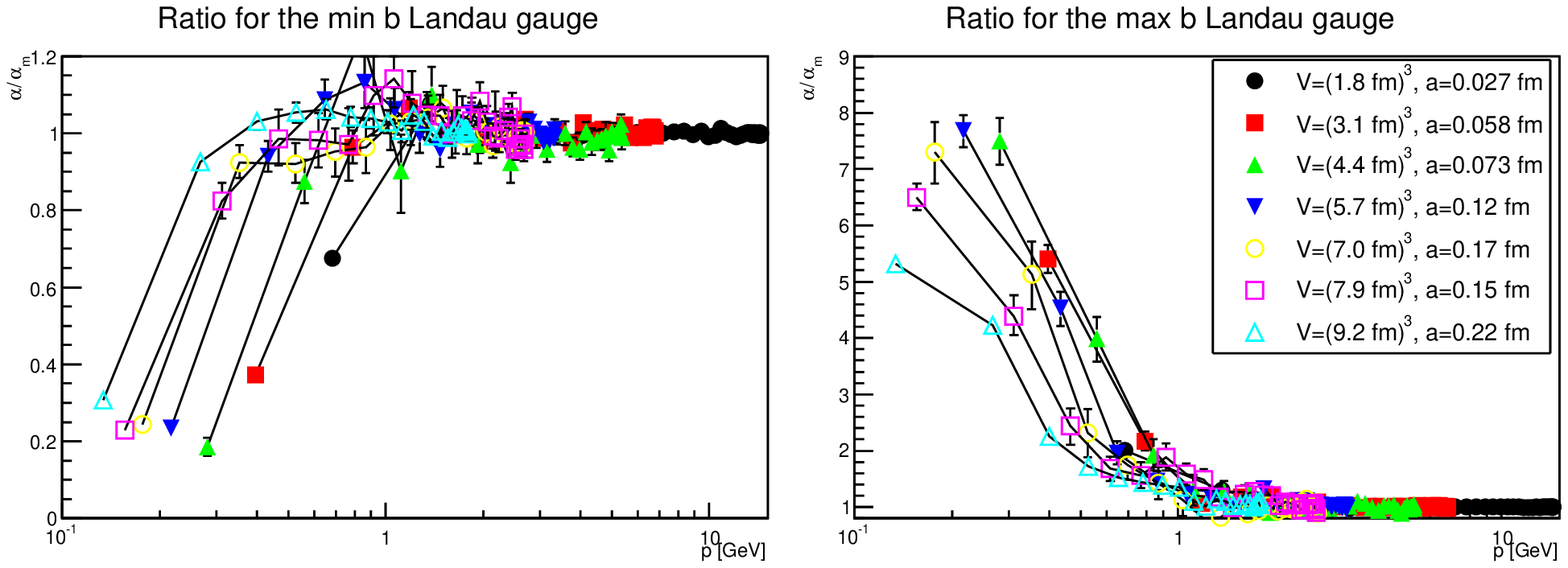}\\
\includegraphics[width=\linewidth]{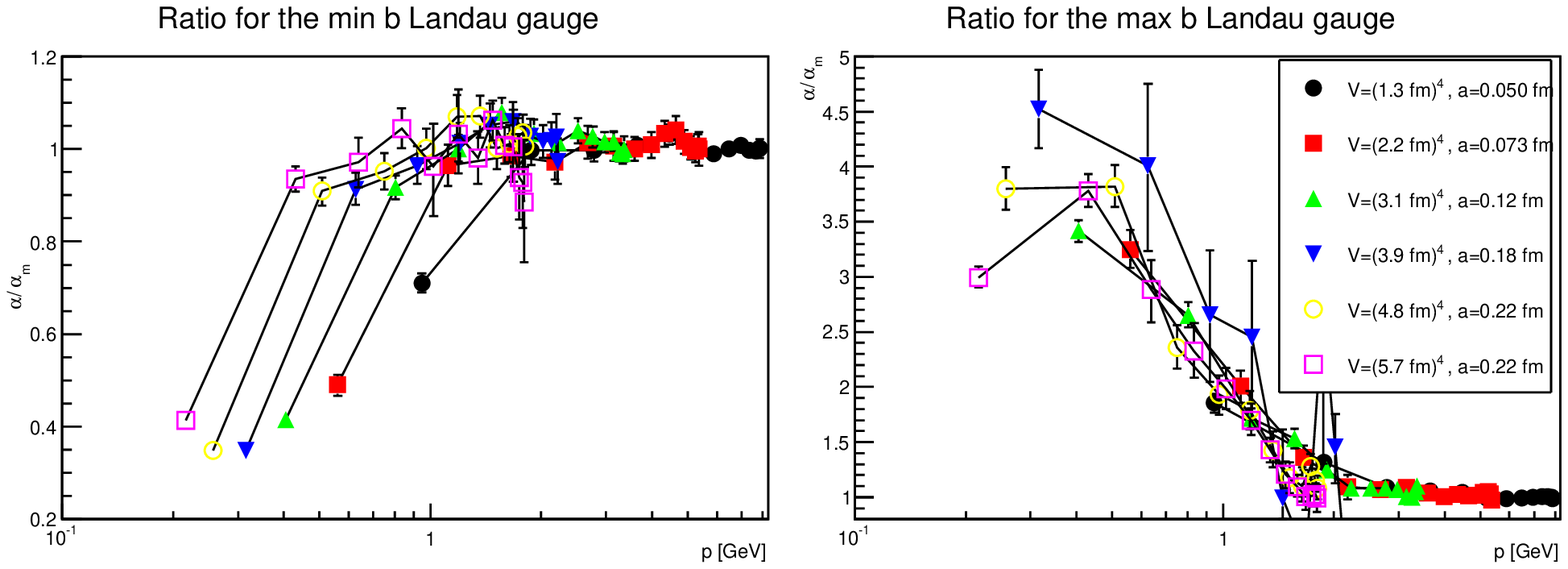}
\caption{\label{fig:alphav-dep} Dependence of the ratio of the running coupling in the min$b$ gauge (left panels) and in the max$b$ gauge (right panels) to the minimal Landau gauge for different physical volumes at the finest available discretizations for every volume. The top, middle, and bottom panels shows results for two, three, and four dimensions, respectively.}
\end{figure}

\begin{figure}
\includegraphics[width=\linewidth]{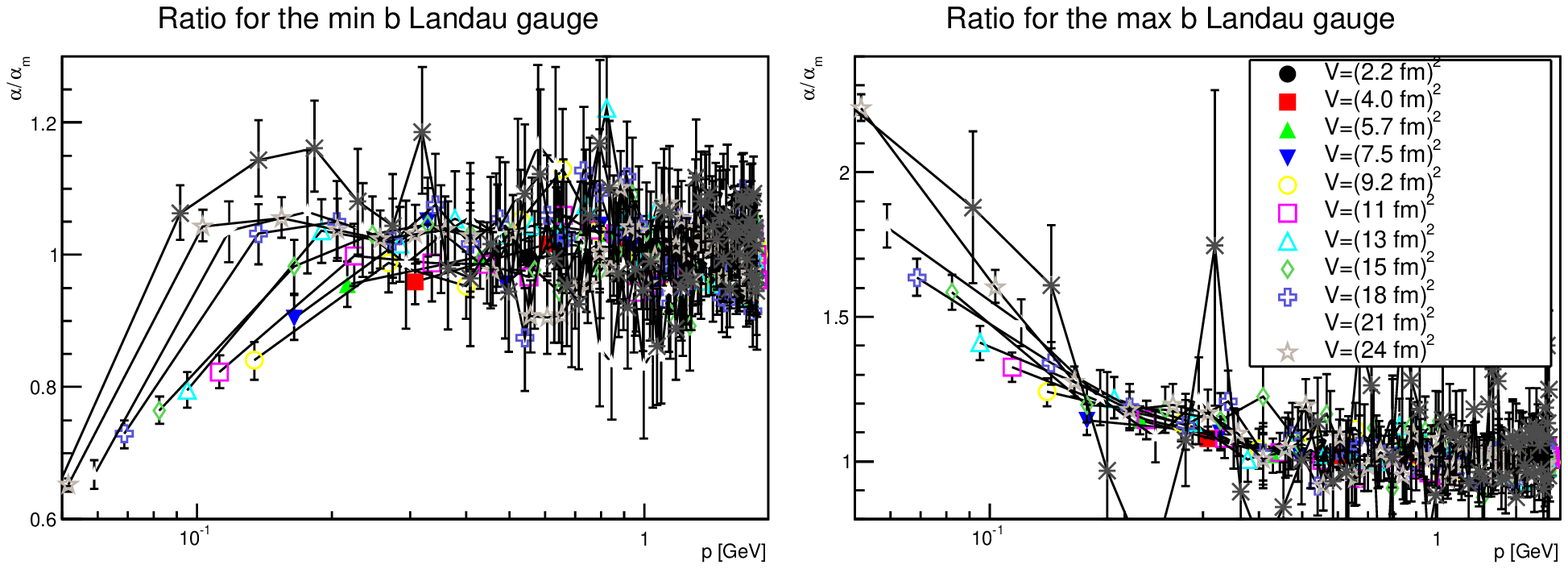}\\
\includegraphics[width=\linewidth]{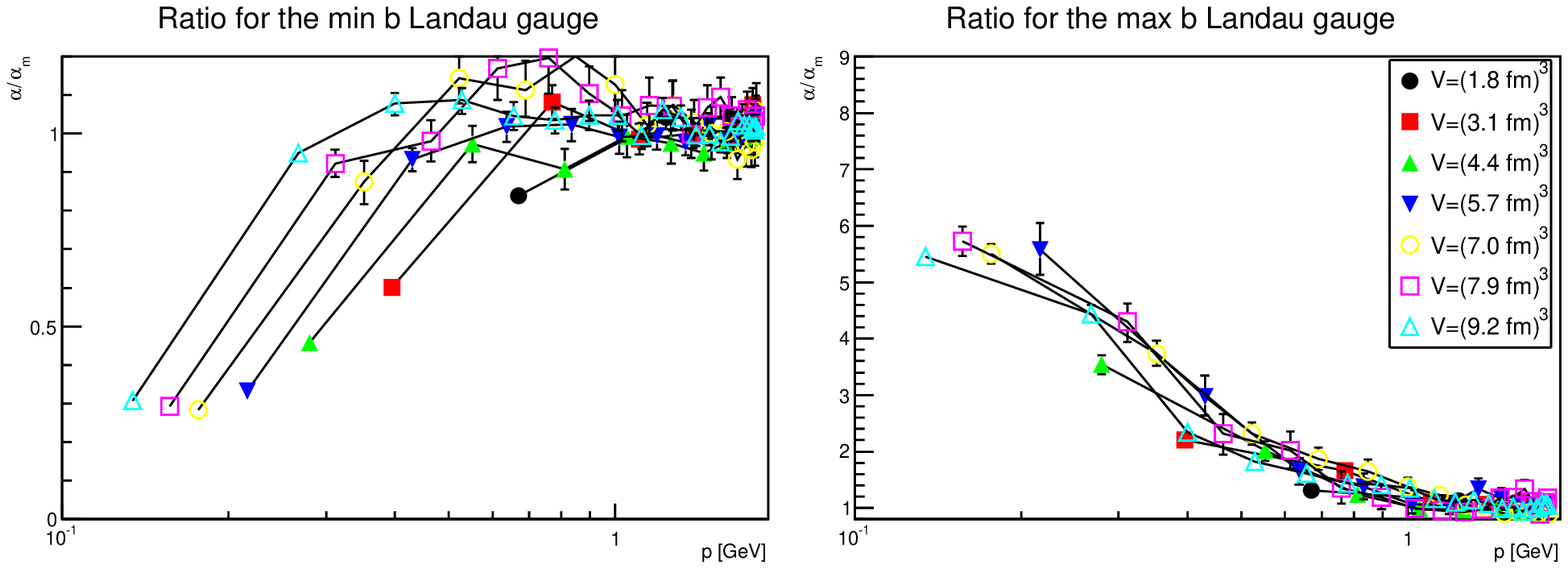}\\
\includegraphics[width=\linewidth]{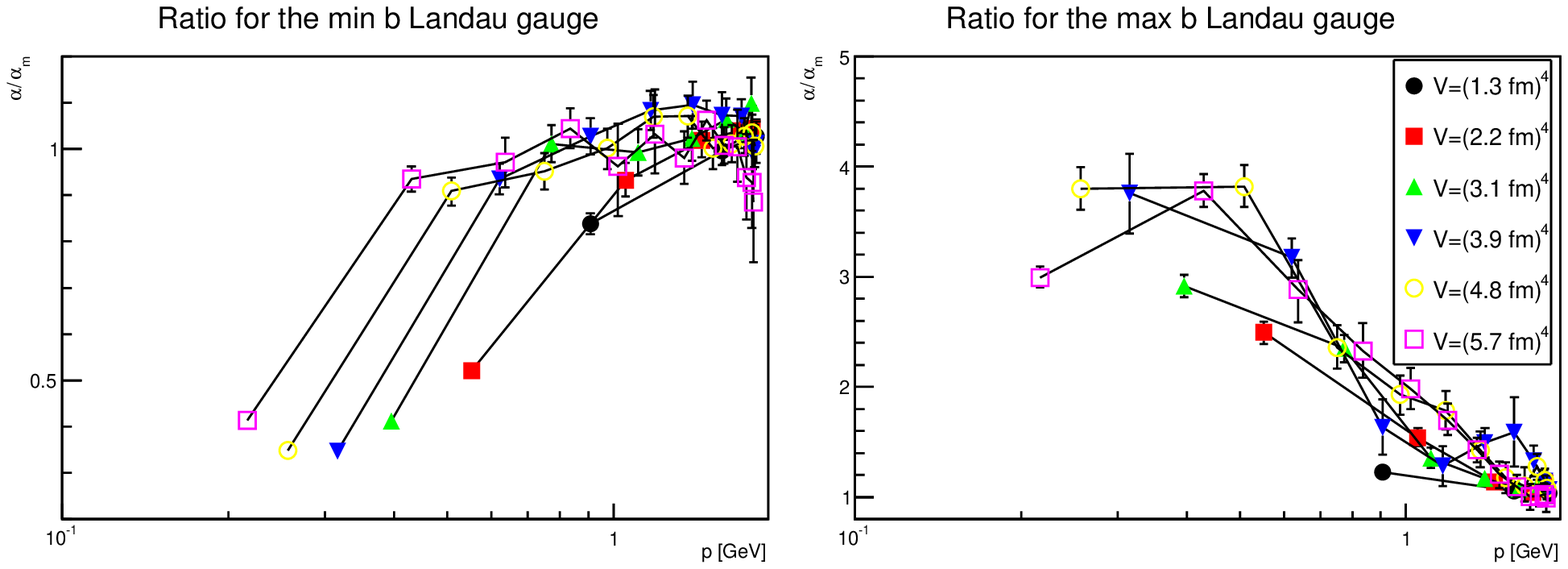}
\caption{\label{fig:alphav-dep-9} Dependence of the ratio of the running coupling in the min$b$ gauge (left panels) and in the max$b$ gauge (right panels) to the minimal Landau gauge for different physical volumes at a discretization of about $a^{-1}\approx 0.9$ GeV. The top, middle, and bottom panels shows results for two, three, and four dimensions, respectively.}
\end{figure}

\begin{figure}
\includegraphics[width=\linewidth]{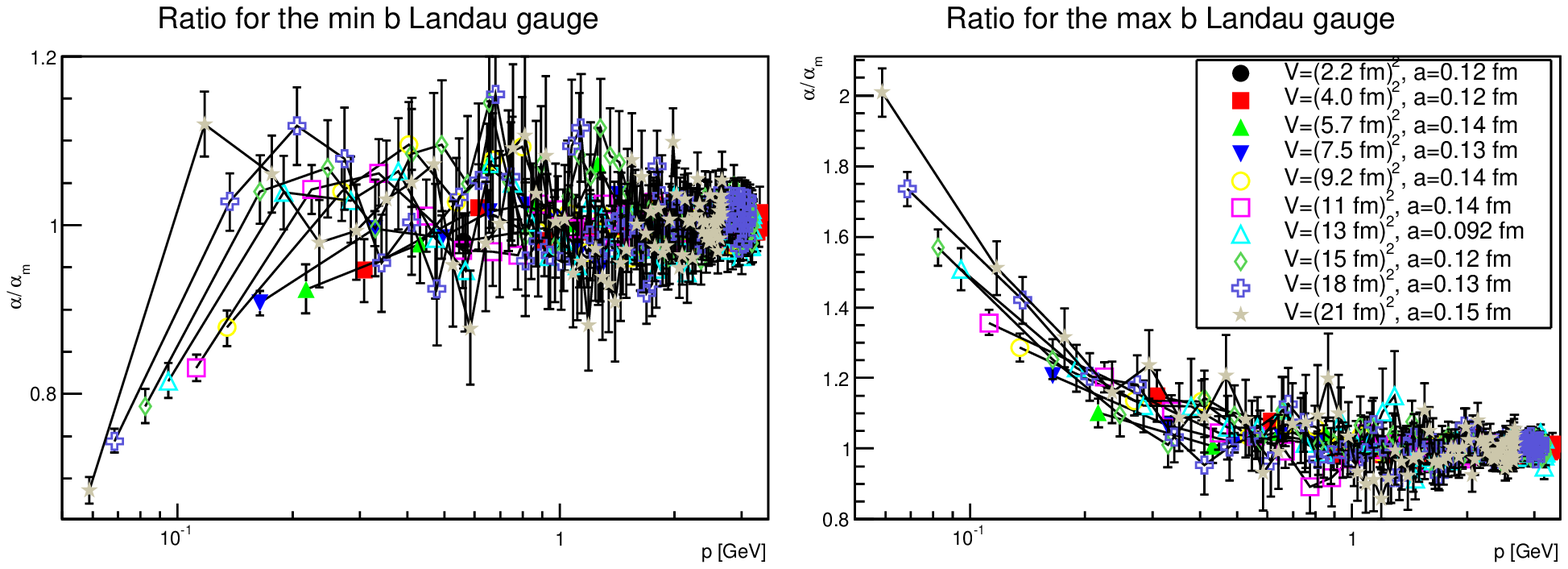}\\
\includegraphics[width=\linewidth]{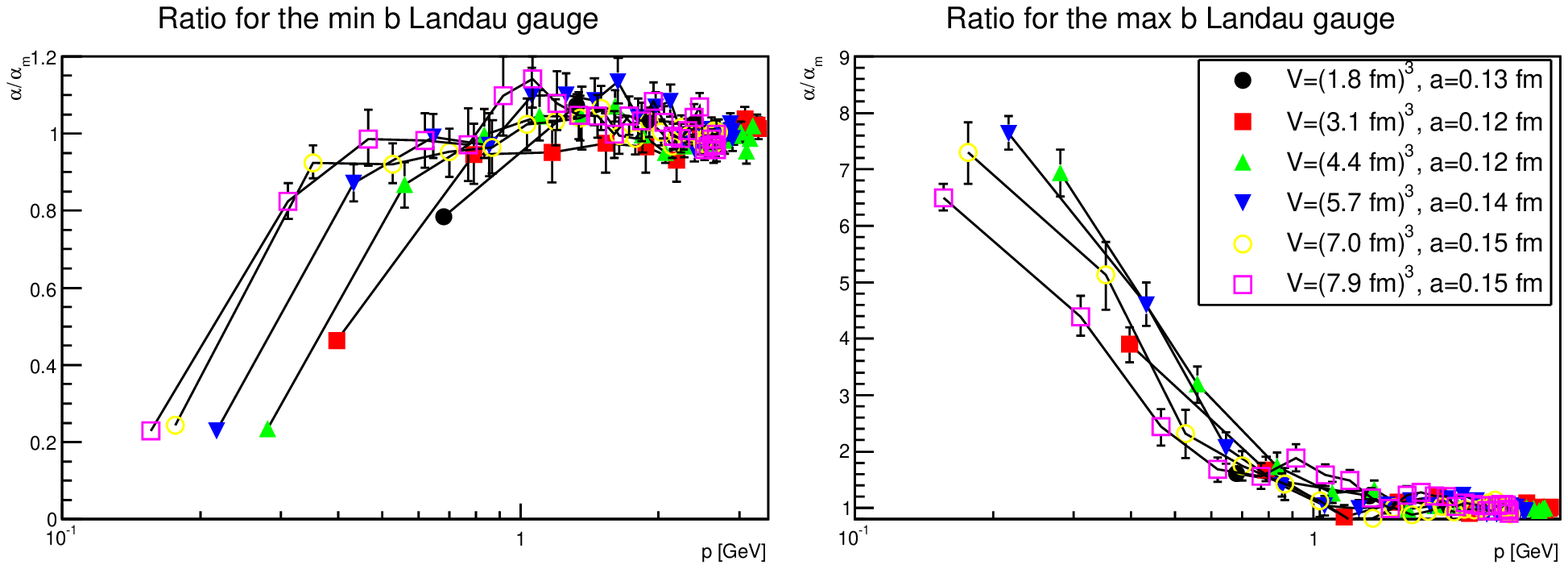}\\
\includegraphics[width=\linewidth]{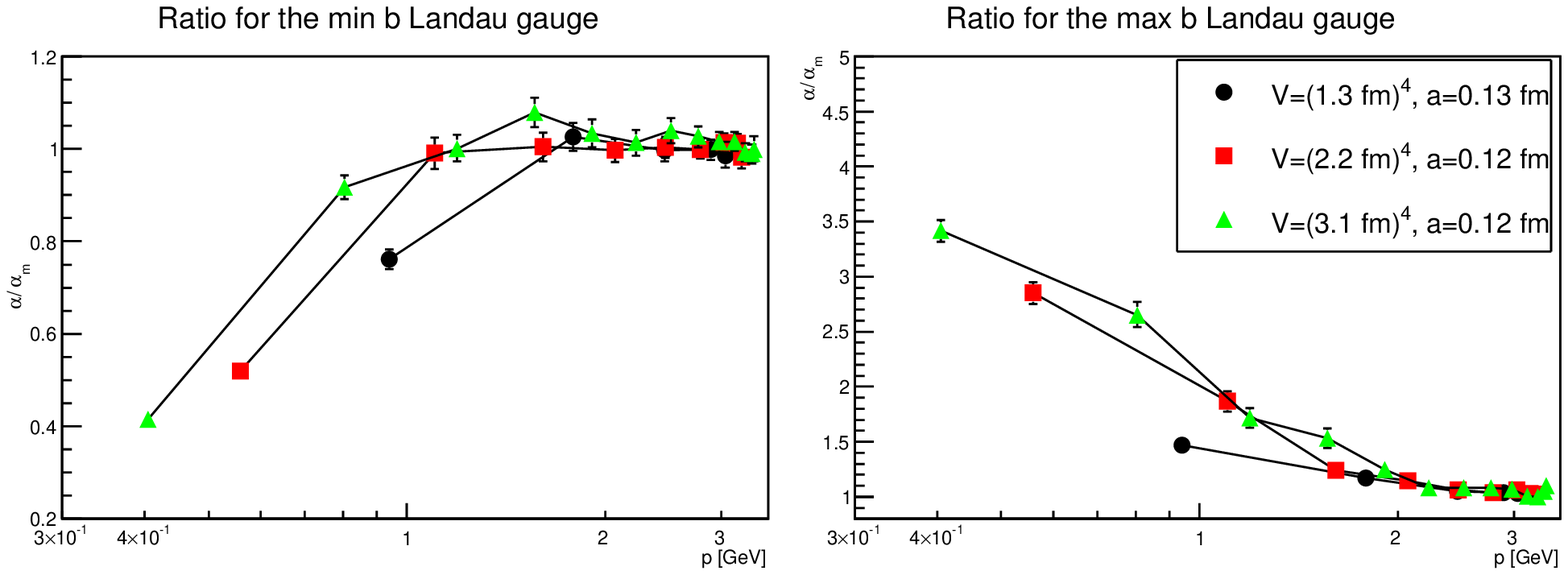}
\caption{\label{fig:alphav-dep-15} Dependence of the ratio of the running coupling in the min$b$ gauge (left panels) and in the max$b$ gauge (right panels) to the minimal Landau gauge for different physical volumes at a discretization of about $a^{-1}\approx 1.5$ GeV. The top, middle, and bottom panels shows results for two, three, and four dimensions, respectively.}
\end{figure}

Conversely, the results of \cite{Maas:2015nva} strongly suggests that the physical volume is the dominating effect, especially the higher the dimension. As will be seen, the volume effects are somewhat contradictory, so an expanded investigation is necessary. Especially, given the possible non-trivial influence of discretization effects studied before. Thus, here three different comparisons are made in figures \ref{fig:alphav-dep}-\ref{fig:alphav-dep-15}. These show the volume-dependence for the finest available discretization for every volume, as well as twice at roughly fixed lattice spacings of $a^{-1}\approx 0.9$ GeV and $a^{-1}\approx 1.5$ GeV, respectively.

The results of figure \ref{fig:alphav-dep} show a strong dependence on the volume. Especially, the impact of the choice of Gribov copies seems to decrease with increasing volume. Taking into account possible discretization effects in figures \ref{fig:alphav-dep-9} and \ref{fig:alphav-dep-15} at fixed discretization shows, however, that this effect is for the max$b$ gauge essentially offset, showing a, more-or-less, stable enhancement compared to the minimal Landau gauge, when changing the volume. This is not the case for the min$b$ gauge. Here, the lowest non-vanishing momentum point is still very strongly affected by the volume. Ignoring this lowest momentum, the suppression is strongly reduced, but at least in three and four dimensions still present.

Thus, both momentum and discretization artifacts play a relevant role, and actually compete with each other. As a consequence, in the main part of the text the lowest non-vanishing momentum point is dropped for the ghost and the running coupling to account for the dominant part of the lattice artifacts. The extent of the influence for the next-to-lowest non-vanishing momentum point is much weaker. In fact, here the systematic effects become comparable to the statistical effects.

\section{Summary}\label{s:sum}

Herein, a systematic study of the dependency of the gluon and ghost propagators on a number of ways to sample the first Gribov region has been presented, for a range of lattice parameters, and for two, three, and four dimensions. All results found are in agreement with the conjectures summarized and reviewed in \cite{Maas:2011se}: While there is a quantitative dependency on the way how to sample Gribov copies, there is no qualitative dependency. However, this is a numerical lattice investigation, and therefore this can only apply to the extent of the lattice parameters sampled. In fact, the detailed investigations in section \ref{s:artifacts} demonstrate that especially very low momenta are more strongly influenced by both volume and discretization than in the usual minimal Landau gauge. But, as shown in the other parts of the text, also a sufficient sampling of Gribov copies is an important additional systematic effect.

Extrapolating from here naturally depends on any bias in the extrapolation procedure. Still, given that the physical volumes are somewhat moderately large enough to see the onset of the large-volume behavior \cite{Maas:2011se,Maas:2014xma,Cucchieri:2007ta,Cucchieri:2007uj,Cucchieri:2016jwg,vonSmekal:2009ae,Sternbeck:2012qs,Boucaud:2011ug}, it appears not completely unlikely that this already gives a reasonable idea of the infinite-volume and continuum behavior. Then, it appears quite possible that the effects remain even when going to the infinite-volume and continuum limit.

It addition, it was shown that many of the more commonly used gauges on the lattice can actually be rewritten in terms of samplings of the path integral \pref{zetaxi}. While this is not yet providing a continuum description of the same gauges, it is certainly a step in this direction. This is especially important when comparing lattice and continuum results, especially from functional methods \cite{Alkofer:2000wg,Fischer:2006ub,Binosi:2009qm,Maas:2011se,Boucaud:2011ug,Vandersickel:2012tg,Pawlowski:2005xe,Gies:2006wv}. Though some conjectures on the implementation of these prescriptions exist, which also give good agreement between functional and lattice results \cite{Maas:2011se,Fischer:2008uz,Maas:2009se,Sternbeck:2012mf,Cyrol:2016tym}, this is by far not a sufficient justification. Thus, formal questions remain.

Given that functional calculations\footnote{Note that for quasi-perturbative calculations like \cite{Vandersickel:2012tg,Reinosa:2013twa,Serreau:2012cg}, this may be of less relevance, as here such subtleties may still drop out. However, also this is not yet fully investigated.} have reached a substantial sophistication \cite{Alkofer:2000wg,Fischer:2006ub,Binosi:2009qm,Maas:2011se,Boucaud:2011ug,Pawlowski:2005xe,Gies:2006wv,Huber:2012td,Dudal:2012td,Huber:2016tvc,Cyrol:2014kca,Huber:2014tva,Blum:2014gna,Eichmann:2014xya,Cyrol:2016tym,Papavassiliou:2010qj,Boucaud:2017obn,Aguilar:2016vin,Binosi:2014kka}, and are also much better suited to probe extreme long-range physics while at the same time covering large hierarchies \cite{Fischer:2007pf} as well as real-time physics \cite{Strauss:2012as,Pawlowski:2015mia}, it is high time to solve this problem. At the current time, strictly speaking, all such comparisons, no matter how well the results agree, are still unjustified in the sense that it is not clear whether the quantities compared are indeed comparable, i.\ e.\ they are in the same gauge. Of course, it may be that they are already \cite{Fischer:2008uz,Maas:2011se}, but we do not yet know for sure. But, eventually, only agreement between multiple (not exact) methods will allow us to conclude that we understand the long-distance features of non-Abelian gauge theories.\\

\no{\bf Acknowledgments}\\

This work was supported by the DFG under grant numbers MA 3935/5-1, MA-3935/8-1 (Heisenberg program) and the FWF under grant number M1099-N16. Simulations were performed on the HPC clusters at the Universities of Jena and Graz. The author is grateful to the HPC teams for the very good performance of the clusters. The ROOT framework \cite{Brun:1997pa} has been used in this project.

\appendix

\section{Technical setup}\label{a:tech}

The simulations have been performed for the SU(2) Wilson action of $d=2,3,4$-dimensional Yang-Mills theories on a lattice of size $N^d$ with bare gauge coupling $\beta$, using a mixture of overrelaxation and heat-bath sweeps \cite{Cucchieri:2006tf}. The lattice spacing is set by assigning the string tension a value of $(440$ MeV$)^2$, as described in \cite{Maas:2014xma}. The employed sets of lattice parameters and number of configurations are listed in table \ref{tcgf}.

\begin{longtable}{|c|c|c|c|c|c|c|c|}
\caption{\label{tcgf}Number and parameters of the configurations used, ordered by dimension, lattice spacing, and physical volume. $p_\text{min}$ is the smallest momentum on the given lattice, and thus the one at which the ghost propagator has been evaluated to determine $b$. In all cases $2(10N+100(d-1))$ thermalization sweeps and $2(N+10(d-1))$ decorrelation sweeps of mixed updates \cite{Cucchieri:2006tf} have been performed, and auto-correlation times of the plaquette have been monitored to be at or below one sweep. The number of configurations were selected such as to have a reasonable small statistical error for the ghost propagator at the momenta used for extracting the effective infrared exponent. The number $N_r$ was chosen such that, given the results from lattices with smaller physical volumes and/or coarser discretizations, the total fraction of identified genuine Gribov copies should be substantial. The number of configurations had to be also chosen large enough such that so-called exceptional gauge orbits with particular extreme Gribov copies \cite{Cucchieri:2006tf,Sternbeck:2005vs,Maas:2015nva}, i.\ e.\ with very large coordinates, were (marginally) sufficiently sampled as well. In total ${\cal O}(10^5)$ configurations have been obtained and ${\cal O}(10^7)$ gauge-fixings have been performed. Entries, where the number of configurations carry an asterisk were statistically not sufficiently well behaved to include them in the main text, but which were nonetheless included in section \ref{s:artifacts} to probe lattice artifacts over a wider range.}\\
\hline
$d$	& $N$	& $\beta$	& $a^{-1}$ [MeV]	& L [fm]	& $p_\text{min}$ [MeV]	& $N_r$	& config.	\endfirsthead
\hline
\multicolumn{8}{|l|}{Table \ref{tcgf} continued}\\
\hline
$d$	& $N$	& $\beta$	& $a^{-1}$ [MeV]	& L [fm]	& $p_\text{min}$ [MeV]	& $N_r$	& config.	\endhead
\hline
\multicolumn{8}{|r|}{Continued on next page}\\
\hline\endfoot
\endlastfoot
\hline
2	& 92	& 6.23		& 863			& 21		& 58.9			& 21	& 2761		\cr
\hline
2	& 106	& 6.33		& 870			& 24		& 51.6			& 22	& 7893		\cr
\hline
2	& 80	& 6.40		& 875			& 18		& 68.7			& 20	& 2634		\cr
\hline
2	& 58	& 6.45		& 879			& 13		& 95.2			& 20	& 2220		\cr
\hline
2	& 18	& 6.55		& 886			& 4.0		& 299			& 20	& 1720		\cr
\hline
2	& 122	& 6.6		& 890			& 27		& 45.8			& 23	& 2664*		\cr
\hline
2	& 34	& 6.64		& 893			& 7.5		& 165			& 20	& 1590		\cr
\hline
2	& 68	& 6.64		& 893			& 15		& 82.5			& 20	& 2503		\cr
\hline
2	& 10	& 6.68		& 895			& 2.2		& 553			& 20	& 2234		\cr
\hline
2	& 50	& 6.68		& 895			& 11		& 112			& 20	& 2107		\cr
\hline
2	& 26	& 6.72		& 898			& 5.7		& 216			& 20	& 1320		\cr
\hline
2	& 42	& 6.73		& 899			& 9.2		& 134			& 20	& 1841		\cr
\hline
2	& 106	& 8.13		& 994			& 21		& 58.9			& 22	& 2478		\cr
\hline
2	& 122	& 8.24		& 1000			& 24		& 51.5			& 23	& 2664*		\cr
\hline
2	& 92	& 8.33		& 1010			& 18		& 69.0			& 21	& 5166		\cr
\hline
2	& 68	& 8.70		& 1030			& 13		& 95.2			& 20	& 4624		\cr
\hline
2	& 58	& 8.83		& 1040			& 11		& 113			& 20	& 2220		\cr
\hline
2	& 80	& 9.03		& 1050			& 15		& 82.4			& 20	& 2388		\cr
\hline
2	& 50	& 9.36		& 1070			& 9.2		& 134			& 20	& 2107		\cr
\hline
2	& 42	& 9.91		& 1100			& 7.5		& 164			& 20	& 1841		\cr
\hline
2	& 122	& 10.6		& 1140			& 21		& 58.7			& 23	& 2806		\cr
\hline
2	& 106	& 10.9		& 1160			& 18		& 68.7			& 22	& 2478		\cr
\hline
2	& 34	& 11.1		& 1170			& 5.7		& 216			& 20	& 1590		\cr
\hline
2	& 92	& 11.7		& 1200			& 15		& 81.9			& 21	& 2489		\cr
\hline
2	& 80	& 11.8		& 1210			& 13		& 95.0			& 20	& 4865		\cr
\hline
2	& 68	& 11.9		& 1210			& 11		& 112			& 20	& 2780		\cr
\hline
2	& 58	& 12.4		& 1240			& 9.2		& 134			& 20	& 2331		\cr
\hline
2	& 26	& 13.1		& 1280			& 4.0		& 309			& 20	& 1320		\cr
\hline
2	& 50	& 13.8		& 1310			& 5.7		& 165			& 20	& 2107		\cr
\hline
2	& 142	& 14.2		& 1330			& 21		& 58.8			& 24	& 3264		\cr
\hline
2	& 122	& 14.3		& 1340			& 18		& 69.0			& 23	& 1536		\cr
\hline
2	& 92	& 15.5		& 1390			& 13		& 94.9			& 21	& 2522		\cr
\hline
2	& 106	& 15.5		& 1390			& 15		& 82.4			& 22	& 2478		\cr
\hline
2	& 80	& 16.3		& 1430			& 11		& 112			& 20	& 4980		\cr
\hline
2	& 42	& 16.8		& 1450			& 5.7		& 217			& 20	& 1829		\cr
\hline
2	& 68	& 16.9		& 1460			& 9.2		& 135			& 20	& 2385		\cr
\hline
2	& 58	& 18.4		& 1520			& 7.5		& 165			& 20	& 8929		\cr
\hline
2	& 142	& 19.2		& 1550			& 18		& 68.6			& 24	& 4658		\cr
\hline
2	& 122	& 20.3		& 1600			& 15		& 82.4			& 23	& 2833		\cr
\hline
2	& 106	& 20.4		& 1600			& 13		& 94.9			& 22	& 2814		\cr
\hline
2	& 18	& 20.6		& 1610			& 2.2		& 559			& 20	& 1720		\cr
\hline
2	& 92	& 21.5		& 1650			& 11		& 113			& 21	& 2583		\cr
\hline
2	& 34	& 22.2		& 1670			& 4.0		& 308			& 20	& 1590		\cr
\hline
2	& 80	& 23.2		& 1710			& 9.2		& 134			& 20	& 2370		\cr
\hline
2	& 50	& 23.6		& 1730			& 5.7		& 217			& 20	& 2103		\cr
\hline
2	& 68	& 25.2		& 1790			& 7.5		& 165			& 20	& 2660		\cr
\hline
2	& 164	& 25.4		& 1790			& 18		& 68.6			& 25	& 1658*		\cr
\hline
2	& 122	& 26.9		& 1850			& 13		& 95.3			& 23	& 1625		\cr
\hline
2	& 142	& 27.4		& 1860			& 15		& 82.3			& 24	& 2927		\cr
\hline
2	& 106	& 28.4		& 1900			& 11		& 113			& 22	& 2508		\cr
\hline
2	& 92	& 30.5		& 1970			& 9.2		& 135			& 21	& 2694		\cr
\hline
2	& 58	& 31.6		& 2000			& 5.7		& 217			& 20	& 2260		\cr
\hline
2	& 42	& 33.6		& 2070			& 4.0		& 309			& 20	& 1814		\cr
\hline
2	& 80	& 34.7		& 2100			& 7.5		& 165			& 20	& 2466		\cr
\hline
2	& 142	& 36.3		& 2150			& 13		& 95.1			& 24	& 3758		\cr
\hline
2	& 122	& 37.4		& 2180			& 11		& 112			& 23	& 2807		\cr
\hline
2	& 106	& 40.4		& 2270			& 9.2		& 135			& 22	& 2604		\cr
\hline
2	& 68	& 43.2		& 2350			& 5.7		& 217			& 20	& 2712		\cr
\hline
2	& 92	& 45.7		& 2420			& 7.5		& 165			& 21	& 2520		\cr
\hline
2	& 26	& 46.5		& 2440			& 2.2		& 588			& 20	& 1320		\cr
\hline
2	& 50	& 47.4		& 2460			& 4.0		& 309			& 20	& 2102		\cr
\hline
2	& 164	& 48.3		& 2480			& 13		& 95.0			& 25	& 2760*		\cr
\hline
2	& 142	& 50.6		& 2540			& 11		& 112			& 24	& 2720		\cr
\hline
2	& 122	& 53.3		& 2610			& 9.2		& 134			& 23	& 2592		\cr
\hline
2	& 80	& 59.7		& 2760			& 5.7		& 217			& 20	& 2650		\cr
\hline
2	& 106	& 60.5		& 2780			& 7.5		& 165			& 22	& 2546		\cr
\hline
2	& 58	& 63.7		& 2860			& 4.0		& 310			& 20	& 3224		\cr
\hline
2	& 142	& 72.1		& 3040			& 9.2		& 135			& 24	& 2833		\cr
\hline
2	& 34	& 72.3		& 3040			& 2.2		& 561			& 20	& 1590		\cr
\hline
2	& 92	& 78.8		& 3180			& 5.7		& 217			& 21	& 2650		\cr
\hline
2	& 122	& 80		& 3200			& 7.5		& 165			& 23	& 2925		\cr
\hline
2	& 68	& 87.3		& 3350			& 4.0		& 309			& 20	& 2790		\cr
\hline
2	& 164	& 96		& 3510			& 9.2		& 134			& 25	& 2774*		\cr
\hline
2	& 106	& 104		& 3650			& 5.7		& 216			& 22	& 2574		\cr
\hline
2	& 42	& 110		& 3760			& 2.2		& 562			& 20	& 1787		\cr
\hline
2	& 80	& 120		& 3930			& 4.0		& 309			& 20	& 2656		\cr
\hline
2	& 122	& 138		& 4210			& 5.7		& 217			& 22	& 2833		\cr
\hline
2	& 50	& 155		& 4470			& 2.2		& 561			& 20	& 2222		\cr
\hline
2	& 92	& 159		& 4520			& 4.0		& 309			& 21	& 2606		\cr
\hline
2	& 142	& 187		& 4910			& 5.7		& 217			& 24	& 2684*		\cr
\hline
2	& 58	& 209		& 5190			& 2.2		& 562			& 20	& 3236		\cr
\hline
2	& 106	& 211		& 5210			& 4.0		& 309			& 22	& 2679		\cr
\hline
2	& 122 	& 280		& 6010			& 4.0		& 310			& 23	& 2624		\cr
\hline
2	& 68	& 287		& 6090			& 2.2		& 563			& 20	& 2940		\cr
\hline
2	& 142	& 379		& 6990			& 4.0		& 309			& 24	& 2684		\cr
\hline
2	& 80	& 398		& 7160			& 2.2		& 562			& 20	& 2646		\cr
\hline
2	& 92	& 526		& 8240			& 2.2		& 563			& 21	& 2557		\cr
\hline
2	& 106	& 698		& 9490			& 2.2		& 562			& 22	& 2520		\cr
\hline
2	& 122	& 925		& 10900			& 2.2		& 561			& 23	& 2623		\cr
\hline
\hline
3	& 8	& 3.40		& 874			& 1.8		& 669			& 20	& 2110		\cr
\hline
3	& 14	& 3.44		& 887			& 3.1		& 395			& 20	& 1650		\cr
\hline
3	& 20	& 3.46		& 894			& 4.4		& 280			& 22	& 1400		\cr
\hline
3	& 26	& 3.47		& 897			& 5.7		& 216			& 39	& 3795		\cr
\hline
3	& 36	& 3.47		& 897			& 7.9		& 156			& 66	& 1688		\cr
\hline
3	& 42	& 3.47		& 897			& 9.2		& 134			& 72	& 11798*	\cr
\hline
3	& 32	& 3.48		& 900			& 7.0		& 176			& 59	& 1627		\cr
\hline
3	& 36	& 3.82		& 1010			& 7.0		& 176			& 66	& 1491		\cr
\hline
3	& 42	& 3.92		& 1070			& 7.9		& 160			& 72	& 1583		\cr
\hline
3	& 32	& 4.10		& 1100			& 5.7		& 216			& 59	& 2791		\cr
\hline
3	& 42	& 4.33		& 1180			& 7.0		& 176			& 71	& 1447		\cr
\hline
3	& 26	& 4.28		& 1160			& 4.4		& 280			& 39	& 1291		\cr
\hline
3	& 48	& 4.38		& 1200			& 7.9		& 157			& 73	& 1590		\cr
\hline
3	& 36	& 4.52		& 1240			& 5.7		& 216			& 66	& 1496		\cr
\hline
3	& 20	& 4.60		& 1270			& 3.1		& 397			& 22	& 1380		\cr
\hline
3	& 54	& 4.83		& 1340			& 7.9		& 156			& 74	& 1774		\cr
\hline
3	& 48	& 4.84		& 1350			& 7.0		& 177			& 72	& 1638*		\cr
\hline
3	& 32	& 5.09		& 1430			& 4.4		& 280			& 57	& 2744		\cr
\hline
3	& 42	& 5.15		& 1450			& 5.7		& 217			& 71	& 1621		\cr
\hline
3	& 14	& 5.39		& 1530			& 1.8		& 680			& 20	& 1720		\cr
\hline
3	& 36	& 5.64		& 1610			& 4.4		& 281			& 63	& 1633		\cr
\hline
3	& 26	& 5.76		& 1650			& 3.1		& 398			& 30	& 1334		\cr
\hline
3	& 48	& 5.78		& 1660			& 5.7		& 217			& 72	& 1963*		\cr
\hline
3	& 42	& 6.45		& 1880			& 4.4		& 281			& 64	& 1707		\cr
\hline
3	& 32	& 6.91		& 2030			& 3.1		& 398			& 40	& 1585		\cr
\hline
3	& 48	& 7.27		& 2150			& 4.4		& 281			& 70	& 1535		\cr
\hline
3	& 20	& 7.39		& 2190			& 1.8		& 685			& 20	& 1450		\cr
\hline
3	& 36	& 7.69		& 2290			& 3.1		& 399			& 45	& 1478		\cr
\hline
3	& 54	& 8.08		& 2420			& 4.4		& 282			& 71	& 1674		\cr
\hline
3	& 42	& 8.84		& 2670			& 3.1		& 399			& 46	& 1592		\cr
\hline
3	& 60	& 8.89		& 2680			& 4.4		& 281			& 72	& 978*		\cr
\hline
3	& 26	& 9.38		& 2840			& 1.8		& 685			& 20	& 1315		\cr
\hline
3	& 48	& 10.0		& 3050			& 3.1		& 399			& 47	& 2037		\cr
\hline
3	& 54	& 11.1		& 3410			& 3.1		& 397			& 48	& 1766		\cr
\hline
3	& 32	& 11.3		& 3480			& 1.8		& 682			& 20	& 1417		\cr
\hline
3	& 36	& 12.7		& 3940			& 1.8		& 687			& 20	& 1370		\cr
\hline
3	& 42	& 14.6		& 4570			& 1.8		& 683			& 20	& 1699		\cr
\hline
3	& 48	& 16.6		& 5220			& 1.8		& 683			& 20	& 1877		\cr
\hline
3	& 54	& 18.6		& 5880			& 1.8		& 682			& 21	& 1832		\cr
\hline
3	& 60	& 20.6		& 6540			& 1.8		& 685			& 22	& 2044		\cr
\hline
3	& 66	& 22.6		& 7200			& 1.8		& 685			& 23	& 4479*		\cr
\hline
\hline
4	& 14	& 2.179		& 889			& 3.1		& 396			& 27	& 1082		\cr
\hline
4	& 10	& 2.181		& 894			& 2.2		& 553			& 20	& 1450		\cr
\hline
4	& 22	& 2.1850	& 902			& 4.8		& 286			& 104	& 1387		\cr
\hline
4	& 6	& 2.188		& 908			& 1.3		& 908			& 20	& 1620		\cr
\hline
4	& 18	& 2.188		& 908			& 3.9		& 315			& 54	& 1505		\cr
\hline
4	& 22	& 2.268		& 1110			& 3.9		& 317			& 108	& 1502*		\cr
\hline
4	& 18	& 2.279		& 1140			& 3.1		& 396			& 54	& 2248		\cr
\hline
4	& 14	& 2.311		& 1250			& 2.2		& 556			& 27	& 1033		\cr
\hline
4	& 22	& 2.349		& 1400			& 3.1		& 398			& 76	& 1488		\cr
\hline
4	& 10	& 2.376		& 1520			& 1.3		& 939			& 20	& 1450		\cr
\hline
4	& 18	& 2.395		& 1610			& 2.2		& 559			& 40	& 1258		\cr
\hline
4	& 26	& 2.403		& 1680			& 3.1		& 406			& 77	& 1565		\cr
\hline
4	& 22	& 2.457		& 1960			& 2.2		& 558			& 50	& 1242		\cr
\hline
4	& 14	& 2.480		& 2120			& 1.3		& 943			& 20	& 1225		\cr
\hline
4	& 26	& 2.507		& 2330			& 2.2		& 563			& 54	& 1392		\cr
\hline
4	& 30	& 2.548		& 2680			& 2.2		& 561			& 55	& 1479*		\cr
\hline
4	& 18	& 2.552		& 2720			& 1.3		& 945			& 20	& 1175		\cr
\hline
4	& 22	& 2.609		& 3330			& 1.3		& 948			& 20	& 1355		\cr
\hline
4	& 26	& 2.656		& 3930			& 1.3		& 947			& 21	& 1335		\cr
\hline
4	& 30	& 2.698		& 4540			& 1.3		& 951			& 22	& 1149*		\cr
\hline
\end{longtable}

\bibliographystyle{bibstyle}
\bibliography{bib}

\end{document}